\documentclass[%
 aip, cha,
 amsmath,amssymb,
reprint,%
superscriptaddress,longbibliography,floatfix,showkeys,nofootinbib
]{revtex4-2}

\usepackage{graphicx}
\usepackage{dcolumn}
\usepackage{bm}

\usepackage{booktabs}
\usepackage[utf8]{inputenc}
\usepackage[T1]{fontenc}
\usepackage{mathptmx}
\usepackage{etoolbox}
\usepackage[normalem]{ulem}
\usepackage{mathtools}

\makeatletter
\def\@email#1#2{%
 \endgroup
 \patchcmd{\titleblock@produce}
  {\frontmatter@RRAPformat}
  {\frontmatter@RRAPformat{\produce@RRAP{*#1\href{mailto:#2}{#2}}}\frontmatter@RRAPformat}
  {}{}
}%
\makeatother

\usepackage[dvipsnames]{xcolor}
\usepackage{amsbsy}
\usepackage{tikz}
\usepackage{amsthm,mathrsfs,amsopn,amsfonts,amsmath,amssymb}

\usepackage[justification=RaggedRight]{caption}

\AtBeginEnvironment{matrix}{\everymath{\displaystyle}}

\usepackage{url}

\usepackage[colorlinks=true,linkcolor=blue,citecolor=blue,urlcolor=blue]{hyperref}

\usepackage[capitalise]{cleveref}
\usepackage{comment}

\newcommand{\T}{^{\mathsf{T}}}
\newcommand{\R}[1]{\mathrm{#1}}
\newcommand{\B}[1]{\if#1\relax\bm{#1}\else\mathbf{#1}\fi} 
\newcommand{\C}[1]{\mathcal{#1}}

\newcommand{\BB}[1]{\mathbb{#1}}

\newcommand{\abs}[1]{\left\lvert #1 \right\rvert}

\begin{document}

\title{When higher-order interactions enhance synchronization: the case of the Kuramoto model}
\author{Riccardo Muolo}
\affiliation{RIKEN Center for Interdisciplinary Theoretical and Mathematical Sciences (iTHEMS), Saitama 351-0198, Japan}
\affiliation{Department of Systems and Control Engineering, Institute of Science Tokyo, Tokyo 152-8552, Japan}
\author{Hiroya Nakao}
\affiliation{Department of Systems and Control Engineering, Institute of Science Tokyo, Tokyo 152-8552, Japan}
\affiliation{Research Center for Autonomous Systems Materialogy, Institute of Science Tokyo, Yokohama 226-8501, Japan}
\author{Marco Coraggio}\email{m.coraggio@ssmeridionale.it}
\affiliation{Scuola Superiore Meridionale, Naples 80138, Italy}

\date{\today}

\begin{abstract}
Synchronization is a fundamental phenomenon in complex systems, observed across a wide range of natural and engineered contexts. 
The Kuramoto model provides a foundational framework for understanding synchronization among coupled oscillators, traditionally assuming pairwise interactions. 
However, many real-world systems exhibit group and many-body interactions, which can be effectively modeled through hypergraphs.
Here we show that the effect of such higher-order interactions on synchronization is non-monotonic.
Through a numerical study of higher-order Kuramoto models on random hypergraphs and on globally coupled systems, we find that the degree of synchronization reached from incoherent initial conditions is maximized at a small but nonzero higher-order coupling strength: weak higher-order interactions enhance synchronization when added to pairwise ones, whereas strong ones work against it, in line with earlier reports of reduced basins and of cluster states.
We further show, through a cost-constrained allocation analysis, that under a constrained budget for interactions a mixed allocation of pairwise and higher-order couplings consistently achieves higher synchronization than relying on  either type alone.
These findings clarify the role of higher-order interactions in shaping collective dynamics and point to design principles for optimizing synchronization in complex systems.
\end{abstract}

\maketitle 

\begin{quotation}
The effect of higher-order interactions on synchronization in the Kuramoto model is non-monotonic.
Starting from incoherent states, we find that the degree of synchronization reached is maximized at a small but nonzero higher-order coupling strength.
We observe this in both random hypergraphs and globally coupled systems, indicating the effect is neither a finite-size artifact nor specific to a particular topology.
We further find that, under a constrained budget for interactions, a mixture of pairwise and higher-order couplings achieves higher synchronization than either interaction type alone, which points to a design principle for engineered oscillator networks.
\end{quotation}

\section{Introduction}
\label{sec:introduction}

Synchronization dynamics is one of the most studied phenomena in the field of complex systems and a paradigmatic example of self-organizing behavior~\cite{pikovsky2001synchronization,arenas2008synchronization,boccaletti2018synchronization}. 
The synchronization capabilities of coupled oscillators were first observed by Christiaan Huygens, who, in the 17th century, noticed that pendulum clocks suspended from a common support tend to synchronize in anti-phase. 
Later research highlighted the occurrence and relevance of synchronization in many natural and artificial systems, ranging from fireflies' blinking and frogs' croaking to rhythmic contraction in cardiac cells and neuronal activity, and from  bridge oscillations to power angles in power grids~\cite{Strogatzbooksync}.

The key factors enabling this collective behavior are the interactions among the oscillators, which can be, e.g., mechanical (bridges, pendulum clocks), electrical (heart, power grids), visual (fireflies), or acoustic (frogs)~\cite{pikovsky2001synchronization,Strogatzbooksync,coraggio2026controlling}.
Indeed, first Winfree recognized that such interactions could be thought of as a perturbation~\cite{winfree1967biological}, then Kuramoto~\cite{kuramoto1975} obtained a simple model in which synchronization emerges as a phase transition triggered by the variation in the coupling strength: the celebrated Kuramoto model~\cite{strogatz2000kuramoto,acebron2005kuramoto}.
In the classic Kuramoto model, two assumptions are made which do not necessarily hold in applications.
First, interactions are assumed to be \emph{pairwise}, i.e., each one involving exactly two oscillators.
Second, an \emph{all-to-all} interaction topology is assumed, where each oscillator interacts with all others with the same strength. 

However, research has shown that, when different complex topologies are considered, the dynamics of the Kuramoto model is much richer, displaying complex patterns and transitions~\cite{rodrigues2016kuramoto}, generally making synchronization more difficult to achieve. 
This shift is motivated by applications, where all-to-all topologies are actually rare, and the interactions are modeled through networks~\cite{newmanbook2,latora_nicosia_russo_2017}.
Despite its versatility, the network approach has the limitation of considering only pairwise interactions. 
Indeed, many-body (rather than pairwise) couplings have been found to be better suited to describe certain kinds of interactions, such as those occurring in social sciences, ecology, and neuroscience~\cite{battiston2020networks,bianconi2021higher,natphys,bick2023higher,boccaletti2023structure,muolo2024turing,millan2025topology}.
Mathematically, such many-body interactions can be modeled via \emph{hypergraphs} and \emph{simplicial complexes}, which are extensions of networks (i.e., graphs).

Over the past few years, it has been shown that higher-order interactions enrich system dynamics, with notable applications in the synchronization of chaotic oscillators~\cite{krawiecki2014chaotic,gambuzza2021stability,gallo2022synchronization,della2023emergence}, chimera states~\cite{kundu2022high,muolo2024phase,djeudjo2026chimera}, swarming and active matter~\cite{anwar2024collective,anwar2025two,leon2025collective}, random walks~\cite{schaub2020random,carletti2020random}, pattern formation~\cite{carletti2020dynamical,muolo2023turing,gao2023turing}, opinion dynamics~\cite{iacopini2019simplicial,neuhauser2020multibody,deville2020consensus}, and pinning control~\cite{de2022pinning,xia2024pinning,muolo2025pinning}, to name a few.
In particular, extensive literature reports a generally detrimental effect of higher-order interactions on synchronization in the Kuramoto model~\cite{tanaka2011multistable,bick2016chaos,skardal2019abrupt,skardal2020higher,millan2020explosive,lucas2020multiorder,adhikari2023synchronization,skardal2023multistability,carballosa2023cluster,leon2024,costa2024bifurcations,huh2024critical,wang2024coexistence,smith2024determining,dai2025higher}. 
Zhang et al.~\cite{zhang2024deeper} showed that higher-order interactions make the attraction basin of the synchronous state smaller but more robust (deeper).
These findings were further supported in Ref.~\onlinecite{fariello2024third}, showing that the critical coupling for synchronization increases, facilitating desynchronization, and in Refs.~\onlinecite{von2024higher,wang2025higher}, where it was demonstrated that, once achieved, synchronization and twisted states become harder to disrupt due to higher-order interactions.
Overall, while it is well known that stronger pairwise interactions enhance the degree of synchronization of phase oscillators in the purely pairwise case~\cite{Kuramoto_book,acebron2005kuramoto,rodrigues2016kuramoto},
the role of higher-order interactions remains far from fully understood.

These recent findings motivate our first research goal, which is to quantify the effects of higher-order interactions on the synchronous state for more general hypergraph topologies. As in previous works, our framework considers the addition of higher-order interactions to the pairwise setting.
To this end, we studied Kuramoto oscillators coupled through $2$- and $3$-body interactions and conducted a numerical study on random hypergraphs of $10$ and $100$ nodes,
a regime relevant for real-world applications~\cite{menara2022functional,simpson2013synchronization,alderisio2017interaction}.
Our analysis confirms that higher-order interactions generally enhance synchronization when the initial conditions are close to the synchronous state; however, incoherent initial states reveal a non-monotonic dependence on the higher-order coupling strength. While the average coherence reached from random incoherent initial conditions decreases at large higher-order coupling strength---consistent with the smaller basins reported in Ref.~\onlinecite{zhang2024deeper}---we find that it first \emph{increases}, so that the system synchronizes best for a small but nonzero higher-order coupling.
This is further validated for the globally (i.e., all-to-all) coupled higher-order Kuramoto model of large size (i.e., $10\,000$ oscillators), showing that this is not a finite-size effect.  

Our second research question is: given a limited amount of resources for connectivity of both pairwise and higher-order interactions, which is the optimal combination to enhance synchronization?
With this investigation, we aim to determine whether higher-order interactions can offer advantages over purely pairwise ones, and whether a mix of both can outperform structures relying exclusively on one type.
This question is relevant in both engineered and natural systems: in the former, it can guide resource allocation for building synchronizable systems; in the latter, it may help explain the interaction patterns that emerge in nature as evolved or self-organized solutions to synchronization demands.

While the problem of optimally allocating links to promote specific collective behaviors, and in particular synchronization, has been extensively studied in complex networks, only a few studies have explored the role of higher-order interactions~\cite{lamata2025hyperedge}.
Much work focused on identifying metrics that promote or inhibit synchronization, showing, e.g., that factors contributing to good synchronizability are low \emph{eigenratio}~\cite{pecora1998master}, high \emph{algebraic connectivity}~\cite{coraggio2018synchronization}, or high \emph{minimum density}~\cite{coraggio2021convergence,coraggio2020distributed}. 
General network structural features also play a critical role: for example, small-world networks synchronize more effectively than random graphs~\cite{barahona2002synchronization}.
Notably, Donetti et al.~\cite{donetti2005entangled,donetti2006optimal} observed that optimally synchronizable networks tend to exhibit ``entangled'' homogeneous structures, while Skardal et al.~\cite{Skardal2014Optimal} found that matching between frequency heterogeneity and network heterogeneity enhances synchronization.
Several approaches have also been proposed, from a network synthesis perspective, to optimally design synchronizable networks, both for oscillators ensembles~\cite{fazlyab2017optimal,lei2023new} and for general dynamical systems~\cite{nishikawa2006synchronization,estrada2010design}; see also Ref.~\onlinecite{coraggio2026controlling} and references therein. 
Importantly, it has been shown that optimal network topologies can vary significantly when individual node dynamics are considered, highlighting a complex interplay between structure and dynamics, and, consequently, the coupling laws governing the interactions~\cite{gorochowski2010evolving,coraggio2024data}.

Many of the mentioned studies focus on chaotic oscillators, whose synchronization is analyzed using the \emph{Master Stability Function}, a linear stability analysis technique first proposed by Fujisaka and Yamada~\cite{fujisaka1983stability} and later extended to general networks by Pecora and Carroll~\cite{pecora1998master}, who also coined its current name.
This approach, however, is not suited to coupled phase oscillators such as the Kuramoto model, as, being a local stability criterion, it provides no information on whether synchronization can emerge from incoherent initial conditions.
On the other hand, the method exploited in Ref.~\onlinecite{coraggio2024data} can also be applied to phase oscillators and partly inspired the numerical approach we employed to answer this question. 
Our analysis on random hypergraphs shows that, when the total budget for interactions (pairwise and higher-order) is limited, synchronization is enhanced by a combination of both types of interactions, regardless of the relative cost of higher-order interactions within that budget. 

\begin{figure*}[tb]
    \centering
    \includegraphics[scale=0.45]{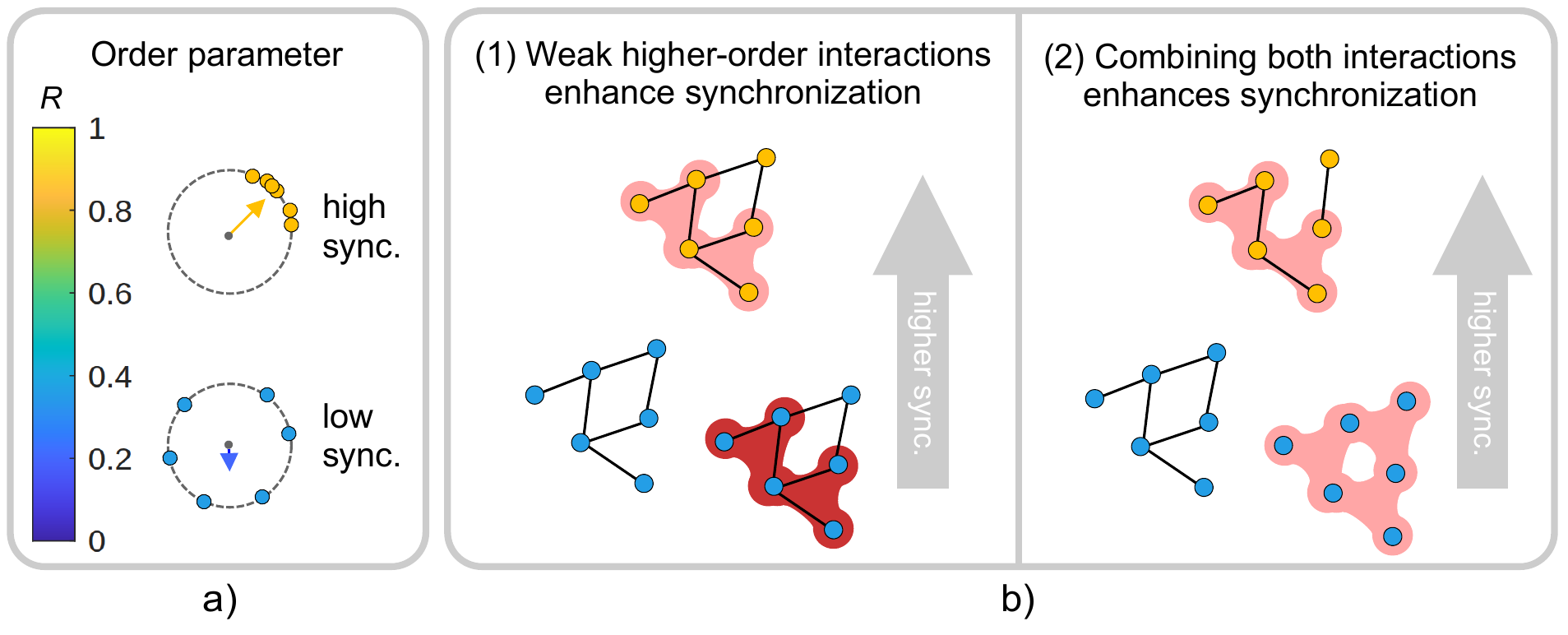}
    \caption{%
    a) Meaning of the order parameter $R$: high values correspond to synchronized states, whereas low values indicate incoherent dynamics.    
    b) Pictorial representation of the main findings of this work.
    (1) The effect of higher-order interactions on synchronization is non-monotonic: adding \emph{weak} higher-order interactions to pairwise-coupled networks enhances synchronization, whereas strong ones hamper it.
    (2) Under a finite budget for interactions, regardless of the relative cost of higher-order interactions, the optimal configuration for synchronization {generally} involves a combination of pairwise and higher-order interactions. 
    Red (resp.~pink) hyperedges denote higher-order interactions with larger (resp.~smaller) coupling strength.
    Yellow (resp.~blue) nodes indicate highly (resp.~weakly) synchronized states, in accordance with panel a).
    }
    \label{fig:main_findings}
\end{figure*}

The two main findings of this work---that (i) weak higher-order interactions enhance synchronization (i.e., increase the order parameter on average), when added to pairwise ones and that (ii) with a finite budget for connections, a combination of pairwise and higher-order interactions optimizes synchronization---are summarized in Figure~\ref{fig:main_findings}. 

The paper is structured as follows: in the next Section, we introduce the framework of higher-order interactions and the higher-order Kuramoto model. In Section~\ref{sec:results1a}, we present our first results, i.e., weak higher-order interactions enhance synchronization. We show this result first for small random hypergraphs, then for large all-to-all coupled systems, showing that it is not a finite-size effect. Finally, in Section~\ref{sec:results2}, we present our second result, i.e., under a finite budget for interactions, the optimal configuration for synchronization {generally} involves a combination of both pairwise and higher-order interactions, before concluding in Section~\ref{sec:conclusion}.

\section{Higher-order interactions and the higher-order Kuramoto model}
\label{sec:HO_Kuramoto}

We consider higher-order Kuramoto models, with interactions described by undirected, unweighted, unsigned hypergraphs.
A \emph{hypergraph} is defined as $\C{H} = (\C{V}, \C{E})$, where $\C{V} = \{1, \dots, N_0\}$ is the set of $N_0$ vertices and $\C{E}$ is the set of hyperedges.
A \emph{hyperedge} $e \in \C{E}$ is a subset of $\C{V}$, representing a $\abs{e}$-body interaction; for example, $e = \{4, 7, 8\}$ is a 3-body interaction between vertices $4$, $7$, and $8$.
Since $j$-body interactions correspond to $(j-1)$-simplices, following the notation used in the framework
of simplicial complexes\cite{bianconi2021higher} %
\footnote{When considering simplicial complexes, the order of the interaction is given by the dimension of the space.
For example, nodes are more formally called $0$-simplices, as they have dimension $0$, links are $1$-simplices, triangles are $2$-simplices, and so on; hence, a $3$-body interaction is encoded by a $2$-simplex.},
we refer to $j$-body interactions as \emph{$(j-1)$-hyperedges} and use the subscript or superscript $j-1$ to denote associated structures.
In this study, we focus on hypergraphs having only $1$-hyperedges (i.e., $2$-body interactions) and $2$-hyperedges (i.e., $3$-body interactions), and denote the number of $1$- and $2$-hyperedges as $N_1$ and $N_2$, respectively.

A hypergraph can be represented algebraically as follows.
The totality of $1$-hyperedges is represented by the \emph{first-order adjacency tensor} $A^{(1)} \in \{0, 1\}^{N_0 \times N_0}$ (also known as the \emph{adjacency matrix}), with $A_{jk}^{(1)} = 1$ if a $1$-hyperedge exists between vertices $j$ and $k$ (i.e., $\{j, k\} \in \C{E}$), or $A_{jk}^{(1)} = 0$ otherwise.
Similarly, the totality of $2$-hyperedges is represented by the \emph{second-order adjacency tensor} $A^{(2)} \in \{0, 1\}^{N_0 \times N_0 \times N_0}$, with $A_{jkl}^{(2)} = 1$ if a $2$-hyperedge exists between vertices $j$, $k$, and $l$ (i.e., $\{j, k, l\} \in \C{E}$), or $A_{jkl}^{(2)} = 0$ otherwise.
$1$-hyperedges can also be represented via the \emph{incidence matrix},%
\footnote{Also known as \emph{boundary operator} in the theory of simplicial complexes~\cite{bianconi2021higher}.}
denoted by $B^{(1)} \in \{-1, 0, 1\}^{N_0 \times N_1}$.
The $k$-th column of $B^{(1)}$ corresponds to the $k$-th $1$-hyperedge, say $\{l, m\}$; this column has $1$ at the $l$-th position, $-1$ at the $m$-th position, and zeros elsewhere.\footnote{Formally, assigning $1$ and $-1$ to the vertices of a $1$-hyperedge requires choosing an arbitrary orientation, that is, an ordering of the node pair. 
This choice does not affect our results.}

\begin{figure*}[tb]
\begin{center}
    \includegraphics[scale=0.4]{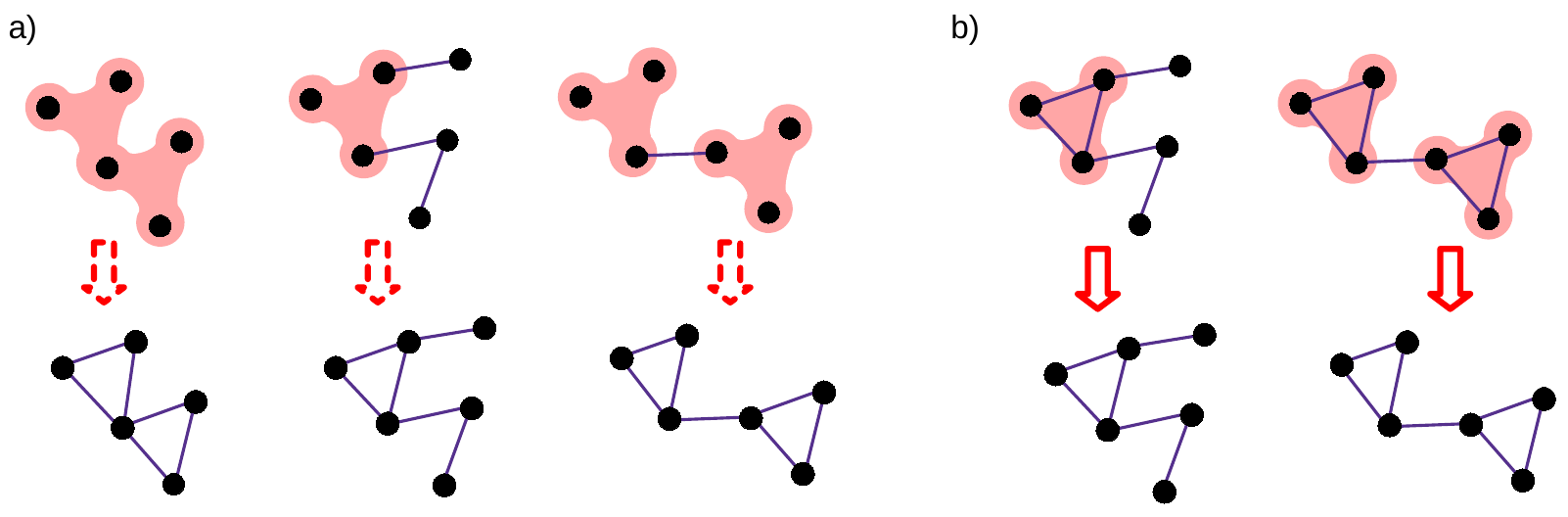}
    \caption{a) Examples of $\mathcal{H}$-connected hypergraphs, i.e., whose projected networks (bottom row) are connected. b) Examples of $1$-connected hypergraphs, i.e., whose underlying networks are already connected.}
    \label{fig:conn_hyper}
    \end{center}
\end{figure*}

We say that a hypergraph is \emph{$\C{H}$-connected} if there exists a path between any pair of vertices using an arbitrary combination of hyperedges of any order.
In contrast, we say that a hypergraph is \emph{$j$-connected} (with $j \in \BB{N}$) if there exists a path between any pair of vertices using only $j$-hyperedges; we use in particular the notion of $1$-connectedness (which corresponds to connection via links only). 
In the literature, a hypergraph that is $\C{H}$-connected is referred to simply as \emph{connected}~\cite{battiston2020networks,nortier2025higher}; however, we adopt this terminology to clearly distinguish it from the notion of $j$-connectedness, which is useful in the analyses presented in this work. These two types of connectivity are illustrated in Figure~\ref{fig:conn_hyper}.
Finally, the \emph{$j$-degree} (with $j \in \BB{N}$) of a vertex $v$, denoted by $d^{(j)}(v)$, is the number of $j$-hyperedges it is a part of.
$\langle d^{(j)} \rangle = \frac{1}{N_0} \sum_{v \in \C{V}} d^{(j)}(v)$ is the \emph{average $j$-degree}. 
Note that $\langle d^{(1)} \rangle$ (related to links) corresponds to the usual average degree of a network.

\subsection{Higher-order Kuramoto models}
\label{sec:methods_higher_order_kuramoto_models}

In the Kuramoto model, higher-order (non-pairwise) interactions naturally emerge from \emph{phase reduction}~\cite{nakao16,pietras2019network,monga2019phase}.
In general, phase reduction theory establishes that it is possible to meaningfully associate a phase oscillator dynamics to a dynamical system in a periodic regime (i.e., whose asymptotic solution is a stable limit cycle)~\cite{Kuramoto_book,Win80,kuramoto2019concept}.
For example, the Kuramoto model itself~\cite{kuramoto1975} can be derived from the complex Ginzburg-Landau equation~\cite{nakao2014complex}, which can be rewritten as a two-dimensional system admitting a stable limit cycle\footnote{Note that the complex Ginzburg-Landau equation (CGLE) is also known as Stuart-Landau equation in the discrete setting; in such a case one more frequently speaks of coupled Stuart-Landau oscillators~\cite{pereti2020stabilizing}.}. 
In high-dimensional networked systems with weak coupling, a similar phase reduction yields a system of phase oscillators interacting via the underlying network topology~\cite{nakao16}.
The classical Kuramoto model is obtained through a first order approximation with respect to the coupling parameter~\cite{kuramoto1975,Kuramoto_book}. 
However, when phase reduction is performed using approximations beyond the first order~\cite{ashwin2016hopf,leon19,gengel2020high,bick2024higher,fujii2026emergence}, non-pairwise interactions emerge.%
\footnote{The term \emph{higher-order} is used ambiguously in the literature, as it can refer both to non-pairwise interactions (many-body coupling) and to higher-order expansions in the coupling parameter. Here, we use \emph{higher-order} to mean non-pairwise interactions exclusively.}
As an example of the emergence of higher-order (i.e., non-pairwise) interactions at the second order, let us consider three oscillators, $j$, $k$ and $l$, coupled so that both $j$ and $l$ are connected to $k$ but not to each other. While first-order interactions exist only between $j$ and $k$, and $l$ and $k$, a small interaction term appears between $j$ and $l$, when second-order phase reduction is performed~\cite{bick2024higher}. These higher-order terms, emerging from a second order phase reduction, are typically small corrections, of the order of the squared coupling strength, and, thus, do not drastically alter the dynamics, but improve the accuracy of the reduced description~\cite{sakaguchi1990breakdown,leon19,mau2024phase}; nonetheless, they can have significant effects near bifurcation points~\cite{bick2024higher} or in phenomena such as remote synchronization~\cite{okuda1991mutual,bergner2012remote}.

The Kuramoto model with higher-order interactions (i.e., non-pairwise interactions of a comparable order as pairwise ones) has been extensively studied. However, these higher-order Kuramoto models were typically not derived via phase reduction from high-dimensional oscillators coupled through higher-order interactions.  Instead, they were constructed by adding higher-order coupling terms to the classical pairwise Kuramoto model~\cite{tanaka2011multistable,skardal2019abrupt,skardal2020higher,lucas2020multiorder}.
León et al.~\cite{leon2024} rigorously derived a higher-order Kuramoto model by means of a first order phase reduction starting from a system of Stuart-Landau oscillators coupled via pairwise and higher-order interactions that respect certain symmetry properties. They found that, for the dynamics of the $j$-th oscillator, the naturally emerging $3$-body interaction takes the form\footnote{More precisely, the derivation also produces a phase lag, as in the Kuramoto-Sakaguchi model~\cite{sakaguchi1986soluble}. 
However, since much of the literature focuses on behavior without phase lag, we focus on that setting here.} $\sin(\theta_k + \theta_l - 2 \theta_j)$. %
This result motivated our choice of the higher-order Kuramoto model, namely,
\begin{equation}\label{eq:ho_kura_leon}
\begin{split}
\dot{\theta}_j
&= \omega_j
+\frac{K_1}{\langle d^{(1)} \rangle}
\sum_{k=1}^{N_0}
A_{jk}^{(1)}
\sin(\theta_k-\theta_j)
\\
&\quad
+\frac{K_2}{2\langle d^{(2)} \rangle}
\sum_{k=1}^{N_0}
\sum_{l=1}^{N_0}
A_{jkl}^{(2)}
\sin(\theta_k+\theta_l-2\theta_j).
\end{split}
\end{equation}

Since the coupling here is not global, the interaction terms are normalized by the average $1$- and $2$-degrees rather than by $N_0$ and $N_0^2$ as in the globally coupled model\cite{leon2024}.

Synchronization can be measured via the Kuramoto order parameter $R = \abs{\frac{1}{N_0}\sum_{j=1}^{N_0} e^{i\theta_j}}$, where $R \approx 1$ indicates synchronization, and lower values of $R$ correspond to increasing incoherence.
To detect two-cluster synchronization, it is possible to use the Kuramoto-Daido order parameter $R_2 = \abs{\frac{1}{N_0}\sum_{j=1}^{N_0} e^{2i\theta_j}}$; in that case, $R$ would be low, while $R_2$ approaches $1$.

We sample natural frequencies from a uniform distribution in $[0, 0.3]$; analysis with a Gaussian distribution with mean $0.15$ and standard deviation $0.1$ yields qualitatively identical results to those presented in the Main Text and is reported in Appendix~\ref{app:C}. The choice is not inconsequential in general: in the thermodynamic limit with global coupling, unimodal distributions give the familiar second-order transition, whereas uniform distributions~\cite{pazo2005thermodynamic} (or unimodal ones with a central plateau~\cite{basnarkov2007phase}) give a first-order one.
This distinction however concerns the nature of the transition and does not affect the non-monotonic dependence on $K_2$ observed here in both settings.

\section{Non-monotonic effect of higher-order interactions}
\label{sec:results1a}

\begin{figure*}[tb]
    \centering
    \includegraphics[scale=1]{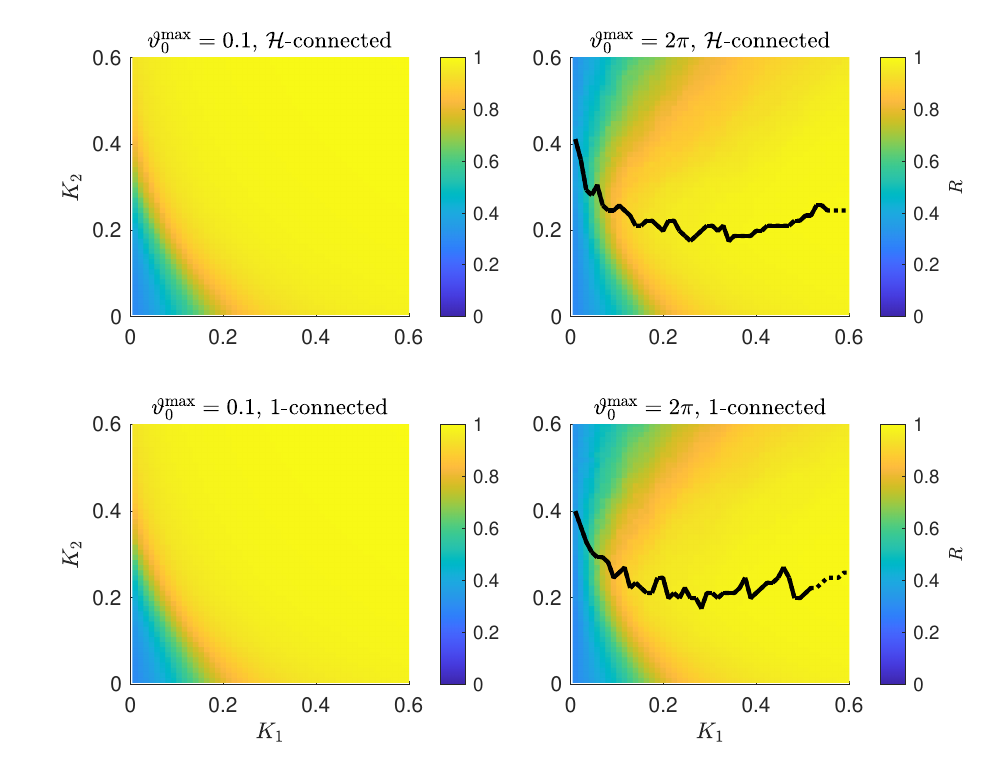}
    \caption{%
    Average Kuramoto order parameter $R$ as a function of the coupling strengths of pairwise ($K_1$) and higher-order $3$-body interactions ($K_2$), computed over $N_{\C{H}}=300$ randomly generated hypergraphs.
    The grid $K_1,K_2$ is $51\times51$.  
    Each hypergraph has $N_0=10$ nodes, $N_1 = 20$ links and $N_2 = 10$ triangles, with oscillator frequencies $\omega_{j}$ uniformly distributed in $[0, 0.3]$, and initial phases in $[0, \theta_0^\R{max}]$.
    Left panels show the cases of initial phases close to synchronization, i.e., $\theta_0^\R{max} = 0.1$;
    right panels show the case of incoherent initial states, i.e., $\theta_0^\R{max} = 2\pi$.
    On the right panels, the black line indicates, for each $K_1$, the $K_2$ yielding the maximum $R$: that value is never zero. {The line is dotted when the 95\textsuperscript{th}–5\textsuperscript{th} percentile spread over the maximization domain is below 0.05}.
    Top panels show the case of $\mathcal{H}$-connected hypergraphs, while bottom panels show the case of $1$-connected hypergraphs: we observe no significant differences between the two.}
    \label{fig:couplings_study_standard_coupl_R1}
\end{figure*}

\begin{figure*}[tb]
    \centering
    \includegraphics[scale=1]{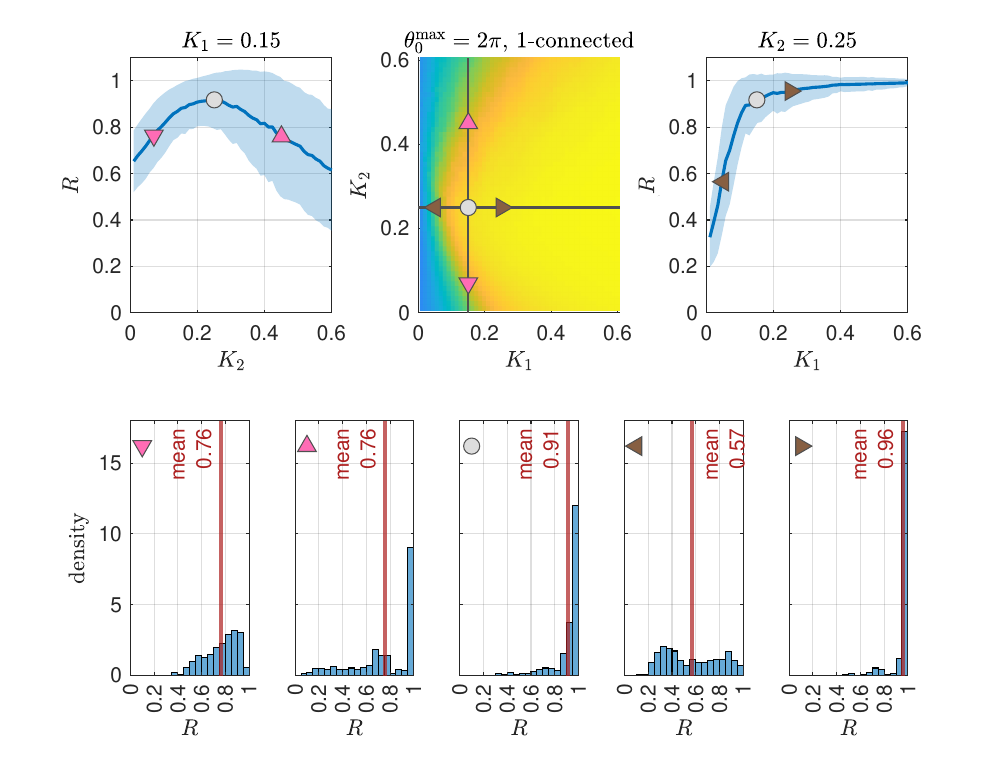}
    \caption{%
    System behavior with different pairwise ($K_1$) and higher-order ($K_2$) coupling strengths, for initially incoherent $1$-connected hypergraphs from Figure~\ref{fig:couplings_study_standard_coupl_R1}.
    The top-center panel replicates the bottom right panel of Figure~\ref{fig:couplings_study_standard_coupl_R1} for comparison (color is the mean order parameter $R$).
    The top-left (resp.~top-right) panel shows the mean value of $R$ (blue line) and its standard deviation (shaded area) as a function of the higher-order coupling strength $K_2$ (resp.~$K_1$) while keeping the pairwise interaction strength $K_1$ (resp.~$K_2$) fixed.
    The values of $K_1$, $K_2$ explored are also indicated in the top-center panel by a vertical (resp.~horizontal) black line.
    Pink (resp.~brown) triangles pointing upward/downward (resp.~leftward/rightward) mark representative pairs $(K_1, K_2)$.
    The bottom panels display the distributions (integral normalized to $1$) of $R$ across the  $N_{\C{H}} = 300$ realizations of hypergraphs, initial conditions, and frequencies.
    For fixed $K_1 = 0.15$, a small $K_2$ (pink downward triangle) produces $R$ values relatively concentrated around the mean, whereas larger $K_2$ (pink upward triangle) leads to a broader distribution, spanning incoherent and strongly coherent regimes.
    For fixed $K_2 = 0.25$, increasing $K_1$ (brown leftward and rightward triangles) causes the distribution to concentrate near $1$, indicating a consistently high level of synchronization.}
    \label{fig:distributions_order_parameter_uniform}
\end{figure*}

\begin{figure*}[tb]
    \centering
    \includegraphics[scale=1]{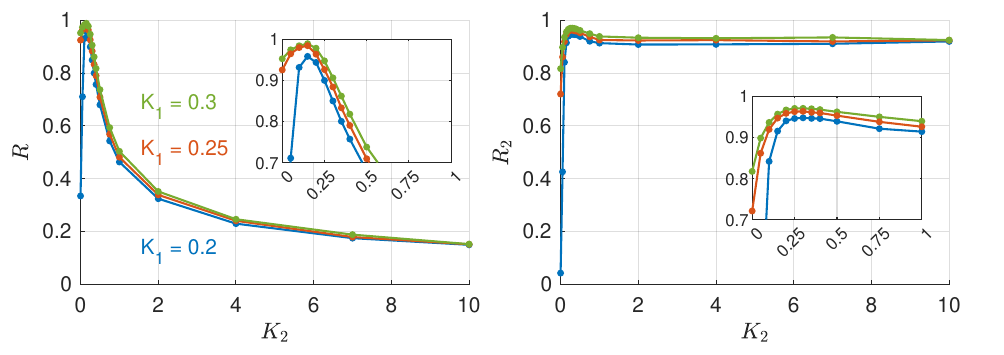}
    \caption{Time-averaged Kuramoto order parameter $R$ (left panel) and Kuramoto-Daido order parameter $R_2$ (right panel) as a function of the coupling strengths of higher-order $3$-body interactions ($K_2$), for different values of the pairwise coupling strength $K_1$ (blue: $K_1 = 0.2$; orange: $K_1 = 0.25$; green: $K_1 = 0.3$), computed over an all-to-all hypergraph with $N_0 = 10\,000$ nodes.
    Oscillator frequencies $\omega_{j}$ are uniformly distributed in $[0, 0.3]$; initial phases are uniformly distributed in $[0, 2\pi]$. We observe that both order parameters have a maximum for low, but non-zero, values of $K_2$.%
    }
    \label{fig:alltoall}
\end{figure*}

Our first question is how the strength of higher-order interactions affects the level of synchronization reached on general hypergraph topologies, and whether the answer depends on how far the initial state is from coherence. To address it, we conducted a systematic numerical study of the higher-order Kuramoto model.
Specifically, we first fixed the total number of nodes ($N_0$), links ($N_1$) and triangles ($N_2$) and then generated $N_\mathcal{H}$ connected hypergraphs randomly allocating all links and triangles.
We considered the two connectivity conditions defined in Section~\ref{sec:HO_Kuramoto}: \emph{$\C{H}$-connectivity}, if the projected network is connected, and \emph{$1$-connectivity}, if the structure is already connected solely with the links.
Then, we ran the simulations varying the coupling strengths of pairwise and higher-order interactions, $K_1$ and $K_2$.
For each pair, we performed $N_{\C{H}} = 300$ integrations over random hypergraphs of $N_0=10$ nodes, initial phases, and natural frequencies, averaging the resulting order parameter $R$ (see Appendix~\ref{app:B} for details). 
Simulations for $N_{\C{H}} = 100$ integrations over random hypergraphs of $N_0=100$ nodes give analogous results, which are presented in Appendix~\ref{app:C}.

\subsection{Adding weak higher-order interactions enhances synchronization}
\label{sec:adding_weak_ho_enhances}

The top panels of Figure~\ref{fig:couplings_study_standard_coupl_R1} show the case of $\C{H}$-connected hypergraphs, while the bottom ones portray the case of $1$-connected hypergraphs. 
We see no significant differences, indicating that the observed effects arise from higher-order interactions, rather than connectivity differences.
The left panels show the case in which the system starts close to the synchronized state.
Higher-order interactions enhance synchronization, as evidenced by the yellow regions (i.e., higher values of the order parameter $R$) widening as $K_2$ increases, in line with earlier reports\cite{zhang2024deeper,von2024higher}.
The right panels, instead, show the case of incoherent initial conditions.
Stronger higher-order interactions ($K_2$) worsen synchronization (lower $R$), as also found in Refs.~\onlinecite{zhang2024deeper,fariello2024third,muolo2025pinning}.
However, for small $K_2$ (weak higher-order interactions) $R$ initially increases before eventually declining. 
To better visualize this effect, we computed, for each $K_1$, the value of $K_2$ that maximizes $R$ in the case where the system starts from incoherence (right panels, black line). 

Since weak higher-order interactions lead to higher values of 
$R$ also when starting from coherent initial conditions (left panels), we conclude that \emph{weak higher-order interactions enhance synchronization} on average when pairwise ones are already present.
We also tested $K_1$ and $K_2$ values up to $1.5$ (results omitted here for brevity), confirming the same trend. 
Appendix~\ref{app:C} reports the same analysis for normally distributed frequencies, for larger hypergraphs, and for the Kuramoto-Daido order parameter $R_2$; the non-monotonic dependence on $K_2$ is confirmed in all cases.
{Let us remark that the average value of $R$ computed numerically is subject to random fluctuations arising from the hypergraph realizations. 
When $R$ takes very similar values across a range of $K_2$, the location of the maximum becomes only weakly determined, and small fluctuations can shift it between non-adjacent values of $K_2$, without any appreciable change in $R$.
The curves indicating the value of $K_2$ maximizing $R$ for each $K_1$ should therefore be read qualitatively rather than as sharply defined boundaries. 
We flag the cases where the whole range of $R$ over the maximization domain is small by marking the curve as dotted whenever the 95\textsuperscript{th}--5\textsuperscript{th} percentile spread of the domain (values of $R$) falls below $0.05$.}

\subsection{Emergence of distinct dynamical regimes}\label{sec:distinct_regimes}

As shown in previous works \cite{tanaka2011multistable,skardal2020higher,leon2024,wang2024coexistence,smith2024determining,leon2025theory,battiston2026collective}, higher-order interactions can induce bistability, i.e., locally stable synchronized states and locally stable incoherent states, in the Kuramoto model. 
Consistently, Figure~\ref{fig:couplings_study_standard_coupl_R1} shows a behavior compatible with bistability for low $K_1$ and high $K_2$: trajectories starting near the synchronous state remain synchronized, whereas incoherent initial conditions do not typically lead to synchronization.

To systematically explore the dependence of the system's behavior on the coupling strengths $K_1$ and $K_2$, Figure~\ref{fig:distributions_order_parameter_uniform} reports the distributions of the order parameter $R$ for representative parameter values across different realizations of hypergraphs, initial phases, and natural frequencies (bottom panels), focusing on initially incoherent $1$-connected hypergraphs (Figure~\ref{fig:couplings_study_standard_coupl_R1}, bottom-right panel).
The top-left and top-right panels of Figure~\ref{fig:distributions_order_parameter_uniform} show bifurcation diagrams of the average $R$ (blue line), together with its standard deviation (shaded area), obtained by varying one coupling strength while keeping the other fixed.
An analogous analysis for Gaussian frequency distributions is provided in Appendix~\ref{app:C}.

For fixed $K_1 = 0.15$, large $K_2$ (pink upward triangle marker) yields a distribution of $R$ spanning both low and high values, which is compatible with bistability~\cite{suman2024finite}.
In contrast, for smaller $K_2$ (pink downward triangle), despite a similar mean value of $R$, the distribution is more concentrated around the mean, indicating a more consistent behavior across realizations.
For fixed $K_2 = 0.25$, increasing $K_1$ leads to distributions of $R$ that become increasingly concentrated near $1$, indicating a progressive stabilization of strongly coherent states across realizations.

\subsection{Analysis of a large all-to-all coupled system}\label{sec:all_to_all}

To show that this non-monotonic effect is not a finite-size one, we analyzed the setting of a large system of oscillators, i.e., $N_0=10\,000$, with all-to-all (i.e., global) coupling.%
\footnote{Note that, in the thermodynamic limit, a random graph coupling can be approximated to an all-to-all one with lower coupling strength, as shown by Restrepo et al.~\cite{restrepo2005onset}. 
This approximation, which works well for large systems and is called annealed mean field approximation, was first shown by Bianconi for scale-free networks~\cite{bianconi2002mean} and justifies the implementation of a mean field analysis~\cite{ichinomiya2004frequency}.}

The results of this analysis are shown in Figure~\ref{fig:alltoall}.
There, we can see that both order parameters, $R$ and $R_2$, have a maximum for low (but non-zero) values of $K_2$. Beyond that maximum, $R$ decreases monotonically with $K_2$. The non-monotonic behavior therefore persists for large, globally coupled systems, and is neither a finite-size effect nor specific to random hypergraphs.

Additionally, for large $K_2$, $R$ falls below $0.2$, while $R_2$ stays near $0.9$, a signature of two-cluster organization\cite{tanaka2011multistable,skardal2020higher,carballosa2023cluster,leon2024}.
For random hypergraphs, instead, the picture depends on size: at large $K_2$ with $N_0 = 10$, $R_2$ tracks $R$ closely (Appendix~\ref{app:C}, Figure~\ref{fig:couplings_study_standard_coupl_R2}), indicating genuine incoherence, and for $N_0 = 100$, $R_2$ remains appreciably above $R$ in the same regime (Appendix~\ref{app:C}, Figure~\ref{fig:couplings_study_standard_coupl_100_nodes_R2}), suggesting a partial clustering of the phases.
{For the sake of completeness, the scenario analyzed in Figure \ref{fig:alltoall} is also portrayed in the $K_1$--$K_2$ plane (similarly to Figure \ref{fig:couplings_study_standard_coupl_R1}) in Figure \ref{fig:alltoall_vary_both_couplings} in Appendix \ref{sec:alltoall_vary_couplings}, confirming the non-monotonic dependence of $R$ on $K_2$ (for incoherent initial phases).}

{The globally coupled system considered here also lends itself to an analytical explanation for the initial increase in $R$ at small $K_2$.
A mean-field self-consistency analysis shows that a fully locked, single-cluster solution of the pairwise model at $K_2=0$ becomes more coherent when continued to infinitesimal positive $3$-body coupling: namely, $\left.\mathrm{d}R/\mathrm{d}K_2\right|_{K_2=0}>0$.
This increase results from a compression of the frequency-induced phase lags, and provides a local mechanism for the initial increase in $R$ once the fully locked branch has been reached.
The complete derivation is reported in Appendix~\ref{sec:mathematical_analysis}.

By contrast, the numerical results show that larger $K_2$ promotes
competing two-cluster states, as indicated in Figure~\ref{fig:alltoall} by $R_2$ remaining high while $R$ decreases.
The competition between local phase compression at weak $K_2$ and cluster formation at stronger $K_2$ provides a qualitative explanation for the non-monotonic dependence of $R$ on $K_2$.
The location of the maximum at which the detrimental effect begins to dominate remains difficult to characterize analytically and is determined here numerically.}

\section{Optimal hyperedge allocation under a finite budget}
\label{sec:results2}

\begin{figure*}[tb]
    \centering
    \includegraphics[scale=1]{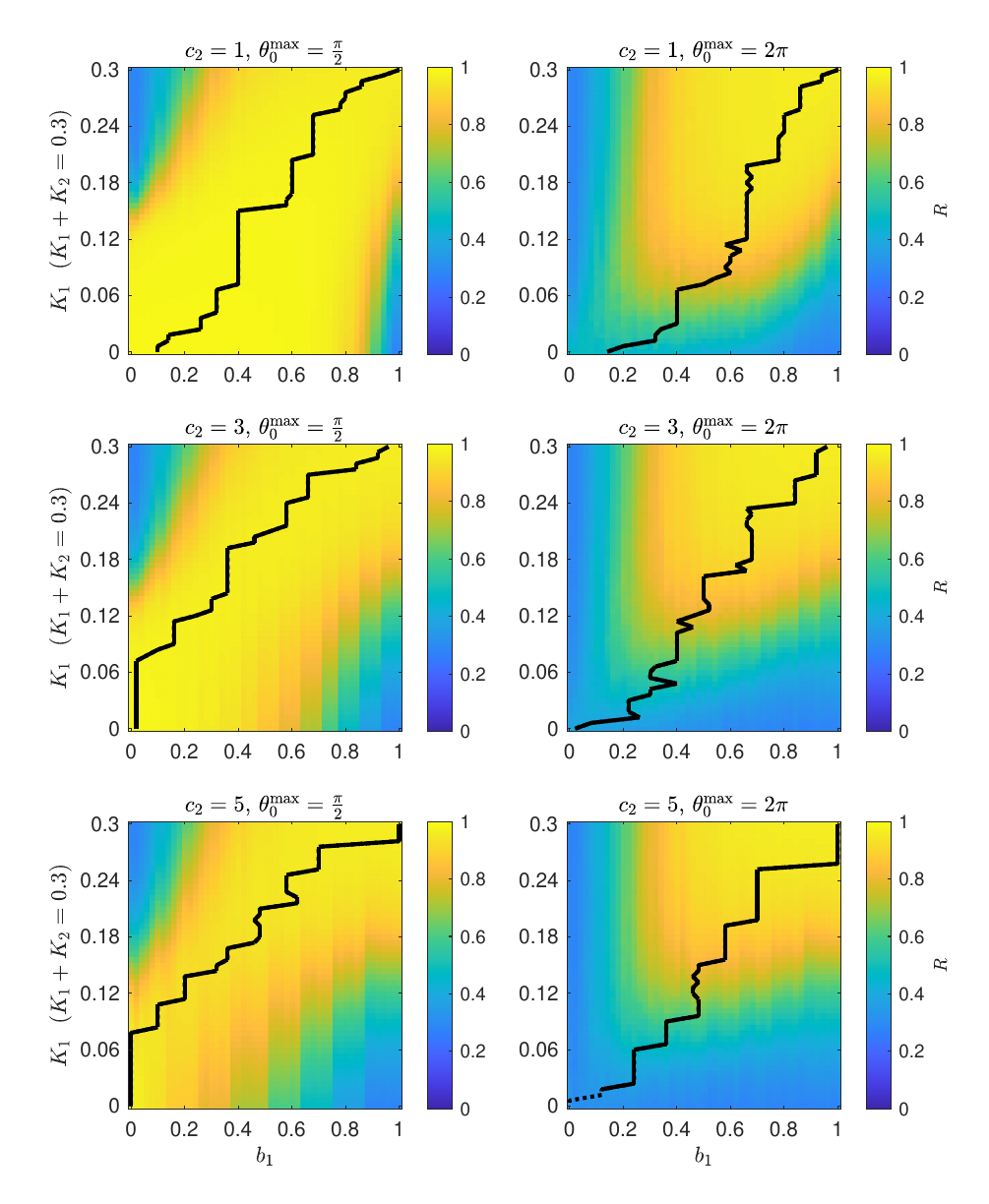}
    \caption{%
    Average Kuramoto order parameter $R$ for different combinations of links and triangles.
    $K_1$ and $K_2$ are the coupling strengths of $2$- and $3$-body interactions.
    The grid $(b_1,K_1)$ is $51\times51$. 
    For each value of the link allocation fraction $b_1$, $N_{\C{H}} = 500$ random hypergraphs were generated, with $N_0=10$ nodes, frequencies drawn uniformly at random from $[0, 0.3]$, and initial phases from $[0, \theta_0^\R{max}]$---$\theta_0^\R{max} = \frac{\pi}{2}$ in the left panels, and $\theta_0^\R{max} = 2\pi$ in the right panels.    
    The relative link cost is fixed at $c_1 = 1$;
    triangle costs $c_2$ are $1$ in top panels, $3$ in middle panels, and $5$ in bottom panels.
    For each $K_1$ (y-axis), the black line marks the $b_1$ (x-axis) yielding the highest $R$ (color). {The line is dotted when the 95\textsuperscript{th}–5\textsuperscript{th} percentile spread over the maximization domain is below 0.05}.
    See Appendix~\ref{app:B} for further details. 
    We observe that combining links and triangles generally yields higher synchronization than using either alone.}
    \label{fig:alloc_study_standard_coupl_omega03_R}
\end{figure*}

Next, we explore how limited resources allocated between pairwise and higher-order interactions influence synchronization.
To model resource constraints, we assume a fixed budget of $J = 40$ arbitrary units, representing energy, material, or financial cost. 
Each link incurs a cost of $c_1 = 1$ arbitrary unit, while triangles cost $c_2$ units, with $c_2 \in \{1, 3, 5\}$.
Resources are allocated proportionally: a fraction $b_1$ of the budget is dedicated to links, while the remaining fraction $1 - b_1$ is allocated to triangles: specifically, the numbers of links and triangles are $N_1 = \lfloor J b_1 / c_1 \rfloor$ and $N_2 = \lfloor J (1 - b_1) / c_2 \rfloor$, respectively; both links and triangles are then chosen randomly, ensuring the resulting hypergraph is $\C{H}$-connected.
This allocation models practical scenarios where different interaction types carry heterogeneous resource costs.
Beyond structural costs, synchronization also depends on coupling strengths $K_1$, $K_2$---associated with links and triangles, respectively---which are typically determined by the system's physics.
Our goal is to identify the optimal allocation strategy
across varying relative costs and a range of coupling strength configurations. 
Additionally, we investigated how the coherence of initial conditions---from partially aligned to fully disordered---impacts the effectiveness of different allocation strategies.
To systematically explore these factors, we evaluated synchronization across a grid of $51 \times 51 = 2{,}601$ experimental conditions, varying coupling strength distributions ($K_1$, $K_2$) and allocation ratios ($b_1$).
Synchronization performance was quantified by the time-averaged Kuramoto order parameter $R$, averaged over $N_\C{H} = 500$ random realizations per experimental condition to capture variability in structure, frequencies, and initial states {(see Appendix \ref{sec:methods_hyperedge_allocation_analysis} for details}).
This exploration was repeated $6$ times in total, varying both the relative cost of triangles ($c_2 \in \{1, 3, 5\}$) and the coherence of initial conditions ($\theta_0^\mathrm{max} = \pi/2$, for more coherent starts and $\theta_0^\mathrm{max} = 2\pi$ for fully incoherent starts).
Figure~\ref{fig:alloc_study_standard_coupl_omega03_R} summarizes the results of this extensive simulation campaign.

Within each panel, the horizontal axis represents the fraction of resources allocated to links, $b_1$, and the vertical axis represents their associated coupling strength, $K_1$, with the coupling strength for triangles set as $K_2 = 0.3 - K_1$; the sum of the two strengths is kept constant to ensure meaningful comparison across conditions.%
\footnote{\label{fn:constraint_gains}We impose this constraint to enable a fair comparison across different allocations. 
In many real-world systems, however, $K_1$ and $K_2$ are set by the physics of the interactions and can be tuned independently, without necessarily trading off against one another. 
For completeness, we study this unconstrained case in
Figure~\ref{fig:alloc_study_free_couplings} of
Appendix~\ref{sec:allocation_study_free_gains}.}
The color of each pixel denotes the order parameter $R$, averaged over both the second half of simulations and the $500$ hypergraph realizations, with brighter colors indicating stronger phase coherence.
{Note that, because $N_1$ and $N_2$ are integers obtained through the floor operation, consecutive $b_1$ values can map to the same number of triangles, and hence to similar expected order parameter $R$.
This produces the vertical bands visible in Figure~\ref{fig:alloc_study_standard_coupl_omega03_R} (and in the analogous figures of Appendix~\ref{app:C}), which become wider as $c_2$ increases---since a larger change in $b_1$ is then required to add or remove a single triangle---and are more pronounced in the lower parts of the panels, where $K_2$ is larger.}
The black line denotes the optimal link allocation fraction $b_1$ for each combination of coupling strengths $K_1$ and $K_2$. 
{As for the previous analysis (Section \ref{sec:results1a}), we note that the optimal allocation identified numerically is subject to random fluctuations arising from the hypergraph realizations.
When the maximization domain is nearly flat (i.e., when $R$ takes very similar values across a range of $b_1$) the location of the maximum becomes only weakly determined, and small fluctuations can shift it between non-adjacent values of $b_1$ or leave it pinned at the same $b_1$ over a range of values of $K_1$ and $K_2$, without any appreciable change in $R$.
The optimal-allocation curves should therefore be read qualitatively.}

\subsection{Combining links and triangles is optimal}
\label{sec:combining_links_trian_optimal}

A key finding from this analysis is that the optimal link allocation fraction $b_1$ is rarely exactly $0$ or $1$, i.e., dedicating all resources exclusively to either links or triangles is almost never optimal.
Trivially, full allocation to links is optimal only when $K_2 \approx 0$; similarly, allocating the entire budget to triangles is optimal only when $K_1 \approx 0$.
Across all panels, the optimal configuration consistently favors a combination of both interaction types.
Notably, the optimal value of $b_1$ strongly depends on the relative coupling strengths $K_1$, $K_2$, with more resources typically allocated to the interaction type associated with the stronger coupling.
{The finding remains confirmed even when $K_1$ and $K_2$ are tuned independently, as illustrated in Figure~\ref{fig:alloc_study_free_couplings} of
Appendix~\ref{sec:allocation_study_free_gains}.}

\subsection{Triangles are less effective but still useful with incoherent initial conditions}

The beneficial effect of triangles persists even when the initial phases are highly incoherent (right panels).
In this scenario, triangles are less effective at enhancing synchronization (see the lower-left corners of the right panels).
Nevertheless, a combination of interaction types remains optimal, as highlighted by the black lines indicating the best $b_1$ for each pair $(K_1$, $K_2)$.
In all cases, including some triangles improves synchronization, with the exact optimal balance depending on the relative values of the coupling strengths $K_1$ and $K_2$.

\subsection{Triangles remain beneficial even when more costly}
\label{sec:triangles_remain_beneficial}

Lower costs for triangles lead to generally higher levels of synchronization (compare the left panels, from top to bottom).
This is because, with a fixed resource budget, a lower cost allows deploying more triangles, resulting in a more connected hypergraph.
Increasing the relative cost $c_2$ of triangles shifts the optimal resource allocation: despite the higher cost, a larger portion of the budget is devoted to triangles, underscoring their strong impact on synchronization.
This is reflected by the black optimal-allocation curves bending slightly left as $c_2$ increases.

These findings indicate that higher-order interactions complement pairwise couplings and can enhance synchronization.
To further validate this, we repeated the analysis for Gaussian rather than uniform frequency distributions, for larger hypergraphs of $N_0 = 100$ nodes, and for the Kuramoto-Daido order parameter $R_2$.
In all cases, the results consistently confirm that a combination of links and triangles is more effective for synchronization than relying on a single interaction type.
These additional results are reported in Appendix~\ref{app:C}.

\section{Conclusion}
\label{sec:conclusion}

In this work, we studied the higher-order Kuramoto model on random hypergraphs---taken as a paradigmatic example of oscillatory dynamics with higher-order interactions---across a wide range of parameter configurations, and on globally coupled hypergraphs.
We found that the effect of higher-order interactions on synchronization is non-monotonic: weak higher-order interactions enhance it when added to pairwise ones, whereas strong ones hamper it, in line with the reduced basins and richer dynamics (cluster synchronization, slow switching, multistability) reported previously~\cite{zhang2024deeper,von2024higher,tanaka2011multistable,skardal2020higher,leon2024}.

We covered two complementary regimes: random hypergraphs of $10$ and $100$ nodes---a range relevant to real-world applications, {such as small subnetworks of interacting brain regions~\cite{menara2022functional}, inverters in islanded microgrids~\cite{simpson2013synchronization}, and small groups of individuals coordinating their movements~\cite{alderisio2017interaction}---}and globally coupled systems of $N_0 = 10\,000$ oscillators. 
{Our choice of $N_0 = 10$ in much of our numerical study is deliberate: besides being application-relevant, this size allows a dense exploration of parameters and conditions ($K_1$, $K_2$, $b_1$, $c_2$, $\theta_0^\mathrm{max}$, frequency distribution) with hundreds of independent hypergraph realizations at a manageable computational cost.
Spanning three orders of magnitude in system size ($10$, $100$, $10\,000$) and finding the same non-monotonic dependence throughout indicates that the effect is not a finite-size artifact but a robust feature of the higher-order Kuramoto dynamics.}

The non-monotonic dependence on the higher-order coupling strength described here was first reported in the preprint version of this work\cite{muolo2025higher}. 
{ A similar effect has since been observed in other works.
Refs.~\onlinecite{wang2026moderate} and \onlinecite{moriame2026efficiency} also report that weak/moderate higher-order (namely, $3$-body) coupling added to pairwise interactions enhances the stability of an existing state (twisted states on hyperrings, and synchronized states on random and hyperring topologies, respectively), corroborating our finding for random and all-to-all topologies. 
Ref.~\onlinecite{skardal2025mixed} shows that mixing higher-order coupling terms (including a $4$-body harmonic) can stabilize new states.
This is related to our mixed-allocation result of Section~\ref{sec:results2}, but concerns mixing within the coupling function rather than in the structural allocation of a fixed resource budget, as we do here (Section~\ref{sec:results2}).} 
That convergent evidence, obtained for different topologies and different target states, indicates the effect is generic rather than specific to the setting considered here.

Since our results on random hypergraphs are obtained by averaging over hundreds of topology realizations, they hold on average; specific hypergraph structures may behave differently, as reflected in the order parameter distributions in Figure \ref{fig:distributions_order_parameter_uniform}. 
Furthermore, our results pertain to the three-body coupling $\sin(\theta_k+\theta_l-2\theta_j)$ of Equation~\eqref{eq:ho_kura_leon}, also called the $(1,1,-2)$ interaction~\cite{namura2025optimal},
which is the form most widely adopted in the literature and the one arising from phase reduction under the symmetry assumptions of Ref.~\onlinecite{leon2024}.
As Le\'on et al.\ have shown~\cite{leon2025theory,leon2026symmetry},
different higher-order interactions emerge under different symmetries of the non-reduced system, notably the $(2,-1,-1)$ interaction $\sin(2\theta_k-\theta_l-\theta_j)$~\cite{namura2025optimal}.
Whether the non-monotonic effect reported here survives for such interactions, for four-body and higher couplings, or for combinations of them, remains an open question and a natural direction for future work, alongside an in-depth exploration of different higher-order topologies, { and considering the effects of noise~\cite{marui2025synchronization}.}
A further promising avenue would be to identify which higher-order topologies most effectively enhance synchronization, by extending the numerical network analysis in Ref.~\onlinecite{coraggio2024data} to the higher-order setting.

\medskip

\noindent \textbf{Acknowledgments:} \\
\noindent  The authors are grateful to Daniele Proverbio and Lauren D. Smith for discussions and feedback on the work. R.M. and M.C. would like to thank Claudio Altafini and all the organizers of the ELLIIT Focus Period on Network Dynamics and Control held at Linköping University in September 2023, where this collaboration started. 

\noindent \textbf{Funding:} \\
\noindent R.M. acknowledges JSPS KAKENHI 24KF0211 for financial support. H.N. acknowledges JSPS KAKENHI 25H01468, 25K03081, and 22H00516 for financial support. 

\noindent \textbf{Author contributions:} \\
R.M. and M.C. contributed equally to this work: they conceptualized the work, developed the methodology, carried out the analysis, wrote the manuscript, curated the code, and the visualization. H.N. provided methodological insights. All authors reviewed and edited the article.

\noindent \textbf{Conflict of interest:} \\
The authors have no conflicts to disclose.

\noindent \textbf{Data availability:} \\
The data that support the findings of this study are available within the article.


\begin{thebibliography}{113}%
\makeatletter
\providecommand \@ifxundefined [1]{%
 \@ifx{#1\undefined}
}%
\providecommand \@ifnum [1]{%
 \ifnum #1\expandafter \@firstoftwo
 \else \expandafter \@secondoftwo
 \fi
}%
\providecommand \@ifx [1]{%
 \ifx #1\expandafter \@firstoftwo
 \else \expandafter \@secondoftwo
 \fi
}%
\providecommand \natexlab [1]{#1}%
\providecommand \enquote  [1]{``#1''}%
\providecommand \bibnamefont  [1]{#1}%
\providecommand \bibfnamefont [1]{#1}%
\providecommand \citenamefont [1]{#1}%
\providecommand \href@noop [0]{\@secondoftwo}%
\providecommand \href [0]{\begingroup \@sanitize@url \@href}%
\providecommand \@href[1]{\@@startlink{#1}\@@href}%
\providecommand \@@href[1]{\endgroup#1\@@endlink}%
\providecommand \@sanitize@url [0]{\catcode `\\12\catcode `\$12\catcode `\&12\catcode `\#12\catcode `\^12\catcode `\_12\catcode `\%12\relax}%
\providecommand \@@startlink[1]{}%
\providecommand \@@endlink[0]{}%
\providecommand \url  [0]{\begingroup\@sanitize@url \@url }%
\providecommand \@url [1]{\endgroup\@href {#1}{\urlprefix }}%
\providecommand \urlprefix  [0]{URL }%
\providecommand \Eprint [0]{\href }%
\providecommand \doibase [0]{https://doi.org/}%
\providecommand \selectlanguage [0]{\@gobble}%
\providecommand \bibinfo  [0]{\@secondoftwo}%
\providecommand \bibfield  [0]{\@secondoftwo}%
\providecommand \translation [1]{[#1]}%
\providecommand \BibitemOpen [0]{}%
\providecommand \bibitemStop [0]{}%
\providecommand \bibitemNoStop [0]{.\EOS\space}%
\providecommand \EOS [0]{\spacefactor3000\relax}%
\providecommand \BibitemShut  [1]{\csname bibitem#1\endcsname}%
\let\auto@bib@innerbib\@empty
\bibitem [{\citenamefont {Pikovsky}, \citenamefont {Kurths},\ and\ \citenamefont {Rosenblum}(2001)}]{pikovsky2001synchronization}%
  \BibitemOpen
  \bibfield  {author} {\bibinfo {author} {\bibfnamefont {A.}~\bibnamefont {Pikovsky}}, \bibinfo {author} {\bibfnamefont {J.}~\bibnamefont {Kurths}},\ and\ \bibinfo {author} {\bibfnamefont {M.}~\bibnamefont {Rosenblum}},\ }\href@noop {} {\emph {\bibinfo {title} {Synchronization: a universal concept in nonlinear sciences}}}\ (\bibinfo  {publisher} {Cambridge University Press},\ \bibinfo {address} {Cambridge},\ \bibinfo {year} {2001})\BibitemShut {NoStop}%
\bibitem [{\citenamefont {Arenas}\ \emph {et~al.}(2008)\citenamefont {Arenas}, \citenamefont {D{\'{i}}az-Guilera}, \citenamefont {Kurths}, \citenamefont {Moreno},\ and\ \citenamefont {Zhou}}]{arenas2008synchronization}%
  \BibitemOpen
  \bibfield  {author} {\bibinfo {author} {\bibfnamefont {A.}~\bibnamefont {Arenas}}, \bibinfo {author} {\bibfnamefont {A.}~\bibnamefont {D{\'{i}}az-Guilera}}, \bibinfo {author} {\bibfnamefont {J.}~\bibnamefont {Kurths}}, \bibinfo {author} {\bibfnamefont {Y.}~\bibnamefont {Moreno}},\ and\ \bibinfo {author} {\bibfnamefont {C.}~\bibnamefont {Zhou}},\ }\bibfield  {title} {\enquote {\bibinfo {title} {{Synchronization in complex networks}},}\ }\href@noop {} {\bibfield  {journal} {\bibinfo  {journal} {Phys. Rep.}\ }\textbf {\bibinfo {volume} {469}},\ \bibinfo {pages} {93--153} (\bibinfo {year} {2008})}\BibitemShut {NoStop}%
\bibitem [{\citenamefont {Boccaletti}\ \emph {et~al.}(2018)\citenamefont {Boccaletti}, \citenamefont {Pisarchik}, \citenamefont {Del~Genio},\ and\ \citenamefont {Amann}}]{boccaletti2018synchronization}%
  \BibitemOpen
  \bibfield  {author} {\bibinfo {author} {\bibfnamefont {S.}~\bibnamefont {Boccaletti}}, \bibinfo {author} {\bibfnamefont {A.~N.}\ \bibnamefont {Pisarchik}}, \bibinfo {author} {\bibfnamefont {C.~I.}\ \bibnamefont {Del~Genio}},\ and\ \bibinfo {author} {\bibfnamefont {A.}~\bibnamefont {Amann}},\ }\href@noop {} {\emph {\bibinfo {title} {Synchronization: from Coupled Systems to Complex Networks}}}\ (\bibinfo  {publisher} {Cambridge University Press},\ \bibinfo {address} {Cambridge},\ \bibinfo {year} {2018})\BibitemShut {NoStop}%
\bibitem [{\citenamefont {Strogatz}(2003)}]{Strogatzbooksync}%
  \BibitemOpen
  \bibfield  {author} {\bibinfo {author} {\bibfnamefont {S.}~\bibnamefont {Strogatz}},\ }\href@noop {} {\emph {\bibinfo {title} {Sync: The Emerging Science of Spontaneous Order}}}\ (\bibinfo  {publisher} {Hyperion},\ \bibinfo {address} {New York},\ \bibinfo {year} {2003})\BibitemShut {NoStop}%
\bibitem [{\citenamefont {Coraggio}, \citenamefont {Salzano},\ and\ \citenamefont {{di Bernardo}}(2026)}]{coraggio2026controlling}%
  \BibitemOpen
  \bibfield  {author} {\bibinfo {author} {\bibfnamefont {M.}~\bibnamefont {Coraggio}}, \bibinfo {author} {\bibfnamefont {D.}~\bibnamefont {Salzano}},\ and\ \bibinfo {author} {\bibfnamefont {M.}~\bibnamefont {{di Bernardo}}},\ }\bibfield  {title} {\enquote {\bibinfo {title} {Controlling complex systems},}\ }\href@noop {} {\bibfield  {journal} {\bibinfo  {journal} {Encyclopedia of Systems and Control Engineering}\ }\textbf {\bibinfo {volume} {1}},\ \bibinfo {pages} {483--497} (\bibinfo {year} {2026})}\BibitemShut {NoStop}%
\bibitem [{\citenamefont {Winfree}(1967)}]{winfree1967biological}%
  \BibitemOpen
  \bibfield  {author} {\bibinfo {author} {\bibfnamefont {A.~T.}\ \bibnamefont {Winfree}},\ }\bibfield  {title} {\enquote {\bibinfo {title} {Biological rhythms and the behavior of populations of coupled oscillators},}\ }\href@noop {} {\bibfield  {journal} {\bibinfo  {journal} {J. Theor. Biol.}\ }\textbf {\bibinfo {volume} {16}},\ \bibinfo {pages} {15--42} (\bibinfo {year} {1967})}\BibitemShut {NoStop}%
\bibitem [{\citenamefont {Kuramoto}(1975)}]{kuramoto1975}%
  \BibitemOpen
  \bibfield  {author} {\bibinfo {author} {\bibfnamefont {Y.}~\bibnamefont {Kuramoto}},\ }\bibfield  {title} {\enquote {\bibinfo {title} {Self-entrainment of a population of coupled non-linear oscillators},}\ }in\ \href@noop {} {\emph {\bibinfo {booktitle} {International Symposium on Mathematical Problems in Theoretical Physics}}},\ \bibinfo {editor} {edited by\ \bibinfo {editor} {\bibfnamefont {H.}~\bibnamefont {Araki}}}\ (\bibinfo  {publisher} {Springer Berlin Heidelberg},\ \bibinfo {address} {Berlin, Heidelberg},\ \bibinfo {year} {1975})\ pp.\ \bibinfo {pages} {420--422}\BibitemShut {NoStop}%
\bibitem [{\citenamefont {Strogatz}(2000)}]{strogatz2000kuramoto}%
  \BibitemOpen
  \bibfield  {author} {\bibinfo {author} {\bibfnamefont {S.}~\bibnamefont {Strogatz}},\ }\bibfield  {title} {\enquote {\bibinfo {title} {From {K}uramoto to {C}rawford: exploring the onset of synchronization in populations of coupled oscillators},}\ }\href@noop {} {\bibfield  {journal} {\bibinfo  {journal} {Physica D}\ }\textbf {\bibinfo {volume} {143}},\ \bibinfo {pages} {1--20} (\bibinfo {year} {2000})}\BibitemShut {NoStop}%
\bibitem [{\citenamefont {Acebr{\'o}n}\ \emph {et~al.}(2005)\citenamefont {Acebr{\'o}n}, \citenamefont {Bonilla}, \citenamefont {Vicente}, \citenamefont {Ritort},\ and\ \citenamefont {Spigler}}]{acebron2005kuramoto}%
  \BibitemOpen
  \bibfield  {author} {\bibinfo {author} {\bibfnamefont {J.~A.}\ \bibnamefont {Acebr{\'o}n}}, \bibinfo {author} {\bibfnamefont {L.~L.}\ \bibnamefont {Bonilla}}, \bibinfo {author} {\bibfnamefont {C.~J.~P.}\ \bibnamefont {Vicente}}, \bibinfo {author} {\bibfnamefont {F.}~\bibnamefont {Ritort}},\ and\ \bibinfo {author} {\bibfnamefont {R.}~\bibnamefont {Spigler}},\ }\bibfield  {title} {\enquote {\bibinfo {title} {The {K}uramoto model: A simple paradigm for synchronization phenomena},}\ }\href@noop {} {\bibfield  {journal} {\bibinfo  {journal} {Rev. Mod. Phys.}\ }\textbf {\bibinfo {volume} {77}},\ \bibinfo {pages} {137} (\bibinfo {year} {2005})}\BibitemShut {NoStop}%
\bibitem [{\citenamefont {Rodrigues}\ \emph {et~al.}(2016)\citenamefont {Rodrigues}, \citenamefont {Peron}, \citenamefont {Ji},\ and\ \citenamefont {Kurths}}]{rodrigues2016kuramoto}%
  \BibitemOpen
  \bibfield  {author} {\bibinfo {author} {\bibfnamefont {F.~A.}\ \bibnamefont {Rodrigues}}, \bibinfo {author} {\bibfnamefont {T.~K.~D.}\ \bibnamefont {Peron}}, \bibinfo {author} {\bibfnamefont {P.}~\bibnamefont {Ji}},\ and\ \bibinfo {author} {\bibfnamefont {J.}~\bibnamefont {Kurths}},\ }\bibfield  {title} {\enquote {\bibinfo {title} {The {K}uramoto model in complex networks},}\ }\href@noop {} {\bibfield  {journal} {\bibinfo  {journal} {Phys. Rep.}\ }\textbf {\bibinfo {volume} {610}},\ \bibinfo {pages} {1--98} (\bibinfo {year} {2016})}\BibitemShut {NoStop}%
\bibitem [{\citenamefont {Newman}(2018)}]{newmanbook2}%
  \BibitemOpen
  \bibfield  {author} {\bibinfo {author} {\bibfnamefont {M.}~\bibnamefont {Newman}},\ }\href@noop {} {\emph {\bibinfo {title} {Networks: An Introduction. Second Edition}}}\ (\bibinfo  {publisher} {Oxford University Press},\ \bibinfo {address} {Oxford},\ \bibinfo {year} {2018})\BibitemShut {NoStop}%
\bibitem [{\citenamefont {Latora}, \citenamefont {Nicosia},\ and\ \citenamefont {Russo}(2017)}]{latora_nicosia_russo_2017}%
  \BibitemOpen
  \bibfield  {author} {\bibinfo {author} {\bibfnamefont {V.}~\bibnamefont {Latora}}, \bibinfo {author} {\bibfnamefont {V.}~\bibnamefont {Nicosia}},\ and\ \bibinfo {author} {\bibfnamefont {G.}~\bibnamefont {Russo}},\ }\href@noop {} {\emph {\bibinfo {title} {Complex Networks: Principles, Methods and Applications}}}\ (\bibinfo  {publisher} {Cambridge University Press},\ \bibinfo {address} {Cambridge},\ \bibinfo {year} {2017})\BibitemShut {NoStop}%
\bibitem [{\citenamefont {Battiston}\ \emph {et~al.}(2020)\citenamefont {Battiston}, \citenamefont {Cencetti}, \citenamefont {Iacopini}, \citenamefont {Latora}, \citenamefont {Lucas}, \citenamefont {Patania}, \citenamefont {Young},\ and\ \citenamefont {Petri}}]{battiston2020networks}%
  \BibitemOpen
  \bibfield  {author} {\bibinfo {author} {\bibfnamefont {F.}~\bibnamefont {Battiston}}, \bibinfo {author} {\bibfnamefont {G.}~\bibnamefont {Cencetti}}, \bibinfo {author} {\bibfnamefont {I.}~\bibnamefont {Iacopini}}, \bibinfo {author} {\bibfnamefont {V.}~\bibnamefont {Latora}}, \bibinfo {author} {\bibfnamefont {M.}~\bibnamefont {Lucas}}, \bibinfo {author} {\bibfnamefont {A.}~\bibnamefont {Patania}}, \bibinfo {author} {\bibfnamefont {J.-G.}\ \bibnamefont {Young}},\ and\ \bibinfo {author} {\bibfnamefont {G.}~\bibnamefont {Petri}},\ }\bibfield  {title} {\enquote {\bibinfo {title} {Networks beyond pairwise interactions: structure and dynamics},}\ }\href@noop {} {\bibfield  {journal} {\bibinfo  {journal} {Phys. Rep.}\ }\textbf {\bibinfo {volume} {874}},\ \bibinfo {pages} {1--92} (\bibinfo {year} {2020})}\BibitemShut {NoStop}%
\bibitem [{\citenamefont {Bianconi}(2021)}]{bianconi2021higher}%
  \BibitemOpen
  \bibfield  {author} {\bibinfo {author} {\bibfnamefont {G.}~\bibnamefont {Bianconi}},\ }\href@noop {} {\emph {\bibinfo {title} {Higher-{O}rder {N}etworks: {A}n introduction to simplicial complexes}}}\ (\bibinfo  {publisher} {Cambridge {U}niversity {P}ress},\ \bibinfo {address} {Cambridge},\ \bibinfo {year} {2021})\BibitemShut {NoStop}%
\bibitem [{\citenamefont {Battiston}\ \emph {et~al.}(2021)\citenamefont {Battiston}, \citenamefont {Amico}, \citenamefont {Barrat}, \citenamefont {Bianconi}, \citenamefont {de~Arruda}, \citenamefont {Franceschiello}, \citenamefont {Iacopini}, \citenamefont {Kéfi}, \citenamefont {Latora}, \citenamefont {Moreno}, \citenamefont {Murray}, \citenamefont {Peixoto}, \citenamefont {Vaccarino},\ and\ \citenamefont {Petri}}]{natphys}%
  \BibitemOpen
  \bibfield  {author} {\bibinfo {author} {\bibfnamefont {F.}~\bibnamefont {Battiston}}, \bibinfo {author} {\bibfnamefont {E.}~\bibnamefont {Amico}}, \bibinfo {author} {\bibfnamefont {A.}~\bibnamefont {Barrat}}, \bibinfo {author} {\bibfnamefont {G.}~\bibnamefont {Bianconi}}, \bibinfo {author} {\bibfnamefont {G.}~\bibnamefont {de~Arruda}}, \bibinfo {author} {\bibfnamefont {B.}~\bibnamefont {Franceschiello}}, \bibinfo {author} {\bibfnamefont {I.}~\bibnamefont {Iacopini}}, \bibinfo {author} {\bibfnamefont {S.}~\bibnamefont {Kéfi}}, \bibinfo {author} {\bibfnamefont {V.}~\bibnamefont {Latora}}, \bibinfo {author} {\bibfnamefont {Y.}~\bibnamefont {Moreno}}, \bibinfo {author} {\bibfnamefont {M.}~\bibnamefont {Murray}}, \bibinfo {author} {\bibfnamefont {T.}~\bibnamefont {Peixoto}}, \bibinfo {author} {\bibfnamefont {F.}~\bibnamefont {Vaccarino}},\ and\ \bibinfo {author} {\bibfnamefont {G.}~\bibnamefont {Petri}},\ }\bibfield  {title} {\enquote {\bibinfo {title} {The physics of higher-order interactions in complex
  systems},}\ }\href@noop {} {\bibfield  {journal} {\bibinfo  {journal} {Nat. Phys.}\ }\textbf {\bibinfo {volume} {17}},\ \bibinfo {pages} {1093–1098} (\bibinfo {year} {2021})}\BibitemShut {NoStop}%
\bibitem [{\citenamefont {Bick}\ \emph {et~al.}(2023)\citenamefont {Bick}, \citenamefont {Gross}, \citenamefont {Harrington},\ and\ \citenamefont {Schaub}}]{bick2023higher}%
  \BibitemOpen
  \bibfield  {author} {\bibinfo {author} {\bibfnamefont {C.}~\bibnamefont {Bick}}, \bibinfo {author} {\bibfnamefont {E.}~\bibnamefont {Gross}}, \bibinfo {author} {\bibfnamefont {H.~A.}\ \bibnamefont {Harrington}},\ and\ \bibinfo {author} {\bibfnamefont {M.~T.}\ \bibnamefont {Schaub}},\ }\bibfield  {title} {\enquote {\bibinfo {title} {What are higher-order networks?}}\ }\href@noop {} {\bibfield  {journal} {\bibinfo  {journal} {SIAM Rev.}\ }\textbf {\bibinfo {volume} {65}},\ \bibinfo {pages} {686--731} (\bibinfo {year} {2023})}\BibitemShut {NoStop}%
\bibitem [{\citenamefont {Boccaletti}\ \emph {et~al.}(2023)\citenamefont {Boccaletti}, \citenamefont {De~Lellis}, \citenamefont {Del~Genio}, \citenamefont {Alfaro-Bittner}, \citenamefont {Criado}, \citenamefont {Jalan},\ and\ \citenamefont {Romance}}]{boccaletti2023structure}%
  \BibitemOpen
  \bibfield  {author} {\bibinfo {author} {\bibfnamefont {S.}~\bibnamefont {Boccaletti}}, \bibinfo {author} {\bibfnamefont {P.}~\bibnamefont {De~Lellis}}, \bibinfo {author} {\bibfnamefont {C.}~\bibnamefont {Del~Genio}}, \bibinfo {author} {\bibfnamefont {K.}~\bibnamefont {Alfaro-Bittner}}, \bibinfo {author} {\bibfnamefont {R.}~\bibnamefont {Criado}}, \bibinfo {author} {\bibfnamefont {S.}~\bibnamefont {Jalan}},\ and\ \bibinfo {author} {\bibfnamefont {M.}~\bibnamefont {Romance}},\ }\bibfield  {title} {\enquote {\bibinfo {title} {The structure and dynamics of networks with higher order interactions},}\ }\href@noop {} {\bibfield  {journal} {\bibinfo  {journal} {Phys. Rep.}\ }\textbf {\bibinfo {volume} {1018}},\ \bibinfo {pages} {1--64} (\bibinfo {year} {2023})}\BibitemShut {NoStop}%
\bibitem [{\citenamefont {Muolo}\ \emph {et~al.}(2024{\natexlab{a}})\citenamefont {Muolo}, \citenamefont {Giambagli}, \citenamefont {Nakao}, \citenamefont {Fanelli},\ and\ \citenamefont {Carletti}}]{muolo2024turing}%
  \BibitemOpen
  \bibfield  {author} {\bibinfo {author} {\bibfnamefont {R.}~\bibnamefont {Muolo}}, \bibinfo {author} {\bibfnamefont {L.}~\bibnamefont {Giambagli}}, \bibinfo {author} {\bibfnamefont {H.}~\bibnamefont {Nakao}}, \bibinfo {author} {\bibfnamefont {D.}~\bibnamefont {Fanelli}},\ and\ \bibinfo {author} {\bibfnamefont {T.}~\bibnamefont {Carletti}},\ }\bibfield  {title} {\enquote {\bibinfo {title} {Turing patterns on discrete topologies: from networks to higher-order structures},}\ }\href@noop {} {\bibfield  {journal} {\bibinfo  {journal} {Proceedings of the Royal Society A}\ }\textbf {\bibinfo {volume} {480}},\ \bibinfo {pages} {20240235} (\bibinfo {year} {2024}{\natexlab{a}})}\BibitemShut {NoStop}%
\bibitem [{\citenamefont {Mill{\'a}n}\ \emph {et~al.}(2025)\citenamefont {Mill{\'a}n}, \citenamefont {Sun}, \citenamefont {Giambagli}, \citenamefont {Muolo}, \citenamefont {Carletti}, \citenamefont {Torres}, \citenamefont {Radicchi}, \citenamefont {Kurths},\ and\ \citenamefont {Bianconi}}]{millan2025topology}%
  \BibitemOpen
  \bibfield  {author} {\bibinfo {author} {\bibfnamefont {A.}~\bibnamefont {Mill{\'a}n}}, \bibinfo {author} {\bibfnamefont {H.}~\bibnamefont {Sun}}, \bibinfo {author} {\bibfnamefont {L.}~\bibnamefont {Giambagli}}, \bibinfo {author} {\bibfnamefont {R.}~\bibnamefont {Muolo}}, \bibinfo {author} {\bibfnamefont {T.}~\bibnamefont {Carletti}}, \bibinfo {author} {\bibfnamefont {J.}~\bibnamefont {Torres}}, \bibinfo {author} {\bibfnamefont {F.}~\bibnamefont {Radicchi}}, \bibinfo {author} {\bibfnamefont {J.}~\bibnamefont {Kurths}},\ and\ \bibinfo {author} {\bibfnamefont {G.}~\bibnamefont {Bianconi}},\ }\bibfield  {title} {\enquote {\bibinfo {title} {Topology shapes dynamics of higher-order networks},}\ }\href@noop {} {\bibfield  {journal} {\bibinfo  {journal} {Nature Physics}\ }\textbf {\bibinfo {volume} {21}},\ \bibinfo {pages} {353--361} (\bibinfo {year} {2025})}\BibitemShut {NoStop}%
\bibitem [{\citenamefont {Krawiecki}(2014)}]{krawiecki2014chaotic}%
  \BibitemOpen
  \bibfield  {author} {\bibinfo {author} {\bibfnamefont {A.}~\bibnamefont {Krawiecki}},\ }\bibfield  {title} {\enquote {\bibinfo {title} {Chaotic synchronization on complex hypergraphs},}\ }\href@noop {} {\bibfield  {journal} {\bibinfo  {journal} {Chaos Solit. Fractals}\ }\textbf {\bibinfo {volume} {65}},\ \bibinfo {pages} {44--50} (\bibinfo {year} {2014})}\BibitemShut {NoStop}%
\bibitem [{\citenamefont {Gambuzza}\ \emph {et~al.}(2021)\citenamefont {Gambuzza}, \citenamefont {Di~Patti}, \citenamefont {Gallo}, \citenamefont {Lepri}, \citenamefont {Romance}, \citenamefont {Criado}, \citenamefont {Frasca}, \citenamefont {Latora},\ and\ \citenamefont {Boccaletti}}]{gambuzza2021stability}%
  \BibitemOpen
  \bibfield  {author} {\bibinfo {author} {\bibfnamefont {L.}~\bibnamefont {Gambuzza}}, \bibinfo {author} {\bibfnamefont {F.}~\bibnamefont {Di~Patti}}, \bibinfo {author} {\bibfnamefont {L.}~\bibnamefont {Gallo}}, \bibinfo {author} {\bibfnamefont {S.}~\bibnamefont {Lepri}}, \bibinfo {author} {\bibfnamefont {M.}~\bibnamefont {Romance}}, \bibinfo {author} {\bibfnamefont {R.}~\bibnamefont {Criado}}, \bibinfo {author} {\bibfnamefont {M.}~\bibnamefont {Frasca}}, \bibinfo {author} {\bibfnamefont {V.}~\bibnamefont {Latora}},\ and\ \bibinfo {author} {\bibfnamefont {S.}~\bibnamefont {Boccaletti}},\ }\bibfield  {title} {\enquote {\bibinfo {title} {Stability of synchronization in simplicial complexes},}\ }\href@noop {} {\bibfield  {journal} {\bibinfo  {journal} {Nat. Comm.}\ }\textbf {\bibinfo {volume} {12}},\ \bibinfo {pages} {1--13} (\bibinfo {year} {2021})}\BibitemShut {NoStop}%
\bibitem [{\citenamefont {Gallo}\ \emph {et~al.}(2022)\citenamefont {Gallo}, \citenamefont {Muolo}, \citenamefont {Gambuzza}, \citenamefont {Latora}, \citenamefont {Frasca},\ and\ \citenamefont {Carletti}}]{gallo2022synchronization}%
  \BibitemOpen
  \bibfield  {author} {\bibinfo {author} {\bibfnamefont {L.}~\bibnamefont {Gallo}}, \bibinfo {author} {\bibfnamefont {R.}~\bibnamefont {Muolo}}, \bibinfo {author} {\bibfnamefont {L.}~\bibnamefont {Gambuzza}}, \bibinfo {author} {\bibfnamefont {V.}~\bibnamefont {Latora}}, \bibinfo {author} {\bibfnamefont {M.}~\bibnamefont {Frasca}},\ and\ \bibinfo {author} {\bibfnamefont {T.}~\bibnamefont {Carletti}},\ }\bibfield  {title} {\enquote {\bibinfo {title} {Synchronization induced by directed higher-order interactions},}\ }\href@noop {} {\bibfield  {journal} {\bibinfo  {journal} {Comm. Phys.}\ }\textbf {\bibinfo {volume} {5}},\ \bibinfo {pages} {236} (\bibinfo {year} {2022})}\BibitemShut {NoStop}%
\bibitem [{\citenamefont {Della~Rossa}\ \emph {et~al.}(2023)\citenamefont {Della~Rossa}, \citenamefont {Liuzza}, \citenamefont {Lo~Iudice},\ and\ \citenamefont {De~Lellis}}]{della2023emergence}%
  \BibitemOpen
  \bibfield  {author} {\bibinfo {author} {\bibfnamefont {F.}~\bibnamefont {Della~Rossa}}, \bibinfo {author} {\bibfnamefont {D.}~\bibnamefont {Liuzza}}, \bibinfo {author} {\bibfnamefont {F.}~\bibnamefont {Lo~Iudice}},\ and\ \bibinfo {author} {\bibfnamefont {P.}~\bibnamefont {De~Lellis}},\ }\bibfield  {title} {\enquote {\bibinfo {title} {Emergence and control of synchronization in networks with directed many-body interactions},}\ }\href@noop {} {\bibfield  {journal} {\bibinfo  {journal} {Phys. Rev. Lett.}\ }\textbf {\bibinfo {volume} {131}},\ \bibinfo {pages} {207401} (\bibinfo {year} {2023})}\BibitemShut {NoStop}%
\bibitem [{\citenamefont {Kundu}\ and\ \citenamefont {Ghosh}(2022)}]{kundu2022high}%
  \BibitemOpen
  \bibfield  {author} {\bibinfo {author} {\bibfnamefont {S.}~\bibnamefont {Kundu}}\ and\ \bibinfo {author} {\bibfnamefont {D.}~\bibnamefont {Ghosh}},\ }\bibfield  {title} {\enquote {\bibinfo {title} {High-order interactions promote chimera states},}\ }\href@noop {} {\bibfield  {journal} {\bibinfo  {journal} {Phys. Rev. E}\ }\textbf {\bibinfo {volume} {105}},\ \bibinfo {pages} {L042202} (\bibinfo {year} {2022})}\BibitemShut {NoStop}%
\bibitem [{\citenamefont {Muolo}\ \emph {et~al.}(2024{\natexlab{b}})\citenamefont {Muolo}, \citenamefont {Njougouo}, \citenamefont {Gambuzza}, \citenamefont {Carletti},\ and\ \citenamefont {Frasca}}]{muolo2024phase}%
  \BibitemOpen
  \bibfield  {author} {\bibinfo {author} {\bibfnamefont {R.}~\bibnamefont {Muolo}}, \bibinfo {author} {\bibfnamefont {T.}~\bibnamefont {Njougouo}}, \bibinfo {author} {\bibfnamefont {L.}~\bibnamefont {Gambuzza}}, \bibinfo {author} {\bibfnamefont {T.}~\bibnamefont {Carletti}},\ and\ \bibinfo {author} {\bibfnamefont {M.}~\bibnamefont {Frasca}},\ }\bibfield  {title} {\enquote {\bibinfo {title} {Phase chimera states on nonlocal hyperrings},}\ }\href@noop {} {\bibfield  {journal} {\bibinfo  {journal} {Phys. Rev. E}\ }\textbf {\bibinfo {volume} {109}},\ \bibinfo {pages} {L022201} (\bibinfo {year} {2024}{\natexlab{b}})}\BibitemShut {NoStop}%
\bibitem [{\citenamefont {Tchinda~Djeudjo}\ \emph {et~al.}(2026)\citenamefont {Tchinda~Djeudjo}, \citenamefont {Carletti}, \citenamefont {Nakao},\ and\ \citenamefont {Muolo}}]{djeudjo2026chimera}%
  \BibitemOpen
  \bibfield  {author} {\bibinfo {author} {\bibfnamefont {R.}~\bibnamefont {Tchinda~Djeudjo}}, \bibinfo {author} {\bibfnamefont {T.}~\bibnamefont {Carletti}}, \bibinfo {author} {\bibfnamefont {H.}~\bibnamefont {Nakao}},\ and\ \bibinfo {author} {\bibfnamefont {R.}~\bibnamefont {Muolo}},\ }\bibfield  {title} {\enquote {\bibinfo {title} {Chimera states on m-directed hypergraphs},}\ }\href@noop {} {\bibfield  {journal} {\bibinfo  {journal} {Physical Review E}\ }\textbf {\bibinfo {volume} {113}} (\bibinfo {year} {2026})}\BibitemShut {NoStop}%
\bibitem [{\citenamefont {Anwar}\ \emph {et~al.}(2024)\citenamefont {Anwar}, \citenamefont {Sar}, \citenamefont {Perc},\ and\ \citenamefont {Ghosh}}]{anwar2024collective}%
  \BibitemOpen
  \bibfield  {author} {\bibinfo {author} {\bibfnamefont {M.~S.}\ \bibnamefont {Anwar}}, \bibinfo {author} {\bibfnamefont {G.~K.}\ \bibnamefont {Sar}}, \bibinfo {author} {\bibfnamefont {M.}~\bibnamefont {Perc}},\ and\ \bibinfo {author} {\bibfnamefont {D.}~\bibnamefont {Ghosh}},\ }\bibfield  {title} {\enquote {\bibinfo {title} {Collective dynamics of swarmalators with higher-order interactions},}\ }\href@noop {} {\bibfield  {journal} {\bibinfo  {journal} {Commun Phys}\ }\textbf {\bibinfo {volume} {7}},\ \bibinfo {pages} {1--11} (\bibinfo {year} {2024})}\BibitemShut {NoStop}%
\bibitem [{\citenamefont {Anwar}\ \emph {et~al.}(2025)\citenamefont {Anwar}, \citenamefont {Sar}, \citenamefont {Carletti},\ and\ \citenamefont {Ghosh}}]{anwar2025two}%
  \BibitemOpen
  \bibfield  {author} {\bibinfo {author} {\bibfnamefont {M.}~\bibnamefont {Anwar}}, \bibinfo {author} {\bibfnamefont {G.}~\bibnamefont {Sar}}, \bibinfo {author} {\bibfnamefont {T.}~\bibnamefont {Carletti}},\ and\ \bibinfo {author} {\bibfnamefont {D.}~\bibnamefont {Ghosh}},\ }\bibfield  {title} {\enquote {\bibinfo {title} {A two-dimensional swarmalator model with higher-order interactions},}\ }\href@noop {} {\bibfield  {journal} {\bibinfo  {journal} {SIAM Journal on Applied Mathematics, in press}\ } (\bibinfo {year} {2025})}\BibitemShut {NoStop}%
\bibitem [{\citenamefont {Le{\'o}n}\ \emph {et~al.}(2025{\natexlab{a}})\citenamefont {Le{\'o}n}, \citenamefont {Muolo}, \citenamefont {Nakao},\ and\ \citenamefont {Taga}}]{leon2025collective}%
  \BibitemOpen
  \bibfield  {author} {\bibinfo {author} {\bibfnamefont {I.}~\bibnamefont {Le{\'o}n}}, \bibinfo {author} {\bibfnamefont {R.}~\bibnamefont {Muolo}}, \bibinfo {author} {\bibfnamefont {H.}~\bibnamefont {Nakao}},\ and\ \bibinfo {author} {\bibfnamefont {K.}~\bibnamefont {Taga}},\ }\bibfield  {title} {\enquote {\bibinfo {title} {Collective dynamics of higher-order {V}icsek model emerging from local conformity interactions},}\ }\href@noop {} {\bibfield  {journal} {\bibinfo  {journal} {arXiv preprint arXiv:2512.19318}\ } (\bibinfo {year} {2025}{\natexlab{a}})}\BibitemShut {NoStop}%
\bibitem [{\citenamefont {Schaub}\ \emph {et~al.}(2020)\citenamefont {Schaub}, \citenamefont {Benson}, \citenamefont {Horn}, \citenamefont {Lippner},\ and\ \citenamefont {Jadbabaie}}]{schaub2020random}%
  \BibitemOpen
  \bibfield  {author} {\bibinfo {author} {\bibfnamefont {M.}~\bibnamefont {Schaub}}, \bibinfo {author} {\bibfnamefont {A.}~\bibnamefont {Benson}}, \bibinfo {author} {\bibfnamefont {P.}~\bibnamefont {Horn}}, \bibinfo {author} {\bibfnamefont {G.}~\bibnamefont {Lippner}},\ and\ \bibinfo {author} {\bibfnamefont {A.}~\bibnamefont {Jadbabaie}},\ }\bibfield  {title} {\enquote {\bibinfo {title} {Random walks on simplicial complexes and the normalized {H}odge 1-{L}aplacian},}\ }\href@noop {} {\bibfield  {journal} {\bibinfo  {journal} {SIAM Rev.}\ }\textbf {\bibinfo {volume} {62}},\ \bibinfo {pages} {353--391} (\bibinfo {year} {2020})}\BibitemShut {NoStop}%
\bibitem [{\citenamefont {Carletti}\ \emph {et~al.}(2020)\citenamefont {Carletti}, \citenamefont {Battiston}, \citenamefont {Cencetti},\ and\ \citenamefont {Fanelli}}]{carletti2020random}%
  \BibitemOpen
  \bibfield  {author} {\bibinfo {author} {\bibfnamefont {T.}~\bibnamefont {Carletti}}, \bibinfo {author} {\bibfnamefont {F.}~\bibnamefont {Battiston}}, \bibinfo {author} {\bibfnamefont {G.}~\bibnamefont {Cencetti}},\ and\ \bibinfo {author} {\bibfnamefont {D.}~\bibnamefont {Fanelli}},\ }\bibfield  {title} {\enquote {\bibinfo {title} {Random walks on hypergraphs},}\ }\href@noop {} {\bibfield  {journal} {\bibinfo  {journal} {Phys. Rev. E}\ }\textbf {\bibinfo {volume} {101}},\ \bibinfo {pages} {022308} (\bibinfo {year} {2020})}\BibitemShut {NoStop}%
\bibitem [{\citenamefont {Carletti}, \citenamefont {Fanelli},\ and\ \citenamefont {Nicoletti}(2020)}]{carletti2020dynamical}%
  \BibitemOpen
  \bibfield  {author} {\bibinfo {author} {\bibfnamefont {T.}~\bibnamefont {Carletti}}, \bibinfo {author} {\bibfnamefont {D.}~\bibnamefont {Fanelli}},\ and\ \bibinfo {author} {\bibfnamefont {S.}~\bibnamefont {Nicoletti}},\ }\bibfield  {title} {\enquote {\bibinfo {title} {Dynamical systems on hypergraphs},}\ }\href@noop {} {\bibfield  {journal} {\bibinfo  {journal} {J. Phys. Complex.}\ }\textbf {\bibinfo {volume} {1}},\ \bibinfo {pages} {035006} (\bibinfo {year} {2020})}\BibitemShut {NoStop}%
\bibitem [{\citenamefont {Muolo}\ \emph {et~al.}(2023)\citenamefont {Muolo}, \citenamefont {Gallo}, \citenamefont {Latora}, \citenamefont {Frasca},\ and\ \citenamefont {Carletti}}]{muolo2023turing}%
  \BibitemOpen
  \bibfield  {author} {\bibinfo {author} {\bibfnamefont {R.}~\bibnamefont {Muolo}}, \bibinfo {author} {\bibfnamefont {L.}~\bibnamefont {Gallo}}, \bibinfo {author} {\bibfnamefont {V.}~\bibnamefont {Latora}}, \bibinfo {author} {\bibfnamefont {M.}~\bibnamefont {Frasca}},\ and\ \bibinfo {author} {\bibfnamefont {T.}~\bibnamefont {Carletti}},\ }\bibfield  {title} {\enquote {\bibinfo {title} {Turing patterns in systems with high-order interactions},}\ }\href@noop {} {\bibfield  {journal} {\bibinfo  {journal} {Chaos, Solitons \& Fractals}\ }\textbf {\bibinfo {volume} {166}},\ \bibinfo {pages} {112912} (\bibinfo {year} {2023})}\BibitemShut {NoStop}%
\bibitem [{\citenamefont {Gao}\ \emph {et~al.}(2023)\citenamefont {Gao}, \citenamefont {Chang}, \citenamefont {Perc},\ and\ \citenamefont {Wang}}]{gao2023turing}%
  \BibitemOpen
  \bibfield  {author} {\bibinfo {author} {\bibfnamefont {S.}~\bibnamefont {Gao}}, \bibinfo {author} {\bibfnamefont {L.}~\bibnamefont {Chang}}, \bibinfo {author} {\bibfnamefont {M.}~\bibnamefont {Perc}},\ and\ \bibinfo {author} {\bibfnamefont {Z.}~\bibnamefont {Wang}},\ }\bibfield  {title} {\enquote {\bibinfo {title} {Turing patterns in simplicial complexes},}\ }\href@noop {} {\bibfield  {journal} {\bibinfo  {journal} {Phys. Rev. E}\ }\textbf {\bibinfo {volume} {107}},\ \bibinfo {pages} {014216} (\bibinfo {year} {2023})}\BibitemShut {NoStop}%
\bibitem [{\citenamefont {Iacopini}\ \emph {et~al.}(2019)\citenamefont {Iacopini}, \citenamefont {Petri}, \citenamefont {Barrat},\ and\ \citenamefont {Latora}}]{iacopini2019simplicial}%
  \BibitemOpen
  \bibfield  {author} {\bibinfo {author} {\bibfnamefont {I.}~\bibnamefont {Iacopini}}, \bibinfo {author} {\bibfnamefont {G.}~\bibnamefont {Petri}}, \bibinfo {author} {\bibfnamefont {A.}~\bibnamefont {Barrat}},\ and\ \bibinfo {author} {\bibfnamefont {V.}~\bibnamefont {Latora}},\ }\bibfield  {title} {\enquote {\bibinfo {title} {Simplicial models of social contagion},}\ }\href@noop {} {\bibfield  {journal} {\bibinfo  {journal} {Nat. Comm.}\ }\textbf {\bibinfo {volume} {10}},\ \bibinfo {pages} {2485} (\bibinfo {year} {2019})}\BibitemShut {NoStop}%
\bibitem [{\citenamefont {Neuh{\"a}user}, \citenamefont {Mellor},\ and\ \citenamefont {Lambiotte}(2020)}]{neuhauser2020multibody}%
  \BibitemOpen
  \bibfield  {author} {\bibinfo {author} {\bibfnamefont {L.}~\bibnamefont {Neuh{\"a}user}}, \bibinfo {author} {\bibfnamefont {A.}~\bibnamefont {Mellor}},\ and\ \bibinfo {author} {\bibfnamefont {R.}~\bibnamefont {Lambiotte}},\ }\bibfield  {title} {\enquote {\bibinfo {title} {Multibody interactions and nonlinear consensus dynamics on networked systems},}\ }\href@noop {} {\bibfield  {journal} {\bibinfo  {journal} {Phys. Rev. E}\ }\textbf {\bibinfo {volume} {101}},\ \bibinfo {pages} {032310} (\bibinfo {year} {2020})}\BibitemShut {NoStop}%
\bibitem [{\citenamefont {DeVille}(2021)}]{deville2020consensus}%
  \BibitemOpen
  \bibfield  {author} {\bibinfo {author} {\bibfnamefont {L.}~\bibnamefont {DeVille}},\ }\bibfield  {title} {\enquote {\bibinfo {title} {Consensus on simplicial complexes: Results on stability and synchronization},}\ }\href@noop {} {\bibfield  {journal} {\bibinfo  {journal} {Chaos}\ }\textbf {\bibinfo {volume} {31}},\ \bibinfo {pages} {023137} (\bibinfo {year} {2021})}\BibitemShut {NoStop}%
\bibitem [{\citenamefont {De~Lellis}\ \emph {et~al.}(2022)\citenamefont {De~Lellis}, \citenamefont {Della~Rossa}, \citenamefont {Iudice},\ and\ \citenamefont {Liuzza}}]{de2022pinning}%
  \BibitemOpen
  \bibfield  {author} {\bibinfo {author} {\bibfnamefont {P.}~\bibnamefont {De~Lellis}}, \bibinfo {author} {\bibfnamefont {F.}~\bibnamefont {Della~Rossa}}, \bibinfo {author} {\bibfnamefont {F.~L.}\ \bibnamefont {Iudice}},\ and\ \bibinfo {author} {\bibfnamefont {D.}~\bibnamefont {Liuzza}},\ }\bibfield  {title} {\enquote {\bibinfo {title} {Pinning control of hypergraphs},}\ }\href@noop {} {\bibfield  {journal} {\bibinfo  {journal} {IEEE Control Syst. Lett.}\ }\textbf {\bibinfo {volume} {7}},\ \bibinfo {pages} {691--696} (\bibinfo {year} {2022})}\BibitemShut {NoStop}%
\bibitem [{\citenamefont {Xia}\ and\ \citenamefont {Xiang}(2024)}]{xia2024pinning}%
  \BibitemOpen
  \bibfield  {author} {\bibinfo {author} {\bibfnamefont {R.}~\bibnamefont {Xia}}\ and\ \bibinfo {author} {\bibfnamefont {L.}~\bibnamefont {Xiang}},\ }\bibfield  {title} {\enquote {\bibinfo {title} {Pinning control of simplicial complexes},}\ }\href@noop {} {\bibfield  {journal} {\bibinfo  {journal} {European Journal of Control}\ }\textbf {\bibinfo {volume} {77}},\ \bibinfo {pages} {100994} (\bibinfo {year} {2024})}\BibitemShut {NoStop}%
\bibitem [{\citenamefont {Muolo}\ \emph {et~al.}(2025)\citenamefont {Muolo}, \citenamefont {Gambuzza}, \citenamefont {Nakao},\ and\ \citenamefont {Frasca}}]{muolo2025pinning}%
  \BibitemOpen
  \bibfield  {author} {\bibinfo {author} {\bibfnamefont {R.}~\bibnamefont {Muolo}}, \bibinfo {author} {\bibfnamefont {L.}~\bibnamefont {Gambuzza}}, \bibinfo {author} {\bibfnamefont {H.}~\bibnamefont {Nakao}},\ and\ \bibinfo {author} {\bibfnamefont {M.}~\bibnamefont {Frasca}},\ }\bibfield  {title} {\enquote {\bibinfo {title} {Pinning control of chimera states in systems with higher-order interactions},}\ }\href@noop {} {\bibfield  {journal} {\bibinfo  {journal} {Nonlinear Dynamics}\ ,\ \bibinfo {pages} {1--23}} (\bibinfo {year} {2025})}\BibitemShut {NoStop}%
\bibitem [{\citenamefont {Tanaka}\ and\ \citenamefont {Aoyagi}(2011)}]{tanaka2011multistable}%
  \BibitemOpen
  \bibfield  {author} {\bibinfo {author} {\bibfnamefont {T.}~\bibnamefont {Tanaka}}\ and\ \bibinfo {author} {\bibfnamefont {T.}~\bibnamefont {Aoyagi}},\ }\bibfield  {title} {\enquote {\bibinfo {title} {Multistable attractors in a network of phase oscillators with three-body interactions},}\ }\href@noop {} {\bibfield  {journal} {\bibinfo  {journal} {Phys. Rev. Lett.}\ }\textbf {\bibinfo {volume} {106}},\ \bibinfo {pages} {224101} (\bibinfo {year} {2011})}\BibitemShut {NoStop}%
\bibitem [{\citenamefont {Bick}, \citenamefont {Ashwin},\ and\ \citenamefont {Rodrigues}(2016)}]{bick2016chaos}%
  \BibitemOpen
  \bibfield  {author} {\bibinfo {author} {\bibfnamefont {C.}~\bibnamefont {Bick}}, \bibinfo {author} {\bibfnamefont {P.}~\bibnamefont {Ashwin}},\ and\ \bibinfo {author} {\bibfnamefont {A.}~\bibnamefont {Rodrigues}},\ }\bibfield  {title} {\enquote {\bibinfo {title} {Chaos in generically coupled phase oscillator networks with nonpairwise interactions},}\ }\href@noop {} {\bibfield  {journal} {\bibinfo  {journal} {Chaos}\ }\textbf {\bibinfo {volume} {26}} (\bibinfo {year} {2016})}\BibitemShut {NoStop}%
\bibitem [{\citenamefont {Skardal}\ and\ \citenamefont {Arenas}(2019)}]{skardal2019abrupt}%
  \BibitemOpen
  \bibfield  {author} {\bibinfo {author} {\bibfnamefont {P.}~\bibnamefont {Skardal}}\ and\ \bibinfo {author} {\bibfnamefont {A.}~\bibnamefont {Arenas}},\ }\bibfield  {title} {\enquote {\bibinfo {title} {Abrupt desynchronization and extensive multistability in globally coupled oscillator simplexes},}\ }\href@noop {} {\bibfield  {journal} {\bibinfo  {journal} {Phys. Rev. Lett.}\ }\textbf {\bibinfo {volume} {122}},\ \bibinfo {pages} {248301} (\bibinfo {year} {2019})}\BibitemShut {NoStop}%
\bibitem [{\citenamefont {Skardal}\ and\ \citenamefont {Arenas}(2020)}]{skardal2020higher}%
  \BibitemOpen
  \bibfield  {author} {\bibinfo {author} {\bibfnamefont {P.}~\bibnamefont {Skardal}}\ and\ \bibinfo {author} {\bibfnamefont {A.}~\bibnamefont {Arenas}},\ }\bibfield  {title} {\enquote {\bibinfo {title} {Higher order interactions in complex networks of phase oscillators promote abrupt synchronization switching},}\ }\href@noop {} {\bibfield  {journal} {\bibinfo  {journal} {Comm. Phys.}\ }\textbf {\bibinfo {volume} {3}},\ \bibinfo {pages} {1--6} (\bibinfo {year} {2020})}\BibitemShut {NoStop}%
\bibitem [{\citenamefont {Millán}, \citenamefont {Torres},\ and\ \citenamefont {Bianconi}(2020)}]{millan2020explosive}%
  \BibitemOpen
  \bibfield  {author} {\bibinfo {author} {\bibfnamefont {A.}~\bibnamefont {Millán}}, \bibinfo {author} {\bibfnamefont {J.}~\bibnamefont {Torres}},\ and\ \bibinfo {author} {\bibfnamefont {G.}~\bibnamefont {Bianconi}},\ }\bibfield  {title} {\enquote {\bibinfo {title} {Explosive higher-order {K}uramoto dynamics on simplicial complexes},}\ }\href@noop {} {\bibfield  {journal} {\bibinfo  {journal} {Phys. Rev. Lett.}\ }\textbf {\bibinfo {volume} {124}},\ \bibinfo {pages} {218301} (\bibinfo {year} {2020})}\BibitemShut {NoStop}%
\bibitem [{\citenamefont {Lucas}, \citenamefont {Cencetti},\ and\ \citenamefont {Battiston}(2020)}]{lucas2020multiorder}%
  \BibitemOpen
  \bibfield  {author} {\bibinfo {author} {\bibfnamefont {M.}~\bibnamefont {Lucas}}, \bibinfo {author} {\bibfnamefont {G.}~\bibnamefont {Cencetti}},\ and\ \bibinfo {author} {\bibfnamefont {F.}~\bibnamefont {Battiston}},\ }\bibfield  {title} {\enquote {\bibinfo {title} {Multiorder laplacian for synchronization in higher-order networks},}\ }\href@noop {} {\bibfield  {journal} {\bibinfo  {journal} {Physical Review Research}\ }\textbf {\bibinfo {volume} {2}},\ \bibinfo {pages} {033410} (\bibinfo {year} {2020})}\BibitemShut {NoStop}%
\bibitem [{\citenamefont {Adhikari}, \citenamefont {Restrepo},\ and\ \citenamefont {Skardal}(2023)}]{adhikari2023synchronization}%
  \BibitemOpen
  \bibfield  {author} {\bibinfo {author} {\bibfnamefont {S.}~\bibnamefont {Adhikari}}, \bibinfo {author} {\bibfnamefont {J.}~\bibnamefont {Restrepo}},\ and\ \bibinfo {author} {\bibfnamefont {P.}~\bibnamefont {Skardal}},\ }\bibfield  {title} {\enquote {\bibinfo {title} {Synchronization of phase oscillators on complex hypergraphs},}\ }\href@noop {} {\bibfield  {journal} {\bibinfo  {journal} {Chaos}\ }\textbf {\bibinfo {volume} {33}} (\bibinfo {year} {2023})}\BibitemShut {NoStop}%
\bibitem [{\citenamefont {Skardal}, \citenamefont {Adhikari},\ and\ \citenamefont {Restrepo}(2023)}]{skardal2023multistability}%
  \BibitemOpen
  \bibfield  {author} {\bibinfo {author} {\bibfnamefont {P.}~\bibnamefont {Skardal}}, \bibinfo {author} {\bibfnamefont {S.}~\bibnamefont {Adhikari}},\ and\ \bibinfo {author} {\bibfnamefont {J.}~\bibnamefont {Restrepo}},\ }\bibfield  {title} {\enquote {\bibinfo {title} {Multistability in coupled oscillator systems with higher-order interactions and community structure},}\ }\href@noop {} {\bibfield  {journal} {\bibinfo  {journal} {Chaos}\ }\textbf {\bibinfo {volume} {33}} (\bibinfo {year} {2023})}\BibitemShut {NoStop}%
\bibitem [{\citenamefont {Carballosa}\ \emph {et~al.}(2023)\citenamefont {Carballosa}, \citenamefont {Mu{\~n}uzuri}, \citenamefont {Boccaletti}, \citenamefont {Torcini},\ and\ \citenamefont {Olmi}}]{carballosa2023cluster}%
  \BibitemOpen
  \bibfield  {author} {\bibinfo {author} {\bibfnamefont {A.}~\bibnamefont {Carballosa}}, \bibinfo {author} {\bibfnamefont {A.~P.}\ \bibnamefont {Mu{\~n}uzuri}}, \bibinfo {author} {\bibfnamefont {S.}~\bibnamefont {Boccaletti}}, \bibinfo {author} {\bibfnamefont {A.}~\bibnamefont {Torcini}},\ and\ \bibinfo {author} {\bibfnamefont {S.}~\bibnamefont {Olmi}},\ }\bibfield  {title} {\enquote {\bibinfo {title} {Cluster states and $\pi$-transition in the {K}uramoto model with higher order interactions},}\ }\href@noop {} {\bibfield  {journal} {\bibinfo  {journal} {Chaos, Solitons \& Fractals}\ }\textbf {\bibinfo {volume} {177}},\ \bibinfo {pages} {114197} (\bibinfo {year} {2023})}\BibitemShut {NoStop}%
\bibitem [{\citenamefont {León}\ \emph {et~al.}(2024)\citenamefont {León}, \citenamefont {Muolo}, \citenamefont {Hata},\ and\ \citenamefont {Nakao}}]{leon2024}%
  \BibitemOpen
  \bibfield  {author} {\bibinfo {author} {\bibfnamefont {I.}~\bibnamefont {León}}, \bibinfo {author} {\bibfnamefont {R.}~\bibnamefont {Muolo}}, \bibinfo {author} {\bibfnamefont {S.}~\bibnamefont {Hata}},\ and\ \bibinfo {author} {\bibfnamefont {H.}~\bibnamefont {Nakao}},\ }\bibfield  {title} {\enquote {\bibinfo {title} {Higher-order interactions induce anomalous transitions to synchrony},}\ }\href@noop {} {\bibfield  {journal} {\bibinfo  {journal} {Chaos}\ }\textbf {\bibinfo {volume} {34}},\ \bibinfo {pages} {013105} (\bibinfo {year} {2024})}\BibitemShut {NoStop}%
\bibitem [{\citenamefont {Costa}, \citenamefont {Novaes},\ and\ \citenamefont {de~Aguiar}(2024)}]{costa2024bifurcations}%
  \BibitemOpen
  \bibfield  {author} {\bibinfo {author} {\bibfnamefont {G.}~\bibnamefont {Costa}}, \bibinfo {author} {\bibfnamefont {M.}~\bibnamefont {Novaes}},\ and\ \bibinfo {author} {\bibfnamefont {M.}~\bibnamefont {de~Aguiar}},\ }\bibfield  {title} {\enquote {\bibinfo {title} {Bifurcations in the {K}uramoto model with external forcing and higher-order interactions},}\ }\href@noop {} {\bibfield  {journal} {\bibinfo  {journal} {Chaos}\ }\textbf {\bibinfo {volume} {34}} (\bibinfo {year} {2024})}\BibitemShut {NoStop}%
\bibitem [{\citenamefont {Huh}\ and\ \citenamefont {Kim}(2024)}]{huh2024critical}%
  \BibitemOpen
  \bibfield  {author} {\bibinfo {author} {\bibfnamefont {H.}~\bibnamefont {Huh}}\ and\ \bibinfo {author} {\bibfnamefont {D.}~\bibnamefont {Kim}},\ }\bibfield  {title} {\enquote {\bibinfo {title} {Critical threshold for synchronizability of high-dimensional {K}uramoto oscillators under higher-order interactions},}\ }\href@noop {} {\bibfield  {journal} {\bibinfo  {journal} {Chaos}\ }\textbf {\bibinfo {volume} {34}} (\bibinfo {year} {2024})}\BibitemShut {NoStop}%
\bibitem [{\citenamefont {Wang}\ \emph {et~al.}(2024)\citenamefont {Wang}, \citenamefont {Li}, \citenamefont {Dai},\ and\ \citenamefont {Yang}}]{wang2024coexistence}%
  \BibitemOpen
  \bibfield  {author} {\bibinfo {author} {\bibfnamefont {X.}~\bibnamefont {Wang}}, \bibinfo {author} {\bibfnamefont {H.}~\bibnamefont {Li}}, \bibinfo {author} {\bibfnamefont {Q.}~\bibnamefont {Dai}},\ and\ \bibinfo {author} {\bibfnamefont {J.}~\bibnamefont {Yang}},\ }\bibfield  {title} {\enquote {\bibinfo {title} {Coexistence of multistable synchronous states in a three-oscillator system with higher-order interaction},}\ }\href@noop {} {\bibfield  {journal} {\bibinfo  {journal} {Phys. Rev. E}\ }\textbf {\bibinfo {volume} {110}},\ \bibinfo {pages} {034311} (\bibinfo {year} {2024})}\BibitemShut {NoStop}%
\bibitem [{\citenamefont {Smith}\ and\ \citenamefont {Liu}(2024)}]{smith2024determining}%
  \BibitemOpen
  \bibfield  {author} {\bibinfo {author} {\bibfnamefont {L.~D.}\ \bibnamefont {Smith}}\ and\ \bibinfo {author} {\bibfnamefont {P.}~\bibnamefont {Liu}},\ }\bibfield  {title} {\enquote {\bibinfo {title} {Determining bifurcations to explosive synchronization for networks of coupled oscillators with higher-order interactions},}\ }\href@noop {} {\bibfield  {journal} {\bibinfo  {journal} {Physical Review E}\ }\textbf {\bibinfo {volume} {109}},\ \bibinfo {pages} {L022202} (\bibinfo {year} {2024})}\BibitemShut {NoStop}%
\bibitem [{\citenamefont {Dai}\ and\ \citenamefont {Kori}(2025)}]{dai2025higher}%
  \BibitemOpen
  \bibfield  {author} {\bibinfo {author} {\bibfnamefont {Q.}~\bibnamefont {Dai}}\ and\ \bibinfo {author} {\bibfnamefont {H.}~\bibnamefont {Kori}},\ }\bibfield  {title} {\enquote {\bibinfo {title} {Higher-order synchronization and multistability in higher-order {K}uramoto phase oscillators: Q. dai, h. kori},}\ }\href@noop {} {\bibfield  {journal} {\bibinfo  {journal} {Nonlinear Dynamics}\ }\textbf {\bibinfo {volume} {113}},\ \bibinfo {pages} {21801--21811} (\bibinfo {year} {2025})}\BibitemShut {NoStop}%
\bibitem [{\citenamefont {Zhang}\ \emph {et~al.}(2024)\citenamefont {Zhang}, \citenamefont {Skardal}, \citenamefont {Battiston}, \citenamefont {Petri},\ and\ \citenamefont {Lucas}}]{zhang2024deeper}%
  \BibitemOpen
  \bibfield  {author} {\bibinfo {author} {\bibfnamefont {Y.}~\bibnamefont {Zhang}}, \bibinfo {author} {\bibfnamefont {P.}~\bibnamefont {Skardal}}, \bibinfo {author} {\bibfnamefont {F.}~\bibnamefont {Battiston}}, \bibinfo {author} {\bibfnamefont {G.}~\bibnamefont {Petri}},\ and\ \bibinfo {author} {\bibfnamefont {M.}~\bibnamefont {Lucas}},\ }\bibfield  {title} {\enquote {\bibinfo {title} {Deeper but smaller: Higher-order interactions increase linear stability but shrink basins},}\ }\href@noop {} {\bibfield  {journal} {\bibinfo  {journal} {Science Advances}\ }\textbf {\bibinfo {volume} {10}},\ \bibinfo {pages} {eado8049} (\bibinfo {year} {2024})}\BibitemShut {NoStop}%
\bibitem [{\citenamefont {Fariello}\ and\ \citenamefont {de~Aguiar}(2024)}]{fariello2024third}%
  \BibitemOpen
  \bibfield  {author} {\bibinfo {author} {\bibfnamefont {R.}~\bibnamefont {Fariello}}\ and\ \bibinfo {author} {\bibfnamefont {M.~A.}\ \bibnamefont {de~Aguiar}},\ }\bibfield  {title} {\enquote {\bibinfo {title} {Third order interactions shift the critical coupling in multidimensional {K}uramoto models},}\ }\href@noop {} {\bibfield  {journal} {\bibinfo  {journal} {Chaos, Solitons \& Fractals}\ }\textbf {\bibinfo {volume} {187}},\ \bibinfo {pages} {115467} (\bibinfo {year} {2024})}\BibitemShut {NoStop}%
\bibitem [{\citenamefont {von~der Gracht}, \citenamefont {Nijholt},\ and\ \citenamefont {Rink}(2024)}]{von2024higher}%
  \BibitemOpen
  \bibfield  {author} {\bibinfo {author} {\bibfnamefont {S.}~\bibnamefont {von~der Gracht}}, \bibinfo {author} {\bibfnamefont {E.}~\bibnamefont {Nijholt}},\ and\ \bibinfo {author} {\bibfnamefont {B.}~\bibnamefont {Rink}},\ }\bibfield  {title} {\enquote {\bibinfo {title} {Higher-order interactions lead to 'reluctant' synchrony breaking},}\ }\bibfield  {booktitle} {\emph {\bibinfo {booktitle} {Proceedings of the Royal Society A}},\ }\href@noop {} {\ \textbf {\bibinfo {volume} {480}},\ \bibinfo {pages} {20230945} (\bibinfo {year} {2024})}\BibitemShut {NoStop}%
\bibitem [{\citenamefont {Wang}\ \emph {et~al.}(2025)\citenamefont {Wang}, \citenamefont {Qi}, \citenamefont {Zhu},\ and\ \citenamefont {Liu}}]{wang2025higher}%
  \BibitemOpen
  \bibfield  {author} {\bibinfo {author} {\bibfnamefont {Z.}~\bibnamefont {Wang}}, \bibinfo {author} {\bibfnamefont {W.}~\bibnamefont {Qi}}, \bibinfo {author} {\bibfnamefont {J.}~\bibnamefont {Zhu}},\ and\ \bibinfo {author} {\bibfnamefont {X.}~\bibnamefont {Liu}},\ }\bibfield  {title} {\enquote {\bibinfo {title} {How do higher-order interactions shape the energy landscape?}}\ }\href@noop {} {\bibfield  {journal} {\bibinfo  {journal} {Phys. Rev. E}\ }\textbf {\bibinfo {volume} {112}},\ \bibinfo {pages} {064217} (\bibinfo {year} {2025})}\BibitemShut {NoStop}%
\bibitem [{\citenamefont {Kuramoto}(1984)}]{Kuramoto_book}%
  \BibitemOpen
  \bibfield  {author} {\bibinfo {author} {\bibfnamefont {Y.}~\bibnamefont {Kuramoto}},\ }\href@noop {} {\emph {\bibinfo {title} {Chemical oscillations, waves, and turbulence}}}\ (\bibinfo  {publisher} {Springer-Verlag},\ \bibinfo {address} {New York},\ \bibinfo {year} {1984})\BibitemShut {NoStop}%
\bibitem [{\citenamefont {Menara}\ \emph {et~al.}(2022)\citenamefont {Menara}, \citenamefont {Baggio}, \citenamefont {Bassett},\ and\ \citenamefont {Pasqualetti}}]{menara2022functional}%
  \BibitemOpen
  \bibfield  {author} {\bibinfo {author} {\bibfnamefont {T.}~\bibnamefont {Menara}}, \bibinfo {author} {\bibfnamefont {G.}~\bibnamefont {Baggio}}, \bibinfo {author} {\bibfnamefont {D.}~\bibnamefont {Bassett}},\ and\ \bibinfo {author} {\bibfnamefont {F.}~\bibnamefont {Pasqualetti}},\ }\bibfield  {title} {\enquote {\bibinfo {title} {Functional control of oscillator networks},}\ }\href@noop {} {\bibfield  {journal} {\bibinfo  {journal} {Nature Communications}\ }\textbf {\bibinfo {volume} {13}},\ \bibinfo {pages} {4721} (\bibinfo {year} {2022})}\BibitemShut {NoStop}%
\bibitem [{\citenamefont {Simpson-Porco}, \citenamefont {D{\"o}rfler},\ and\ \citenamefont {Bullo}(2013)}]{simpson2013synchronization}%
  \BibitemOpen
  \bibfield  {author} {\bibinfo {author} {\bibfnamefont {J.~W.}\ \bibnamefont {Simpson-Porco}}, \bibinfo {author} {\bibfnamefont {F.}~\bibnamefont {D{\"o}rfler}},\ and\ \bibinfo {author} {\bibfnamefont {F.}~\bibnamefont {Bullo}},\ }\bibfield  {title} {\enquote {\bibinfo {title} {Synchronization and power sharing for droop-controlled inverters in islanded microgrids},}\ }\href@noop {} {\bibfield  {journal} {\bibinfo  {journal} {Automatica}\ }\textbf {\bibinfo {volume} {49}},\ \bibinfo {pages} {2603--2611} (\bibinfo {year} {2013})}\BibitemShut {NoStop}%
\bibitem [{\citenamefont {Alderisio}\ \emph {et~al.}(2017)\citenamefont {Alderisio}, \citenamefont {Fiore}, \citenamefont {Salesse}, \citenamefont {Bardy},\ and\ \citenamefont {di~Bernardo}}]{alderisio2017interaction}%
  \BibitemOpen
  \bibfield  {author} {\bibinfo {author} {\bibfnamefont {F.}~\bibnamefont {Alderisio}}, \bibinfo {author} {\bibfnamefont {G.}~\bibnamefont {Fiore}}, \bibinfo {author} {\bibfnamefont {R.~N.}\ \bibnamefont {Salesse}}, \bibinfo {author} {\bibfnamefont {B.~G.}\ \bibnamefont {Bardy}},\ and\ \bibinfo {author} {\bibfnamefont {M.}~\bibnamefont {di~Bernardo}},\ }\bibfield  {title} {\enquote {\bibinfo {title} {Interaction patterns and individual dynamics shape the way we move in synchrony},}\ }\href@noop {} {\bibfield  {journal} {\bibinfo  {journal} {Scientific Reports}\ }\textbf {\bibinfo {volume} {7}},\ \bibinfo {pages} {6846} (\bibinfo {year} {2017})}\BibitemShut {NoStop}%
\bibitem [{\citenamefont {Lamata-Ot{\'\i}n}\ \emph {et~al.}(2025)\citenamefont {Lamata-Ot{\'\i}n}, \citenamefont {Malizia}, \citenamefont {Latora}, \citenamefont {Frasca},\ and\ \citenamefont {G{\'o}mez-Garde{\~n}es}}]{lamata2025hyperedge}%
  \BibitemOpen
  \bibfield  {author} {\bibinfo {author} {\bibfnamefont {S.}~\bibnamefont {Lamata-Ot{\'\i}n}}, \bibinfo {author} {\bibfnamefont {F.}~\bibnamefont {Malizia}}, \bibinfo {author} {\bibfnamefont {V.}~\bibnamefont {Latora}}, \bibinfo {author} {\bibfnamefont {M.}~\bibnamefont {Frasca}},\ and\ \bibinfo {author} {\bibfnamefont {J.}~\bibnamefont {G{\'o}mez-Garde{\~n}es}},\ }\bibfield  {title} {\enquote {\bibinfo {title} {Hyperedge overlap drives synchronizability of systems with higher-order interactions},}\ }\href@noop {} {\bibfield  {journal} {\bibinfo  {journal} {Phys. Rev. E}\ }\textbf {\bibinfo {volume} {111}},\ \bibinfo {pages} {034302} (\bibinfo {year} {2025})}\BibitemShut {NoStop}%
\bibitem [{\citenamefont {Pecora}\ and\ \citenamefont {Carroll}(1998)}]{pecora1998master}%
  \BibitemOpen
  \bibfield  {author} {\bibinfo {author} {\bibfnamefont {L.~M.}\ \bibnamefont {Pecora}}\ and\ \bibinfo {author} {\bibfnamefont {T.~L.}\ \bibnamefont {Carroll}},\ }\bibfield  {title} {\enquote {\bibinfo {title} {Master stability functions for synchronized coupled systems},}\ }\href@noop {} {\bibfield  {journal} {\bibinfo  {journal} {Phys. Rev. Lett.}\ }\textbf {\bibinfo {volume} {80}},\ \bibinfo {pages} {2109--2112} (\bibinfo {year} {1998})}\BibitemShut {NoStop}%
\bibitem [{\citenamefont {Coraggio}\ \emph {et~al.}(2018)\citenamefont {Coraggio}, \citenamefont {De~Lellis}, \citenamefont {Hogan},\ and\ \citenamefont {{di Bernardo}}}]{coraggio2018synchronization}%
  \BibitemOpen
  \bibfield  {author} {\bibinfo {author} {\bibfnamefont {M.}~\bibnamefont {Coraggio}}, \bibinfo {author} {\bibfnamefont {P.}~\bibnamefont {De~Lellis}}, \bibinfo {author} {\bibfnamefont {S.~J.}\ \bibnamefont {Hogan}},\ and\ \bibinfo {author} {\bibfnamefont {M.}~\bibnamefont {{di Bernardo}}},\ }\bibfield  {title} {\enquote {\bibinfo {title} {Synchronization of networks of piecewise-smooth systems},}\ }\href@noop {} {\bibfield  {journal} {\bibinfo  {journal} {IEEE Control Systems Letters}\ }\textbf {\bibinfo {volume} {2}},\ \bibinfo {pages} {653--658} (\bibinfo {year} {2018})}\BibitemShut {NoStop}%
\bibitem [{\citenamefont {Coraggio}, \citenamefont {De~Lellis},\ and\ \citenamefont {{di Bernardo}}(2021)}]{coraggio2021convergence}%
  \BibitemOpen
  \bibfield  {author} {\bibinfo {author} {\bibfnamefont {M.}~\bibnamefont {Coraggio}}, \bibinfo {author} {\bibfnamefont {P.}~\bibnamefont {De~Lellis}},\ and\ \bibinfo {author} {\bibfnamefont {M.}~\bibnamefont {{di Bernardo}}},\ }\bibfield  {title} {\enquote {\bibinfo {title} {Convergence and synchronization in networks of piecewise-smooth systems via distributed discontinuous coupling},}\ }\href@noop {} {\bibfield  {journal} {\bibinfo  {journal} {Automatica}\ }\textbf {\bibinfo {volume} {129}},\ \bibinfo {pages} {109596} (\bibinfo {year} {2021})}\BibitemShut {NoStop}%
\bibitem [{\citenamefont {Coraggio}, \citenamefont {DeLellis},\ and\ \citenamefont {{di Bernardo}}(2020)}]{coraggio2020distributed}%
  \BibitemOpen
  \bibfield  {author} {\bibinfo {author} {\bibfnamefont {M.}~\bibnamefont {Coraggio}}, \bibinfo {author} {\bibfnamefont {P.}~\bibnamefont {DeLellis}},\ and\ \bibinfo {author} {\bibfnamefont {M.}~\bibnamefont {{di Bernardo}}},\ }\bibfield  {title} {\enquote {\bibinfo {title} {Distributed discontinuous coupling for convergence in heterogeneous networks},}\ }\href@noop {} {\bibfield  {journal} {\bibinfo  {journal} {IEEE Control Systems Letters}\ }\textbf {\bibinfo {volume} {5}},\ \bibinfo {pages} {1037--1042} (\bibinfo {year} {2020})}\BibitemShut {NoStop}%
\bibitem [{\citenamefont {Barahona}\ and\ \citenamefont {Pecora}(2002)}]{barahona2002synchronization}%
  \BibitemOpen
  \bibfield  {author} {\bibinfo {author} {\bibfnamefont {M.}~\bibnamefont {Barahona}}\ and\ \bibinfo {author} {\bibfnamefont {L.}~\bibnamefont {Pecora}},\ }\bibfield  {title} {\enquote {\bibinfo {title} {Synchronization in small-world systems},}\ }\href@noop {} {\bibfield  {journal} {\bibinfo  {journal} {Phys. Rev. Lett.}\ }\textbf {\bibinfo {volume} {89}},\ \bibinfo {pages} {054101} (\bibinfo {year} {2002})}\BibitemShut {NoStop}%
\bibitem [{\citenamefont {Donetti}, \citenamefont {Hurtado},\ and\ \citenamefont {Mu{\~n}oz}(2005)}]{donetti2005entangled}%
  \BibitemOpen
  \bibfield  {author} {\bibinfo {author} {\bibfnamefont {L.}~\bibnamefont {Donetti}}, \bibinfo {author} {\bibfnamefont {P.~I.}\ \bibnamefont {Hurtado}},\ and\ \bibinfo {author} {\bibfnamefont {M.~A.}\ \bibnamefont {Mu{\~n}oz}},\ }\bibfield  {title} {\enquote {\bibinfo {title} {Entangled networks, synchronization, and optimal network topology},}\ }\href@noop {} {\bibfield  {journal} {\bibinfo  {journal} {Phys. Rev. Lett.}\ }\textbf {\bibinfo {volume} {95}},\ \bibinfo {pages} {188701} (\bibinfo {year} {2005})}\BibitemShut {NoStop}%
\bibitem [{\citenamefont {Donetti}, \citenamefont {Neri},\ and\ \citenamefont {Mu{\~n}oz}(2006)}]{donetti2006optimal}%
  \BibitemOpen
  \bibfield  {author} {\bibinfo {author} {\bibfnamefont {L.}~\bibnamefont {Donetti}}, \bibinfo {author} {\bibfnamefont {F.}~\bibnamefont {Neri}},\ and\ \bibinfo {author} {\bibfnamefont {M.~A.}\ \bibnamefont {Mu{\~n}oz}},\ }\bibfield  {title} {\enquote {\bibinfo {title} {Optimal network topologies: Expanders, cages, ramanujan graphs, entangled networks and all that},}\ }\href@noop {} {\bibfield  {journal} {\bibinfo  {journal} {Journal of Statistical Mechanics: Theory and Experiment}\ }\textbf {\bibinfo {volume} {2006}},\ \bibinfo {pages} {P08007} (\bibinfo {year} {2006})}\BibitemShut {NoStop}%
\bibitem [{\citenamefont {Skardal}, \citenamefont {Taylor},\ and\ \citenamefont {Sun}(2014)}]{Skardal2014Optimal}%
  \BibitemOpen
  \bibfield  {author} {\bibinfo {author} {\bibfnamefont {P.~S.}\ \bibnamefont {Skardal}}, \bibinfo {author} {\bibfnamefont {D.}~\bibnamefont {Taylor}},\ and\ \bibinfo {author} {\bibfnamefont {J.}~\bibnamefont {Sun}},\ }\bibfield  {title} {\enquote {\bibinfo {title} {Optimal synchronization of complex networks},}\ }\href@noop {} {\bibfield  {journal} {\bibinfo  {journal} {Physical Review Letters}\ }\textbf {\bibinfo {volume} {113}},\ \bibinfo {pages} {144101} (\bibinfo {year} {2014})}\BibitemShut {NoStop}%
\bibitem [{\citenamefont {Fazlyab}, \citenamefont {D{\"o}rfler},\ and\ \citenamefont {Preciado}(2017)}]{fazlyab2017optimal}%
  \BibitemOpen
  \bibfield  {author} {\bibinfo {author} {\bibfnamefont {M.}~\bibnamefont {Fazlyab}}, \bibinfo {author} {\bibfnamefont {F.}~\bibnamefont {D{\"o}rfler}},\ and\ \bibinfo {author} {\bibfnamefont {V.~M.}\ \bibnamefont {Preciado}},\ }\bibfield  {title} {\enquote {\bibinfo {title} {Optimal network design for synchronization of coupled oscillators},}\ }\href@noop {} {\bibfield  {journal} {\bibinfo  {journal} {Automatica}\ }\textbf {\bibinfo {volume} {84}},\ \bibinfo {pages} {181--189} (\bibinfo {year} {2017})}\BibitemShut {NoStop}%
\bibitem [{\citenamefont {Lei}\ \emph {et~al.}(2023)\citenamefont {Lei}, \citenamefont {Xu}, \citenamefont {Wang}, \citenamefont {Zou},\ and\ \citenamefont {Kurths}}]{lei2023new}%
  \BibitemOpen
  \bibfield  {author} {\bibinfo {author} {\bibfnamefont {Y.}~\bibnamefont {Lei}}, \bibinfo {author} {\bibfnamefont {X.-J.}\ \bibnamefont {Xu}}, \bibinfo {author} {\bibfnamefont {X.}~\bibnamefont {Wang}}, \bibinfo {author} {\bibfnamefont {Y.}~\bibnamefont {Zou}},\ and\ \bibinfo {author} {\bibfnamefont {J.}~\bibnamefont {Kurths}},\ }\bibfield  {title} {\enquote {\bibinfo {title} {A new criterion for optimizing synchrony of coupled oscillators},}\ }\href@noop {} {\bibfield  {journal} {\bibinfo  {journal} {Chaos, Solitons \& Fractals}\ }\textbf {\bibinfo {volume} {168}},\ \bibinfo {pages} {113192} (\bibinfo {year} {2023})}\BibitemShut {NoStop}%
\bibitem [{\citenamefont {Nishikawa}\ and\ \citenamefont {Motter}(2006)}]{nishikawa2006synchronization}%
  \BibitemOpen
  \bibfield  {author} {\bibinfo {author} {\bibfnamefont {T.}~\bibnamefont {Nishikawa}}\ and\ \bibinfo {author} {\bibfnamefont {A.~E.}\ \bibnamefont {Motter}},\ }\bibfield  {title} {\enquote {\bibinfo {title} {Synchronization is optimal in nondiagonalizable networks},}\ }\href@noop {} {\bibfield  {journal} {\bibinfo  {journal} {Phys. Rev. E}\ }\textbf {\bibinfo {volume} {73}},\ \bibinfo {pages} {065106} (\bibinfo {year} {2006})}\BibitemShut {NoStop}%
\bibitem [{\citenamefont {Estrada}, \citenamefont {Gago},\ and\ \citenamefont {Caporossi}(2010)}]{estrada2010design}%
  \BibitemOpen
  \bibfield  {author} {\bibinfo {author} {\bibfnamefont {E.}~\bibnamefont {Estrada}}, \bibinfo {author} {\bibfnamefont {S.}~\bibnamefont {Gago}},\ and\ \bibinfo {author} {\bibfnamefont {G.}~\bibnamefont {Caporossi}},\ }\bibfield  {title} {\enquote {\bibinfo {title} {Design of highly synchronizable and robust networks},}\ }\href@noop {} {\bibfield  {journal} {\bibinfo  {journal} {Automatica}\ }\textbf {\bibinfo {volume} {46}},\ \bibinfo {pages} {1835--1842} (\bibinfo {year} {2010})}\BibitemShut {NoStop}%
\bibitem [{\citenamefont {Gorochowski}, \citenamefont {{di Bernardo}},\ and\ \citenamefont {Grierson}(2010)}]{gorochowski2010evolving}%
  \BibitemOpen
  \bibfield  {author} {\bibinfo {author} {\bibfnamefont {T.~E.}\ \bibnamefont {Gorochowski}}, \bibinfo {author} {\bibfnamefont {M.}~\bibnamefont {{di Bernardo}}},\ and\ \bibinfo {author} {\bibfnamefont {C.~S.}\ \bibnamefont {Grierson}},\ }\bibfield  {title} {\enquote {\bibinfo {title} {Evolving enhanced topologies for the synchronization of dynamical complex networks},}\ }\href@noop {} {\bibfield  {journal} {\bibinfo  {journal} {Phys. Rev. E}\ }\textbf {\bibinfo {volume} {81}},\ \bibinfo {pages} {056212} (\bibinfo {year} {2010})}\BibitemShut {NoStop}%
\bibitem [{\citenamefont {Coraggio}\ and\ \citenamefont {di~Bernardo}(2024)}]{coraggio2024data}%
  \BibitemOpen
  \bibfield  {author} {\bibinfo {author} {\bibfnamefont {M.}~\bibnamefont {Coraggio}}\ and\ \bibinfo {author} {\bibfnamefont {M.}~\bibnamefont {di~Bernardo}},\ }\bibfield  {title} {\enquote {\bibinfo {title} {Data-driven design of complex network structures to promote synchronization},}\ }in\ \href@noop {} {\emph {\bibinfo {booktitle} {2024 American Control Conference (ACC)}}}\ (\bibinfo {organization} {IEEE},\ \bibinfo {year} {2024})\ pp.\ \bibinfo {pages} {4396--4401}\BibitemShut {NoStop}%
\bibitem [{\citenamefont {Fujisaka}\ and\ \citenamefont {Yamada}(1983)}]{fujisaka1983stability}%
  \BibitemOpen
  \bibfield  {author} {\bibinfo {author} {\bibfnamefont {H.}~\bibnamefont {Fujisaka}}\ and\ \bibinfo {author} {\bibfnamefont {T.}~\bibnamefont {Yamada}},\ }\bibfield  {title} {\enquote {\bibinfo {title} {Stability theory of synchronized motion in coupled-oscillator systems},}\ }\href@noop {} {\bibfield  {journal} {\bibinfo  {journal} {Prog. Theor. Phys.}\ }\textbf {\bibinfo {volume} {69}},\ \bibinfo {pages} {32--47} (\bibinfo {year} {1983})}\BibitemShut {NoStop}%
\bibitem [{\citenamefont {Nortier}, \citenamefont {Dobson},\ and\ \citenamefont {Battiston}(2025)}]{nortier2025higher}%
  \BibitemOpen
  \bibfield  {author} {\bibinfo {author} {\bibfnamefont {B.~L.}\ \bibnamefont {Nortier}}, \bibinfo {author} {\bibfnamefont {S.}~\bibnamefont {Dobson}},\ and\ \bibinfo {author} {\bibfnamefont {F.}~\bibnamefont {Battiston}},\ }\bibfield  {title} {\enquote {\bibinfo {title} {Higher-order shortest paths in hypergraphs},}\ }\href@noop {} {\bibfield  {journal} {\bibinfo  {journal} {Physical Review E}\ }\textbf {\bibinfo {volume} {112}},\ \bibinfo {pages} {054302} (\bibinfo {year} {2025})}\BibitemShut {NoStop}%
\bibitem [{\citenamefont {Nakao}(2016)}]{nakao16}%
  \BibitemOpen
  \bibfield  {author} {\bibinfo {author} {\bibfnamefont {H.}~\bibnamefont {Nakao}},\ }\bibfield  {title} {\enquote {\bibinfo {title} {Phase reduction approach to synchronisation of nonlinear oscillators},}\ }\href@noop {} {\bibfield  {journal} {\bibinfo  {journal} {Contemp. Phys.}\ }\textbf {\bibinfo {volume} {57}},\ \bibinfo {pages} {188--214} (\bibinfo {year} {2016})}\BibitemShut {NoStop}%
\bibitem [{\citenamefont {Pietras}\ and\ \citenamefont {Daffertshofer}(2019)}]{pietras2019network}%
  \BibitemOpen
  \bibfield  {author} {\bibinfo {author} {\bibfnamefont {B.}~\bibnamefont {Pietras}}\ and\ \bibinfo {author} {\bibfnamefont {A.}~\bibnamefont {Daffertshofer}},\ }\bibfield  {title} {\enquote {\bibinfo {title} {Network dynamics of coupled oscillators and phase reduction techniques},}\ }\href@noop {} {\bibfield  {journal} {\bibinfo  {journal} {Physics Reports}\ }\textbf {\bibinfo {volume} {819}},\ \bibinfo {pages} {1--105} (\bibinfo {year} {2019})}\BibitemShut {NoStop}%
\bibitem [{\citenamefont {Monga}\ \emph {et~al.}(2019)\citenamefont {Monga}, \citenamefont {Wilson}, \citenamefont {Matchen},\ and\ \citenamefont {Moehlis}}]{monga2019phase}%
  \BibitemOpen
  \bibfield  {author} {\bibinfo {author} {\bibfnamefont {B.}~\bibnamefont {Monga}}, \bibinfo {author} {\bibfnamefont {D.}~\bibnamefont {Wilson}}, \bibinfo {author} {\bibfnamefont {T.}~\bibnamefont {Matchen}},\ and\ \bibinfo {author} {\bibfnamefont {J.}~\bibnamefont {Moehlis}},\ }\bibfield  {title} {\enquote {\bibinfo {title} {Phase reduction and phase-based optimal control for biological systems: a tutorial},}\ }\href@noop {} {\bibfield  {journal} {\bibinfo  {journal} {Biological cybernetics}\ }\textbf {\bibinfo {volume} {113}},\ \bibinfo {pages} {11--46} (\bibinfo {year} {2019})}\BibitemShut {NoStop}%
\bibitem [{\citenamefont {Winfree}(1980)}]{Win80}%
  \BibitemOpen
  \bibfield  {author} {\bibinfo {author} {\bibfnamefont {A.~T.}\ \bibnamefont {Winfree}},\ }\href@noop {} {\emph {\bibinfo {title} {The Geometry of Biological Time}}}\ (\bibinfo  {publisher} {Springer},\ \bibinfo {address} {New York},\ \bibinfo {year} {1980})\BibitemShut {NoStop}%
\bibitem [{\citenamefont {Kuramoto}\ and\ \citenamefont {Nakao}(2019)}]{kuramoto2019concept}%
  \BibitemOpen
  \bibfield  {author} {\bibinfo {author} {\bibfnamefont {Y.}~\bibnamefont {Kuramoto}}\ and\ \bibinfo {author} {\bibfnamefont {H.}~\bibnamefont {Nakao}},\ }\bibfield  {title} {\enquote {\bibinfo {title} {On the concept of dynamical reduction: the case of coupled oscillators},}\ }\href@noop {} {\bibfield  {journal} {\bibinfo  {journal} {Phil. Trans. R. Soc. A}\ }\textbf {\bibinfo {volume} {377}},\ \bibinfo {pages} {20190041} (\bibinfo {year} {2019})}\BibitemShut {NoStop}%
\bibitem [{\citenamefont {Nakao}(2014)}]{nakao2014complex}%
  \BibitemOpen
  \bibfield  {author} {\bibinfo {author} {\bibfnamefont {H.}~\bibnamefont {Nakao}},\ }\bibfield  {title} {\enquote {\bibinfo {title} {Complex {G}inzburg-{L}andau equation on networks and its non-uniform dynamics},}\ }\href@noop {} {\bibfield  {journal} {\bibinfo  {journal} {Eur. Phys. J.: Spec. Top.}\ }\textbf {\bibinfo {volume} {223}},\ \bibinfo {pages} {2411--2421} (\bibinfo {year} {2014})}\BibitemShut {NoStop}%
\bibitem [{\citenamefont {Pereti}\ and\ \citenamefont {Fanelli}(2020)}]{pereti2020stabilizing}%
  \BibitemOpen
  \bibfield  {author} {\bibinfo {author} {\bibfnamefont {C.}~\bibnamefont {Pereti}}\ and\ \bibinfo {author} {\bibfnamefont {D.}~\bibnamefont {Fanelli}},\ }\bibfield  {title} {\enquote {\bibinfo {title} {Stabilizing {S}tuart-{L}andau oscillators via time-varying networks},}\ }\href@noop {} {\bibfield  {journal} {\bibinfo  {journal} {Chaos, Solitons \& Fractals}\ }\textbf {\bibinfo {volume} {133}},\ \bibinfo {pages} {109587} (\bibinfo {year} {2020})}\BibitemShut {NoStop}%
\bibitem [{\citenamefont {Ashwin}\ and\ \citenamefont {Rodrigues}(2016)}]{ashwin2016hopf}%
  \BibitemOpen
  \bibfield  {author} {\bibinfo {author} {\bibfnamefont {P.}~\bibnamefont {Ashwin}}\ and\ \bibinfo {author} {\bibfnamefont {A.}~\bibnamefont {Rodrigues}},\ }\bibfield  {title} {\enquote {\bibinfo {title} {Hopf normal form with sn symmetry and reduction to systems of nonlinearly coupled phase oscillators},}\ }\href@noop {} {\bibfield  {journal} {\bibinfo  {journal} {Physica D: Nonlinear Phenomena}\ }\textbf {\bibinfo {volume} {325}},\ \bibinfo {pages} {14--24} (\bibinfo {year} {2016})}\BibitemShut {NoStop}%
\bibitem [{\citenamefont {Le\'on}\ and\ \citenamefont {Paz\'o}(2019)}]{leon19}%
  \BibitemOpen
  \bibfield  {author} {\bibinfo {author} {\bibfnamefont {I.}~\bibnamefont {Le\'on}}\ and\ \bibinfo {author} {\bibfnamefont {D.}~\bibnamefont {Paz\'o}},\ }\bibfield  {title} {\enquote {\bibinfo {title} {Phase reduction beyond the first order: The case of the mean-field complex {G}inzburg-{L}andau equation},}\ }\href@noop {} {\bibfield  {journal} {\bibinfo  {journal} {Phys. Rev. E}\ }\textbf {\bibinfo {volume} {100}},\ \bibinfo {pages} {012211} (\bibinfo {year} {2019})}\BibitemShut {NoStop}%
\bibitem [{\citenamefont {Gengel}\ \emph {et~al.}(2020)\citenamefont {Gengel}, \citenamefont {Teichmann}, \citenamefont {Rosenblum},\ and\ \citenamefont {Pikovsky}}]{gengel2020high}%
  \BibitemOpen
  \bibfield  {author} {\bibinfo {author} {\bibfnamefont {E.}~\bibnamefont {Gengel}}, \bibinfo {author} {\bibfnamefont {E.}~\bibnamefont {Teichmann}}, \bibinfo {author} {\bibfnamefont {M.}~\bibnamefont {Rosenblum}},\ and\ \bibinfo {author} {\bibfnamefont {A.}~\bibnamefont {Pikovsky}},\ }\bibfield  {title} {\enquote {\bibinfo {title} {High-order phase reduction for coupled oscillators},}\ }\href@noop {} {\bibfield  {journal} {\bibinfo  {journal} {Journal of Physics: Complexity}\ }\textbf {\bibinfo {volume} {2}},\ \bibinfo {pages} {015005} (\bibinfo {year} {2020})}\BibitemShut {NoStop}%
\bibitem [{\citenamefont {Bick}, \citenamefont {B{\"o}hle},\ and\ \citenamefont {Kuehn}(2024)}]{bick2024higher}%
  \BibitemOpen
  \bibfield  {author} {\bibinfo {author} {\bibfnamefont {C.}~\bibnamefont {Bick}}, \bibinfo {author} {\bibfnamefont {T.}~\bibnamefont {B{\"o}hle}},\ and\ \bibinfo {author} {\bibfnamefont {C.}~\bibnamefont {Kuehn}},\ }\bibfield  {title} {\enquote {\bibinfo {title} {Higher-order network interactions through phase reduction for oscillators with phase-dependent amplitude},}\ }\href@noop {} {\bibfield  {journal} {\bibinfo  {journal} {Journal of Nonlinear Science}\ }\textbf {\bibinfo {volume} {34}},\ \bibinfo {pages} {77} (\bibinfo {year} {2024})}\BibitemShut {NoStop}%
\bibitem [{\citenamefont {Fujii}\ \emph {et~al.}(2026)\citenamefont {Fujii}, \citenamefont {Taga}, \citenamefont {Muolo}, \citenamefont {Rink},\ and\ \citenamefont {Nakao}}]{fujii2026emergence}%
  \BibitemOpen
  \bibfield  {author} {\bibinfo {author} {\bibfnamefont {N.}~\bibnamefont {Fujii}}, \bibinfo {author} {\bibfnamefont {K.}~\bibnamefont {Taga}}, \bibinfo {author} {\bibfnamefont {R.}~\bibnamefont {Muolo}}, \bibinfo {author} {\bibfnamefont {B.}~\bibnamefont {Rink}},\ and\ \bibinfo {author} {\bibfnamefont {H.}~\bibnamefont {Nakao}},\ }\bibfield  {title} {\enquote {\bibinfo {title} {Emergence of higher-order interactions in systems of coupled {K}uramoto oscillators with time delay},}\ }\href@noop {} {\bibfield  {journal} {\bibinfo  {journal} {Physical Review E}\ }\textbf {\bibinfo {volume} {114}} (\bibinfo {year} {2026})}\BibitemShut {NoStop}%
\bibitem [{\citenamefont {Sakaguchi}(1990)}]{sakaguchi1990breakdown}%
  \BibitemOpen
  \bibfield  {author} {\bibinfo {author} {\bibfnamefont {H.}~\bibnamefont {Sakaguchi}},\ }\bibfield  {title} {\enquote {\bibinfo {title} {Breakdown of the phase dynamics},}\ }\href@noop {} {\bibfield  {journal} {\bibinfo  {journal} {Progress of theoretical physics}\ }\textbf {\bibinfo {volume} {84}},\ \bibinfo {pages} {792--800} (\bibinfo {year} {1990})}\BibitemShut {NoStop}%
\bibitem [{\citenamefont {Mau}, \citenamefont {Omel'chenko},\ and\ \citenamefont {Rosenblum}(2024)}]{mau2024phase}%
  \BibitemOpen
  \bibfield  {author} {\bibinfo {author} {\bibfnamefont {E.}~\bibnamefont {Mau}}, \bibinfo {author} {\bibfnamefont {O.~E.}\ \bibnamefont {Omel'chenko}},\ and\ \bibinfo {author} {\bibfnamefont {M.}~\bibnamefont {Rosenblum}},\ }\bibfield  {title} {\enquote {\bibinfo {title} {Phase reduction explains chimera shape: When multibody interaction matters},}\ }\href@noop {} {\bibfield  {journal} {\bibinfo  {journal} {Physical Review E}\ }\textbf {\bibinfo {volume} {110}},\ \bibinfo {pages} {L022201} (\bibinfo {year} {2024})}\BibitemShut {NoStop}%
\bibitem [{\citenamefont {Okuda}\ and\ \citenamefont {Kuramoto}(1991)}]{okuda1991mutual}%
  \BibitemOpen
  \bibfield  {author} {\bibinfo {author} {\bibfnamefont {K.}~\bibnamefont {Okuda}}\ and\ \bibinfo {author} {\bibfnamefont {Y.}~\bibnamefont {Kuramoto}},\ }\bibfield  {title} {\enquote {\bibinfo {title} {Mutual entrainment between populations of coupled oscillators},}\ }\href@noop {} {\bibfield  {journal} {\bibinfo  {journal} {Progress of theoretical physics}\ }\textbf {\bibinfo {volume} {86}},\ \bibinfo {pages} {1159--1176} (\bibinfo {year} {1991})}\BibitemShut {NoStop}%
\bibitem [{\citenamefont {Bergner}\ \emph {et~al.}(2012)\citenamefont {Bergner}, \citenamefont {Frasca}, \citenamefont {Sciuto}, \citenamefont {Buscarino}, \citenamefont {Ngamga}, \citenamefont {Fortuna},\ and\ \citenamefont {Kurths}}]{bergner2012remote}%
  \BibitemOpen
  \bibfield  {author} {\bibinfo {author} {\bibfnamefont {A.}~\bibnamefont {Bergner}}, \bibinfo {author} {\bibfnamefont {M.}~\bibnamefont {Frasca}}, \bibinfo {author} {\bibfnamefont {G.}~\bibnamefont {Sciuto}}, \bibinfo {author} {\bibfnamefont {A.}~\bibnamefont {Buscarino}}, \bibinfo {author} {\bibfnamefont {E.~J.}\ \bibnamefont {Ngamga}}, \bibinfo {author} {\bibfnamefont {L.}~\bibnamefont {Fortuna}},\ and\ \bibinfo {author} {\bibfnamefont {J.}~\bibnamefont {Kurths}},\ }\bibfield  {title} {\enquote {\bibinfo {title} {Remote synchronization in star networks},}\ }\href@noop {} {\bibfield  {journal} {\bibinfo  {journal} {Phys. Rev. E}\ }\textbf {\bibinfo {volume} {85}},\ \bibinfo {pages} {026208} (\bibinfo {year} {2012})}\BibitemShut {NoStop}%
\bibitem [{\citenamefont {Sakaguchi}\ and\ \citenamefont {Kuramoto}(1986)}]{sakaguchi1986soluble}%
  \BibitemOpen
  \bibfield  {author} {\bibinfo {author} {\bibfnamefont {H.}~\bibnamefont {Sakaguchi}}\ and\ \bibinfo {author} {\bibfnamefont {Y.}~\bibnamefont {Kuramoto}},\ }\bibfield  {title} {\enquote {\bibinfo {title} {A soluble active rotater model showing phase transitions via mutual entertainment},}\ }\href@noop {} {\bibfield  {journal} {\bibinfo  {journal} {Progress of Theoretical Physics}\ }\textbf {\bibinfo {volume} {76}},\ \bibinfo {pages} {576--581} (\bibinfo {year} {1986})}\BibitemShut {NoStop}%
\bibitem [{\citenamefont {Paz{\'o}}(2005)}]{pazo2005thermodynamic}%
  \BibitemOpen
  \bibfield  {author} {\bibinfo {author} {\bibfnamefont {D.}~\bibnamefont {Paz{\'o}}},\ }\bibfield  {title} {\enquote {\bibinfo {title} {Thermodynamic limit of the first-order phase transition in the {K}uramoto model},}\ }\href@noop {} {\bibfield  {journal} {\bibinfo  {journal} {Phys. Rev. E}\ }\textbf {\bibinfo {volume} {72}},\ \bibinfo {pages} {046211} (\bibinfo {year} {2005})}\BibitemShut {NoStop}%
\bibitem [{\citenamefont {Basnarkov}\ and\ \citenamefont {Urumov}(2007)}]{basnarkov2007phase}%
  \BibitemOpen
  \bibfield  {author} {\bibinfo {author} {\bibfnamefont {L.}~\bibnamefont {Basnarkov}}\ and\ \bibinfo {author} {\bibfnamefont {V.}~\bibnamefont {Urumov}},\ }\bibfield  {title} {\enquote {\bibinfo {title} {Phase transitions in the {K}uramoto model},}\ }\href@noop {} {\bibfield  {journal} {\bibinfo  {journal} {Phys. Rev. E}\ }\textbf {\bibinfo {volume} {76}},\ \bibinfo {pages} {057201} (\bibinfo {year} {2007})}\BibitemShut {NoStop}%
\bibitem [{\citenamefont {Le{\'o}n}\ \emph {et~al.}(2025{\natexlab{b}})\citenamefont {Le{\'o}n}, \citenamefont {Muolo}, \citenamefont {Hata},\ and\ \citenamefont {Nakao}}]{leon2025theory}%
  \BibitemOpen
  \bibfield  {author} {\bibinfo {author} {\bibfnamefont {I.}~\bibnamefont {Le{\'o}n}}, \bibinfo {author} {\bibfnamefont {R.}~\bibnamefont {Muolo}}, \bibinfo {author} {\bibfnamefont {S.}~\bibnamefont {Hata}},\ and\ \bibinfo {author} {\bibfnamefont {H.}~\bibnamefont {Nakao}},\ }\bibfield  {title} {\enquote {\bibinfo {title} {Theory of phase reduction from hypergraphs to simplicial complexes: a general route to higher-order {K}uramoto models},}\ }\href@noop {} {\bibfield  {journal} {\bibinfo  {journal} {Physica D}\ } (\bibinfo {year} {2025}{\natexlab{b}})}\BibitemShut {NoStop}%
\bibitem [{\citenamefont {Battiston}\ \emph {et~al.}(2026)\citenamefont {Battiston}, \citenamefont {Bick}, \citenamefont {Lucas}, \citenamefont {Mill{\'a}n}, \citenamefont {Skardal},\ and\ \citenamefont {Zhang}}]{battiston2026collective}%
  \BibitemOpen
  \bibfield  {author} {\bibinfo {author} {\bibfnamefont {F.}~\bibnamefont {Battiston}}, \bibinfo {author} {\bibfnamefont {C.}~\bibnamefont {Bick}}, \bibinfo {author} {\bibfnamefont {M.}~\bibnamefont {Lucas}}, \bibinfo {author} {\bibfnamefont {A.~P.}\ \bibnamefont {Mill{\'a}n}}, \bibinfo {author} {\bibfnamefont {P.~S.}\ \bibnamefont {Skardal}},\ and\ \bibinfo {author} {\bibfnamefont {Y.}~\bibnamefont {Zhang}},\ }\bibfield  {title} {\enquote {\bibinfo {title} {Collective dynamics on higher-order networks},}\ }\href@noop {} {\bibfield  {journal} {\bibinfo  {journal} {Nature Reviews Physics}\ ,\ \bibinfo {pages} {1--14}} (\bibinfo {year} {2026})}\BibitemShut {NoStop}%
\bibitem [{\citenamefont {Suman}\ and\ \citenamefont {Jalan}(2024)}]{suman2024finite}%
  \BibitemOpen
  \bibfield  {author} {\bibinfo {author} {\bibfnamefont {A.}~\bibnamefont {Suman}}\ and\ \bibinfo {author} {\bibfnamefont {S.}~\bibnamefont {Jalan}},\ }\bibfield  {title} {\enquote {\bibinfo {title} {Finite-size effect in {K}uramoto oscillators with higher-order interactions},}\ }\href@noop {} {\bibfield  {journal} {\bibinfo  {journal} {Chaos: An Interdisciplinary Journal of Nonlinear Science}\ }\textbf {\bibinfo {volume} {34}} (\bibinfo {year} {2024})}\BibitemShut {NoStop}%
\bibitem [{\citenamefont {Restrepo}, \citenamefont {Ott},\ and\ \citenamefont {Hunt}(2005)}]{restrepo2005onset}%
  \BibitemOpen
  \bibfield  {author} {\bibinfo {author} {\bibfnamefont {J.~G.}\ \bibnamefont {Restrepo}}, \bibinfo {author} {\bibfnamefont {E.}~\bibnamefont {Ott}},\ and\ \bibinfo {author} {\bibfnamefont {B.~R.}\ \bibnamefont {Hunt}},\ }\bibfield  {title} {\enquote {\bibinfo {title} {Onset of synchronization in large networks of coupled oscillators},}\ }\href@noop {} {\bibfield  {journal} {\bibinfo  {journal} {Physical Review E—Statistical, Nonlinear, and Soft Matter Physics}\ }\textbf {\bibinfo {volume} {71}},\ \bibinfo {pages} {036151} (\bibinfo {year} {2005})}\BibitemShut {NoStop}%
\bibitem [{\citenamefont {Bianconi}(2002)}]{bianconi2002mean}%
  \BibitemOpen
  \bibfield  {author} {\bibinfo {author} {\bibfnamefont {G.}~\bibnamefont {Bianconi}},\ }\bibfield  {title} {\enquote {\bibinfo {title} {Mean field solution of the ising model on a {B}arab{\'a}si--{A}lbert network},}\ }\href@noop {} {\bibfield  {journal} {\bibinfo  {journal} {Physics Letters A}\ }\textbf {\bibinfo {volume} {303}},\ \bibinfo {pages} {166--168} (\bibinfo {year} {2002})}\BibitemShut {NoStop}%
\bibitem [{\citenamefont {Ichinomiya}(2004)}]{ichinomiya2004frequency}%
  \BibitemOpen
  \bibfield  {author} {\bibinfo {author} {\bibfnamefont {T.}~\bibnamefont {Ichinomiya}},\ }\bibfield  {title} {\enquote {\bibinfo {title} {Frequency synchronization in a random oscillator network},}\ }\href@noop {} {\bibfield  {journal} {\bibinfo  {journal} {Physical Review E—Statistical, Nonlinear, and Soft Matter Physics}\ }\textbf {\bibinfo {volume} {70}},\ \bibinfo {pages} {026116} (\bibinfo {year} {2004})}\BibitemShut {NoStop}%
\bibitem [{\citenamefont {Muolo}, \citenamefont {Nakao},\ and\ \citenamefont {Coraggio}(2025)}]{muolo2025higher}%
  \BibitemOpen
  \bibfield  {author} {\bibinfo {author} {\bibfnamefont {R.}~\bibnamefont {Muolo}}, \bibinfo {author} {\bibfnamefont {H.}~\bibnamefont {Nakao}},\ and\ \bibinfo {author} {\bibfnamefont {M.}~\bibnamefont {Coraggio}},\ }\href@noop {} {\enquote {\bibinfo {title} {When higher-order interactions enhance synchronization: the case of the {K}uramoto model},}\ } (\bibinfo {year} {2025}),\ \Eprint {https://arxiv.org/abs/2508.10992} {arXiv:2508.10992} \BibitemShut {NoStop}%
\bibitem [{\citenamefont {Wang}, \citenamefont {Zhu},\ and\ \citenamefont {Liu}(2026)}]{wang2026moderate}%
  \BibitemOpen
  \bibfield  {author} {\bibinfo {author} {\bibfnamefont {Z.}~\bibnamefont {Wang}}, \bibinfo {author} {\bibfnamefont {J.}~\bibnamefont {Zhu}},\ and\ \bibinfo {author} {\bibfnamefont {X.}~\bibnamefont {Liu}},\ }\bibfield  {title} {\enquote {\bibinfo {title} {Moderate higher-order interactions enhance stability while preserving basin structure},}\ }\href@noop {} {\bibfield  {journal} {\bibinfo  {journal} {Chaos, Solitons \& Fractals}\ }\textbf {\bibinfo {volume} {208}},\ \bibinfo {pages} {118069} (\bibinfo {year} {2026})}\BibitemShut {NoStop}%
\bibitem [{\citenamefont {Moriam{\'e}}\ \emph {et~al.}(2026)\citenamefont {Moriam{\'e}}, \citenamefont {Muolo}, \citenamefont {Carletti},\ and\ \citenamefont {Lucas}}]{moriame2026efficiency}%
  \BibitemOpen
  \bibfield  {author} {\bibinfo {author} {\bibfnamefont {M.}~\bibnamefont {Moriam{\'e}}}, \bibinfo {author} {\bibfnamefont {R.}~\bibnamefont {Muolo}}, \bibinfo {author} {\bibfnamefont {T.}~\bibnamefont {Carletti}},\ and\ \bibinfo {author} {\bibfnamefont {M.}~\bibnamefont {Lucas}},\ }\bibfield  {title} {\enquote {\bibinfo {title} {On the efficiency of pairwise {H}amiltonian control to desynchronize the higher-order {K}uramoto model},}\ }\href@noop {} {\bibfield  {journal} {\bibinfo  {journal} {Chaos: An Interdisciplinary Journal of Nonlinear Science}\ }\textbf {\bibinfo {volume} {36}} (\bibinfo {year} {2026})}\BibitemShut {NoStop}%
\bibitem [{\citenamefont {Skardal}\ \emph {et~al.}(2025)\citenamefont {Skardal}, \citenamefont {Battiston}, \citenamefont {Lucas}, \citenamefont {Mizuhara}, \citenamefont {Petri},\ and\ \citenamefont {Zhang}}]{skardal2025mixed}%
  \BibitemOpen
  \bibfield  {author} {\bibinfo {author} {\bibfnamefont {P.}~\bibnamefont {Skardal}}, \bibinfo {author} {\bibfnamefont {F.}~\bibnamefont {Battiston}}, \bibinfo {author} {\bibfnamefont {M.}~\bibnamefont {Lucas}}, \bibinfo {author} {\bibfnamefont {M.}~\bibnamefont {Mizuhara}}, \bibinfo {author} {\bibfnamefont {G.}~\bibnamefont {Petri}},\ and\ \bibinfo {author} {\bibfnamefont {Y.}~\bibnamefont {Zhang}},\ }\bibfield  {title} {\enquote {\bibinfo {title} {Mixed higher-order coupling stabilizes new states},}\ }\href@noop {} {\bibfield  {journal} {\bibinfo  {journal} {arXiv preprint arXiv:2510.09387}\ } (\bibinfo {year} {2025})}\BibitemShut {NoStop}%
\bibitem [{\citenamefont {Namura}, \citenamefont {Muolo},\ and\ \citenamefont {Nakao}(2026)}]{namura2025optimal}%
  \BibitemOpen
  \bibfield  {author} {\bibinfo {author} {\bibfnamefont {N.}~\bibnamefont {Namura}}, \bibinfo {author} {\bibfnamefont {R.}~\bibnamefont {Muolo}},\ and\ \bibinfo {author} {\bibfnamefont {H.}~\bibnamefont {Nakao}},\ }\bibfield  {title} {\enquote {\bibinfo {title} {Optimal interaction functions realizing higher-order {K}uramoto dynamics with arbitrary limit-cycle oscillators},}\ }\href@noop {} {\bibfield  {journal} {\bibinfo  {journal} {Chaos: An Interdisciplinary Journal of Nonlinear Science}\ }\textbf {\bibinfo {volume} {36}},\ \bibinfo {pages} {023120} (\bibinfo {year} {2026})}\BibitemShut {NoStop}%
\bibitem [{\citenamefont {Le{\'o}n}\ \emph {et~al.}(2026)\citenamefont {Le{\'o}n}, \citenamefont {Muolo}, \citenamefont {Zhang},\ and\ \citenamefont {Lucas}}]{leon2026symmetry}%
  \BibitemOpen
  \bibfield  {author} {\bibinfo {author} {\bibfnamefont {I.}~\bibnamefont {Le{\'o}n}}, \bibinfo {author} {\bibfnamefont {R.}~\bibnamefont {Muolo}}, \bibinfo {author} {\bibfnamefont {Y.}~\bibnamefont {Zhang}},\ and\ \bibinfo {author} {\bibfnamefont {M.}~\bibnamefont {Lucas}},\ }\bibfield  {title} {\enquote {\bibinfo {title} {Symmetry-based selection rules for higher-order interactions in coupled oscillators},}\ }\href@noop {} {\bibfield  {journal} {\bibinfo  {journal} {arXiv preprint arXiv:2606.04904}\ } (\bibinfo {year} {2026})}\BibitemShut {NoStop}%
\bibitem [{\citenamefont {Marui}\ and\ \citenamefont {Kori}(2025)}]{marui2025synchronization}%
  \BibitemOpen
  \bibfield  {author} {\bibinfo {author} {\bibfnamefont {Y.}~\bibnamefont {Marui}}\ and\ \bibinfo {author} {\bibfnamefont {H.}~\bibnamefont {Kori}},\ }\bibfield  {title} {\enquote {\bibinfo {title} {Synchronization and its slow decay in noisy oscillators with simplicial interactions},}\ }\href@noop {} {\bibfield  {journal} {\bibinfo  {journal} {Physical Review E}\ }\textbf {\bibinfo {volume} {111}},\ \bibinfo {pages} {014223} (\bibinfo {year} {2025})}\BibitemShut {NoStop}%
\bibitem [{\citenamefont {{The MathWorks Inc.}}(2024)}]{Matlab}%
  \BibitemOpen
  \bibfield  {author} {\bibinfo {author} {\bibnamefont {{The MathWorks Inc.}}},\ }\href@noop {} {\enquote {\bibinfo {title} {Matlab: 24.1 (r2024a)},}\ } (\bibinfo {year} {2024})\BibitemShut {NoStop}%
\end{thebibliography}

%

\appendix
\onecolumngrid

\section{Computationally efficient simulation of the higher-order Kuramoto model}\label{app:A}

To enable more efficient and scalable numerical simulation of \eqref{eq:ho_kura_leon} in the case of random hypergraphs, we rewrite it in vector form.
Let $z^{(1)}, z^{(2)} \in \BB{R}^{N_0}$ be the vectors collecting all coupling terms associated with $1$- and $2$-hyperedges, respectively, with their $j$-th components given by
\begin{equation*}
    z_j^{(1)} = \sum_{k=1}^{N_0} A_{jk}^{(1)}\sin(\theta_k-\theta_j),
    \quad
    z_j^{(2)} = \sum_{k=1}^{N_0} \sum_{l=1}^{N_0} A_{jkl}^{(2)}\sin(\theta_k+\theta_l-2\theta_j).
\end{equation*}
For $1$-hyperedges, we first note that $z^{(1)} = B^{(1)} \sin(- (B^{(1)})\T \theta)$, where $B^{(1)}$ is the incidence matrix.
Then, for $2$-hyperedges, we define the two auxiliary matrices $Y^{(2)}_{\R{a}}, Y^{(2)}_{\R{s}} \in \BB{Z}^{N_0 \times 3 N_2}$; the former is used to assemble linear combinations of the phases and the latter to select specific combinations contributing to the dynamics of each oscillator.
In $Y^{(2)}_{\R{a}}$, given some $k \in \{1, \dots, N_2\}$, the $3$ columns from the $(3k-2)$-th to the $3k$-th correspond to the $k$-th $2$-hyperedge, say $\{l, m, p\}$; 
the first of these columns has $2$ in position $l$, $-1$ in positions $m$ and $p$, and zeros elsewhere;
the second has $2$ in $m$, and $-1$ in $l$ and $p$; 
the third has $2$ in $p$, and $-1$ in $l$ and $m$.
In $Y^{(2)}_{\R{s}}$, the $3$ columns from the $(3k-2)$-th to the $3k$-th also correspond to the $k$-th $2$-hyperedge, say $\{l, m, p\}$;
these three columns are different scaled indicator vectors with $2$ in positions $l$, $m$ and $p$, respectively, and zeros elsewhere.
An example set of the auxiliary matrices $Y^{(2)}_{\R{a}}$, $Y^{(2)}_{\R{s}}$ for a hypergraph with $N_0 = 4$ vertices and hyperedges $\{1, 2, 3\}$, $\{1, 3, 4\}$ is
\begin{equation*}
    Y^{(2)}_{\R{a}} = 
    \begin{bmatrix}
    2 & -1 & -1 & & 2 & -1 & -1 \\
    -1 & 2 & -1 & & 0 & 0 & 0 \\
    -1 & -1 & 2 & & -1 & 2 & -1 \\
    0 & 0 & 0  & &  -1 & -1 & 2
    \end{bmatrix}, \quad
    Y^{(2)}_{\R{s}} = 
    \begin{bmatrix}
    2 & 0 & 0 & & 2 & 0 & 0 \\
    0 & 2 & 0 & & 0 & 0 & 0 \\
    0 & 0 & 2 & & 0 & 2 & 0 \\
    0 & 0 & 0 & & 0 & 0 & 2
    \end{bmatrix}.
\end{equation*}
Then, it is immediate to verify that $z^{(2)} = Y^{(2)}_{\R{s}} \sin(- (Y^{(2)}_{\R{a}})\T \theta)$.

In conclusion, it is possible to rewrite \eqref{eq:ho_kura_leon} in a vector form as
\begin{equation}\label{eq:higher_order_kuramoto_efficient}
    \dot{\theta} = \omega + \frac{K_1}{\langle d^{(1)} \rangle} B^{(1)} \sin\left(-(B^{(1)})\T \theta \right) + \frac{K_2}{2\langle d^{(2)} \rangle} Y^{(2)}_{\R{s}} \sin\left(-(Y^{(2)}_{\R{a}})\T \theta \right),
\end{equation}
where $\theta$ and $\omega$ are $N_0$-dimensional column vectors, collecting all $\theta_j$'s and $\omega_j$'s, respectively.

\section{Description of the numerical methods}\label{app:B}

Here, we illustrate the setup of the numerical simulations performed in this work.
Integration of the higher-order Kuramoto model was performed using the forward Euler method with time step $0.05$ for $600$ time units.
{We verified that results remained qualitatively unchanged by using an explicit fourth-order Runge–Kutta method and a smaller time step of $0.005$, and by integrating for $2\,000$ time units.}
The Kuramoto order parameter $R$ and the Kuramoto-Daido order parameter $R_2$ shown in the figures are time averaged over the second half of the simulation and, for random hypergraphs, multiple hypergraph realizations 
(time averaging is necessary as, even when starting close to the synchronous state, heterogeneity in frequencies and interactions can prevent the solution from becoming stationary).
All simulations were implemented in Matlab~\cite{Matlab}.

\subsection{Coupling strength analysis}

In the analysis in Section \ref{sec:results1a} (Figure~\ref{fig:couplings_study_standard_coupl_R1}), we fix $N_0 = 10$ nodes, $N_1 = 20$ links, and $N_2 = 10$  triangles for the hypergraphs to be generated.
Then, we define the coupling strength ranges for pairwise and higher-order interactions, $K_1$ and $K_2$, both spanning from $0$ to $0.6$. 
A grid of $51 \times 51$ points is created over these ranges, and for each grid point, we integrate
\eqref{eq:ho_kura_leon} over $N_{\C{H}}=300$ randomly generated hypergraphs with the prescribed $(N_0,N_1,N_2)$, ensuring either $\C{H}$-connectivity or $1$-connectivity {(we verified that increasing $N_{\mathcal{H}}$ further to $800$ showed no significant impact)}.
For each hypergraph, frequencies $\omega_j$ and initial phases $\theta_{0,j}$, for $j = 1, \dots, N_0$, are drawn uniformly from $[0, 0.3]$ and $[0, \theta_0^{\R{max}}]$, respectively, with $\theta_0^{\R{max}} \in \{0.1, 2\pi\}$.
For each integration, we compute the Kuramoto order parameter $R$ and the Kuramoto-Daido order parameter $R_2$, averaging their values over the hypergraph samples at each grid point as described above.
When starting from incoherent initial conditions (i.e., $\theta_0^{\mathrm{max}} = 2\pi$), we also mark, for each $K_1$, the value of $K_2$ yielding the maximum order parameters.

We also performed the analysis with different numbers of links and triangles, namely $N_1=40$, $N_2=6$ and $N_1 = 15$, $N_2=6$, observing no meaningful difference (results not shown). 

For the analysis in Figure~\ref{fig:alltoall}, we used the same setup described above, applied to the all-to-all hypergraph with $N_0 = 10\,000$ nodes, $\theta_0^{\R{max}}=2\pi$, and the same uniform distribution of natural frequencies.
Given the system size, a single realization per parameter value suffices, as the order parameters are self-averaging; the reported values are time averages over the second half of each run.

In Appendix~\ref{app:C}, we show analogous results for the Kuramoto-Daido order parameter $R_2$ of the model in~\eqref{eq:ho_kura_leon}, for normally distributed frequencies, and for larger structures.
All configurations exhibit the same qualitative behavior shown in the Main Text.

\subsection{Hyperedge allocation analysis}
\label{sec:methods_hyperedge_allocation_analysis}

For the analysis in Section \ref{sec:results2} (Figure~\ref{fig:alloc_study_standard_coupl_omega03_R}),
we considered Kuramoto models with $N_0 = 10$ oscillators.
We assigned a fixed cost of $c_1 = 1$ (arbitrary units) to $1$-hyperedges, while $2$-hyperedges had a variable cost $c_2 \in \{1, 3, 5\}$.
Interpreting a $2$-hyperedge as equivalent to three projected $1$-hyperedges, we consider $c_2 = 3$ to represent cost parity, $c_2 = 1$ to reflect a lower relative cost of $2$-hyperedges, and $c_2 = 5$ to reflect a higher relative cost.
The total resource budget was fixed at $J = 40$, imposing the constraint 
\begin{equation}\label{eq:budget_constraint}
    c_1 N_1 + c_2 N_2 \le J.
\end{equation}
The value of $J$ was chosen to ensure that 
(i) hypergraphs can always be made $\C{H}$-connected (otherwise synchronization would be impossible) and 
(ii) the entire resource budget is effectively used, avoiding cases where all possible hyperedges are deployed but unused budget remains.
To fulfill these conditions, we note the following facts: with $N_0 = 10$ vertices,
the minimum number of $1$-hyperedges for $1$-connectivity is $N_1^\R{min} = N_0-1 = 9$;
the minimum number of $2$-hyperedges for $2$-connectivity is $N_2^\R{min} = \lceil (N_0-1)/2 \rceil = 5$;
the maximum number of distinct $1$-hyperedges is $N_1^\R{max} = N_0 (N_0-1) / 2 = 45$;
the maximum number of distinct $2$-hyperedges is $N_2^\R{max} = N_0 (N_0-1) (N_0-2) / 6 = 120$.
Hence, condition (i) requires selecting $J \ge \max \{N_1^\R{min}, 5 N_2^\R{min}\}$ ($5$ is the largest value of $c_2$);
condition (ii) requires that $J \le \min \{N_1^\R{max}, 1 N_2^\R{max}\}$ ($1$ is the smallest value of $c_2$).
This yields an admissible range $J \in [25, 45]$, from which we choose $J = 40$.
To enforce constraint \eqref{eq:budget_constraint}, we define $b_1 \in [0, 1]$ as the fraction of resources to be allocated to $1$-hyperedges, and assign $N_1 = \lfloor J b_1 / c_1 \rfloor$ and 
$N_2 = \lfloor J (1 - b_1) / c_2 \rfloor$.

The experimental conditions were generated by varying coupling strengths $K_1, K_2$ in \eqref{eq:ho_kura_leon}, keeping their sum fixed, and the link allocation fraction $b_1$.

Namely, we selected $K_1 \in [0, K^{\R{sum}}]$, with $K^{\R{sum}}= 0.3$, using $51$ linearly spaced points, and consequently set $K_2 = K^{\R{sum}} - K_1$.
The allocation parameter $b_1$ was varied over $[0, 1]$, also using $51$ linearly spaced points, resulting in $51 \times 51 = 2\,601$ experimental conditions.
For each value of $b_1$, we generated $N_\C{H} = 500$ random hypergraphs {(increasing $N_{\mathcal{H}}$ to $800$ did not yield significantly different results)}; {for each one, we} sampled random initial conditions $\theta_0$ uniformly from $[0, \theta_0^\R{max}]^{N_0}$, as well as natural frequencies uniformly from $[0, 0.3]^{N_0}$, with $\theta_0^\R{max}$ selected as either $\pi/2$ or $2\pi$.
When generating the hypergraphs, the $N_1$ and $N_2$ hyperedges are selected randomly from all possible ones, each with equal probability. 
If the resulting hypergraph is not $\C{H}$-connected, it is discarded and a new one is generated in its place, repeating this process until an $\C{H}$-connected hypergraph is obtained.

Also for this analysis, in Appendix~\ref{app:C}, we show analogous results for the Kuramoto-Daido order parameter $R_2$, for normally distributed frequencies, and for larger structures.
All configurations exhibit the same qualitative behavior shown here.

{
\section{Beneficial effect on $R$ of small $K_2$ for all-to-all systems}
\label{sec:mathematical_analysis}

Because a common frequency can be removed by a change of rotating frame, we assume, without loss of generality, that
$\frac{1}{N_0}\sum_{j=1}^{N_0}\omega_j=0$.
Then, we consider the all-to-all network model
\begin{equation}\label{eq:global_model_weakK2}
     \dot\theta_j
     =\omega_j
     +\frac{K_1}{N_0}\sum_{k=1}^{N_0}\sin(\theta_k-\theta_j)
     +\frac{K_2}{N_0^2}\sum_{k,l=1}^{N_0}
           \sin(\theta_k+\theta_l-2\theta_j).
\end{equation}
By defining $Z_1:=\frac{1}{N_0}\sum_{j=1}^{N_0}e^{i\theta_j} = R e^{i\Phi}$, we can then rewrite the two global sums in \eqref{eq:global_model_weakK2} in terms of $Z_1$, which gives
\begin{equation}\label{eq:exact_global_field_weakK2}
     \dot\theta_j
     =\omega_j
     +K_1\operatorname{Im}\!\left(Z_1e^{-i\theta_j}\right)
     +K_2\operatorname{Im}\!\left(Z_1^2e^{-2i\theta_j}\right).
\end{equation}
In the continuum limit ($N_0 \to \infty$), the state of the population is described by a density $f(\theta, \omega, t)$, with frequency marginal density $g(\omega)=\int_0^{2\pi} f(\theta,\omega,t)\,\mathrm{d}\theta$.
The density $f$ obeys the transport equation\cite{strogatz2000kuramoto} $\partial_t f+\partial_\theta(f\,v)=0$, where, from \eqref{eq:exact_global_field_weakK2}, $v(\theta,\omega,t)= \omega + K_1\operatorname{Im}\!\bigl(Z_1e^{-i\theta}\bigr)+ K_2\operatorname{Im}\!\bigl(Z_1^2e^{-2i\theta}\bigr)$, with $Z_1(t) = \int_\mathbb{R} \int_0^{2\pi} e^{i\theta} f(\theta, \omega, t)\,\mathrm{d}\theta \,\mathrm{d}\omega$.
We fix the phase origin so that $Z_1$ is real and nonnegative at the initial time ($Z_1(t=0)=R(t=0)\ge 0$, $\Phi(t=0)=0$); with $Z_1(t=0)$ real, the velocity field satisfies $v(\theta,\omega,0)=-v(-\theta,-\omega,0)$.
Therefore, an initial density $f$ that is symmetric under the joint reflection $(\theta,\omega)\to(-\theta,-\omega)$ (which requires $g$ to be even) retains this symmetry for all $t$, so that $Z_1(t)$ remains real for all $t$. We consider such symmetric initial conditions throughout (as is the case, e.g., for any initial phase distribution independent of $\omega$); consequently, the phase-locked stationary state analyzed below inherits this symmetry, i.e., $\theta(-\omega)=-\theta(\omega)$.
Each oscillator then obeys
\begin{equation}
\label{eq:rotating_field_weakK2}
 \dot\theta = v
 =\omega-K_1R\sin\theta-K_2 R^2\sin2\theta.
\end{equation}

Now, consider a stationary phase-locked solution, i.e., one in which every
oscillator's phase settles to a time-independent value $\theta(\omega)$
(the collective frequency being zero), with
all oscillators forming a single cluster centered at $\theta=0$. Each such
$\theta(\omega)$ is, by definition, the fixed point of~(C3) for the
oscillator with natural frequency $\omega$:
\begin{equation}
0 = \omega - K_1 R \sin\theta(\omega) - K_2 R^2 \sin 2\theta(\omega).
\label{eq:stationary}
\end{equation}
In this stationary state the density is
$f(\theta,\omega) = \delta(\theta-\theta(\omega))\, g(\omega)$, so that
\begin{equation}\label{eq:Z1}
Z_1 = \int_{\mathbb R}\int_0^{2\pi} e^{i\theta}
      f(\theta,\omega,t)\, d\theta\, d\omega
    = \int_{\mathbb R} e^{i\theta(\omega)} g(\omega)\, d\omega.
\end{equation}
Since $Z_1 = R$ is real and non-negative by the choice of phase origin,
self-consistency reads
\begin{equation} \label{eq:self_consistency_weakK2b}
R = \int_{\mathbb R} e^{i\theta(\omega)} g(\omega)\, d\omega. 
\end{equation}

Now, let us denote by $R_0$ and $\theta_0(\omega)$ the values of $R$ and $\theta(\omega)$ in the phase-locked state, when $K_2 = 0$, and let
$s_0(\omega):=\sin\theta_0(\omega)=\frac{\omega}{K_1R_0}$,%
\footnote{Full phase lock thus requires $\sup_{\omega\in\operatorname{supp}g}|\omega|<K_1R_0$.}
and
$c_0(\omega):=\cos\theta_0(\omega) = \sqrt{1-s_0(\omega)^2}>0$.
From \eqref{eq:self_consistency_weakK2b}, we have
$R_0=\int_{\mathbb R}c_0(\omega)g(\omega)\,\mathrm{d}\omega$.

We now determine how this stationary branch changes under an infinitesimal positive $K_2$.  
We define $R_0' := \frac{\mathrm{d}R}{\mathrm{d}K_2}\big|_{K_2=0}$ and $\theta_0'(\omega) := \frac{\partial \theta}{\partial K_2}\big|_{K_2=0}(\omega)$.
Differentiating~\eqref{eq:stationary}, evaluated along the branch $R(K_2)$,
$\theta(\omega;K_2)$, with respect to $K_2$ at $K_2=0$ gives
$0= K_1 R'_0 s_0 + K_1 R_0 c_0 \theta'_0 + R_0^2 \sin2 \theta_0$.
Using $\sin2\theta_0 = 2 s_0 c_0$, we obtain
\begin{equation}
\label{eq:theta_prime_weakK2}
 \theta'_0(\omega) = -\frac{R'_0}{R_0}\frac{s_0(\omega)}{c_0(\omega)} -\frac{2 R_0}{K_1}s_0(\omega).
\end{equation}
Differentiating \eqref{eq:self_consistency_weakK2b} with respect to $K_2$ yields 
$R'_0=-\int_{\mathbb R} s_0(\omega) \theta'_0(\omega) g(\omega) \, \mathrm{d}\omega$;
substitution of Eq.~\eqref{eq:theta_prime_weakK2} gives
\begin{equation}
\label{eq:abcde}
 R'_0 = \frac{R'_0}{R_0} \int_{\mathbb R} \frac{s_0(\omega)^2}{c_0(\omega)} g(\omega) \, \mathrm{d}\omega + \frac{2R_0}{K_1}
   \int_{\mathbb R}s_0(\omega)^2g(\omega) \, \mathrm{d}\omega.
\end{equation}

Next, define 
$D \coloneqq 1 - \frac{1}{R_0} \int_{\mathbb R} \frac{s_0(\omega)^2}{c_0(\omega)} g(\omega)\, \mathrm{d}\omega$.
Solving \eqref{eq:abcde} for $R'_0$ yields
\begin{equation*}
 R_0' = \left.\frac{\mathrm{d}R}{\mathrm{d} K_2}\right|_{K_2=0}
 =\frac{\displaystyle \frac{2 R_0}{K_1} \int_{\mathbb R} s_0(\omega)^2 g(\omega) \, \mathrm{d} \omega}
      {D}.
\end{equation*}
If $D>0$ (see below), then 
$\left.\frac{\mathrm{d}R}{ \mathrm{d}K_2}\right|_{K_2=0}>0$
whenever the frequency distribution has nonzero variance.  
Moreover, from \eqref{eq:theta_prime_weakK2}, 
the fact that $R_0'>0$ yields $\operatorname{sign} \theta'_0(\omega) = - \operatorname{sign} \theta_0(\omega)$.
This provides a local mean-field mechanism for the enhancement of coherence observed numerically: on a fully locked single-cluster branch satisfying $D>0$, an infinitesimal positive higher-order coupling $K_2$ increases $R$ and compresses the locked phases toward the center of the cluster.%
\footnote{{Note that this calculation does not determine how the measure of the attraction basin of the single-cluster branch varies with $K_2$, nor the probability that this branch is reached from incoherent initial conditions.}}

Two sufficient conditions for $D>0$ are a uniform frequency distribution with strict full locking and a sufficiently concentrated locked cluster.
Indeed, using
$R_0=\int_{\mathbb R}c_0(\omega)g(\omega)\,\mathrm{d}\omega$ and
$c_0^2 = 1 - s_0^2$, we can rewrite
\begin{equation}\label{eq:form_D}
 D
 =\frac{1}{R_0}\int_{\mathbb R}
 \left(c_0-\frac{s_0^2}{c_0}\right)g(\omega)\,\mathrm{d}\omega
 =\frac{1}{R_0}\int_{\mathbb R}
 \frac{1-2s_0^2}{c_0}g(\omega)\,\mathrm{d}\omega.
\end{equation}

For a uniform distribution on $[-\gamma,\gamma]$, set
$a:=\gamma/(K_1R_0)<1$, where $a<1$ is the strict full-locking
condition, and let $x:=\omega/(K_1R_0)$. Since
$\frac{\mathrm{d}}{\mathrm{d}x} [x\sqrt{1-x^2}]=(1-2x^2)/\sqrt{1-x^2}$, we obtain from \eqref{eq:form_D} that
\begin{equation}
 D =\frac{1}{2aR_0}\int_{-a}^{a}
 \frac{1-2x^2}{\sqrt{1-x^2}}\,\mathrm{d}x =\frac{\sqrt{1-a^2}}{R_0}>0.
\end{equation}

Alternatively, if $|\theta_0(\omega)|<\pi/4$ for $g$-almost every $\omega$, then $1-2s_0(\omega)^2=\cos(2\theta_0(\omega))>0$ and $c_0(\omega)>0$. Hence, the integrand in \eqref{eq:form_D} is positive and $D>0$.
}

\section{Complementary results}\label{app:C}

In the Main Text, we presented numerical analyses of the higher-order Kuramoto model on small hypergraphs ($N_0=10$) and with uniform frequency distribution. Here, we provide the simulation results obtained with frequencies distributed normally as well as those for larger structures. 
We also report the Kuramoto-Daido order parameter $R_2$, which serves both
to check that the non-monotonic optimum is not specific to the choice of
order parameter and to distinguish incoherence from two-cluster states.
The main findings of the Main Text are confirmed throughout.

All figures presenting simulation results, including those in the Main Text, are summarized in Tables~\ref{tab:coupl} and~\ref{tab:hyper_alloc}, for the coupling strength and the hyperedge allocation analyses, respectively.

\begin{table}[t]
    \centering
    \setlength{\tabcolsep}{6pt}
    \caption{{Review of figures concerning the coupling strength analysis.}}
    \label{tab:coupl}
    \begin{tabular}{@{}lllll@{}}
    \toprule
    Figure & Discussed in & $N_0$ & Metric & Frequencies \\
    \midrule
    Fig.~\ref{fig:couplings_study_standard_coupl_R1} & Sec.~\ref{sec:adding_weak_ho_enhances} & $10$ & $R$ & uniform \\
    Fig.~\ref{fig:distributions_order_parameter_uniform} & Sec.~\ref{sec:distinct_regimes} & $10$ & $R$ & uniform \\
    Fig.~\ref{fig:alltoall} & Sec.~\ref{sec:all_to_all} & $10\,000$ & $R$, $R_2$ & uniform \\
    \addlinespace 
    Fig.~\ref{fig:couplings_study_standard_coupl_gaussian_R1} & App.~\ref{sec:gaussian_frequencies} & $10$ & $R$ & Gaussian \\
    Fig.~\ref{fig:distributions_order_parameter_gaussian} & App.~\ref{sec:gaussian_frequencies} & $10$ & $R$ & Gaussian \\
    Fig.~\ref{fig:couplings_study_standard_coupl_100_nodes_R1} & App.~\ref{sec:larger_networks} & $100$ & $R$ & uniform, Gaussian \\
    \addlinespace
    
    Fig.~\ref{fig:couplings_study_standard_coupl_R2} & App.~\ref{sec:kuramoto_daido} & $10$ & $R_2$ & uniform \\
    Fig.~\ref{fig:couplings_study_standard_coupl_100_nodes_R2} & App.~\ref{sec:kuramoto_daido} & $100$ & $R_2$ & uniform, Gaussian \\
     Fig.~\ref{fig:alltoall_vary_both_couplings} & App.~\ref{sec:alltoall_vary_couplings} & $10\,000$ & $R, R_2$ & uniform \\
    \bottomrule
    \end{tabular}
\end{table}

\begin{table}[t]
    \centering
    \setlength{\tabcolsep}{6pt}
    \caption{{Review of figures concerning the hyperedge allocation analysis.}}
    \label{tab:hyper_alloc}
    \begin{tabular}{@{}lllll@{}}
    \toprule
    Figure & Discussed in & $N_0$ & Metric & Frequencies \\
    \midrule
    Fig.~\ref{fig:alloc_study_standard_coupl_omega03_R}, \ref{fig:alloc_study_free_couplings} & Secs.~\ref{sec:results2}, \ref{sec:allocation_study_free_gains} & $10$ & $R$ & uniform \\
    \addlinespace
    Fig.~\ref{fig:alloc_study_standard_coupl_gaussian_R1} & App.~\ref{sec:gaussian_frequencies} & $10$ & $R$ & Gaussian \\
    Fig.~\ref{fig:alloc_study_standard_coupl_uniform_100_nodes_R1} & App.~\ref{sec:larger_networks} & $100$ & $R$ & uniform \\
    Fig.~\ref{fig:alloc_study_standard_coupl_gaussian_100_nodes_R1} & App.~\ref{sec:larger_networks} & $100$ & $R$ & Gaussian \\
    \addlinespace
    Fig.~\ref{fig:alloc_study_standard_coupl_omega03_R2} & App.~\ref{sec:kuramoto_daido} & $10$ & $R_2$ & uniform \\
    Fig.~\ref{fig:alloc_study_standard_coupl_uniform_100_nodes_R2} & App.~\ref{sec:kuramoto_daido} & $100$ & $R_2$ & uniform \\
    \addlinespace
     Fig.~\ref{fig:alloc_study_free_couplings} & App.~\ref{sec:allocation_study_free_gains} & $10$ & $R$ & uniform \\
    \bottomrule
    \end{tabular}
\end{table}

\subsection{Gaussian frequencies}\label{sec:gaussian_frequencies}

\begin{figure}[p]
    \centering
    \includegraphics[scale=1]{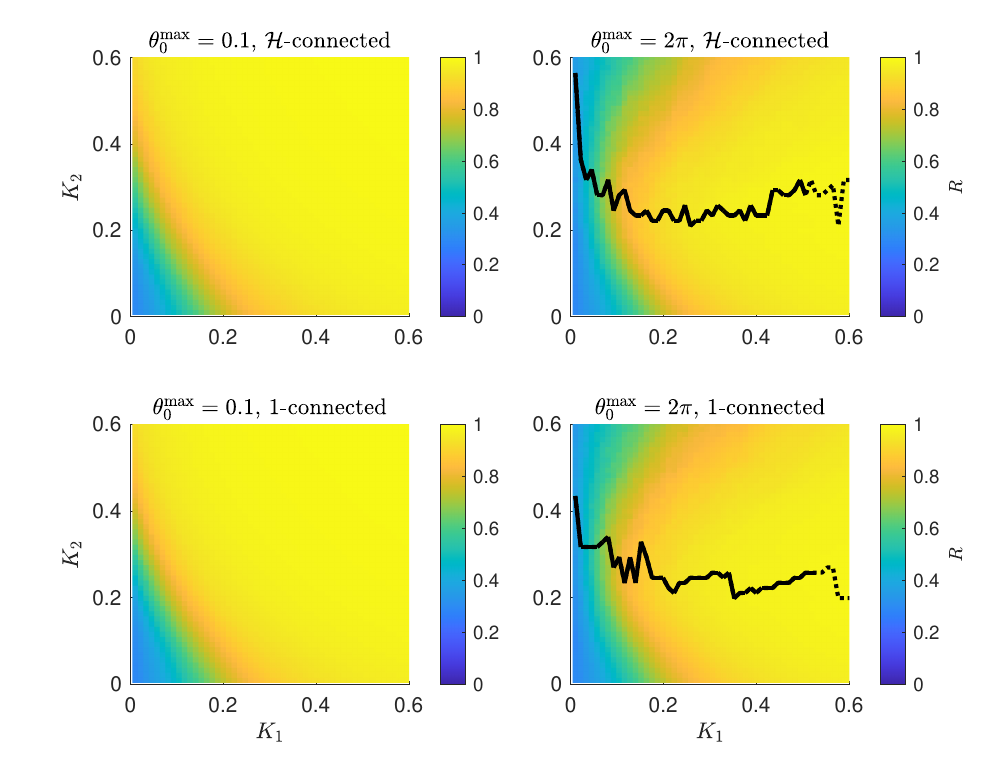}
    \caption{
    Average Kuramoto order parameter $R$ as a function of the coupling strengths of pairwise ($K_1$) and higher-order interactions ($K_2$), computed over $N_{\C{H}}=300$ randomly generated hypergraphs. 
    The grid $K_1,K_2$ is $51\times51$. 
    Each hypergraph has $N_0=10$ nodes, $N_1 = 20$ links and $N_2 = 10$ triangles, with oscillator frequencies $\omega_{j}$ distributed normally with average $0.15$ and standard deviation $0.1$, and initial phases distributed uniformly in $[0, \theta_0^\R{max}]$.
    Left panels show the cases of initial phases close to synchronization, i.e., $\theta_0^\R{max} = 0.1$;
    right panels show the case of incoherent initial states, i.e., $\theta_0^\R{max} = 2\pi$.
    On the right panels, the black line indicates, for each $K_1$, the $K_2$ yielding the maximum $R$. {The line is dotted when the 95\textsuperscript{th}–5\textsuperscript{th} percentile spread over the maximization domain is below 0.05.  {This Figure should be compared with Figure~\ref{fig:couplings_study_standard_coupl_R1} of the Main Text, for which simulations were performed with uniformly distributed frequencies}}.
    }\label{fig:couplings_study_standard_coupl_gaussian_R1}
\end{figure}

\begin{figure}[p]
    \centering
    \includegraphics[scale=1]{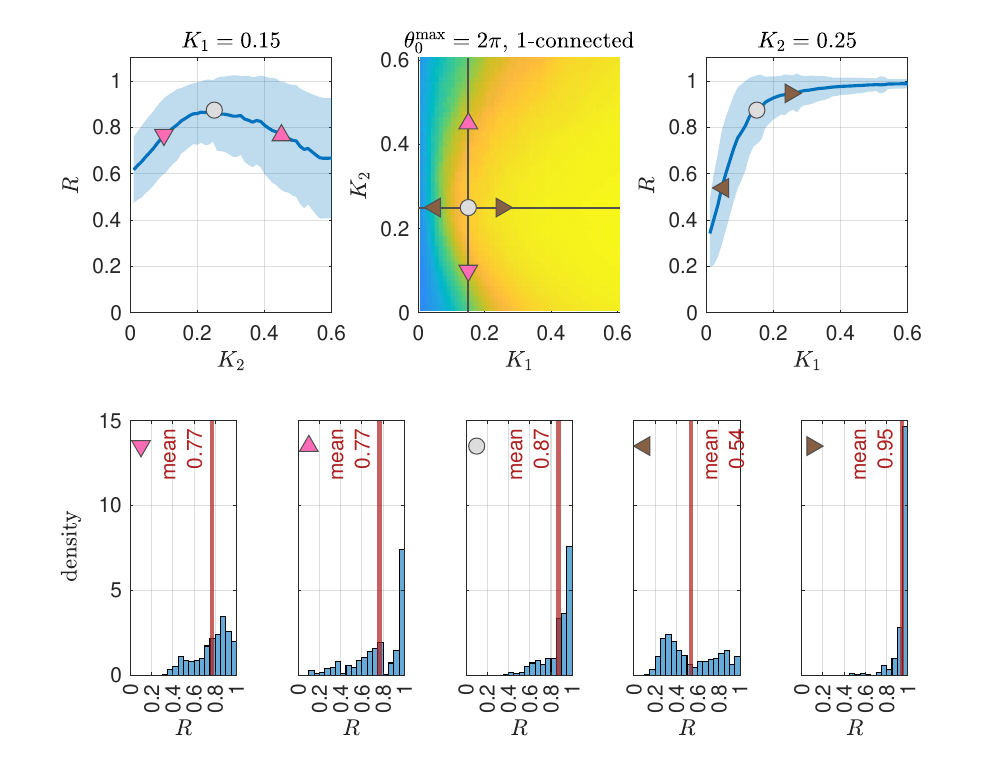}
    \caption{
    System response to different pairwise ($K_1$) and higher-order ($K_2$) coupling strengths, for initially incoherent $1$-connected hypergraphs from Figure~\ref{fig:couplings_study_standard_coupl_gaussian_R1}.
    The top-center panel replicates the bottom right panel of Figure~\ref{fig:couplings_study_standard_coupl_gaussian_R1} for comparison (color is the mean order parameter $R$).
    The top-left (resp.~top-right) panel shows the mean value of $R$ (blue line) and its standard deviation (shaded area) as a function of the higher-order coupling strength $K_2$ (resp.~$K_1$) while keeping the pairwise interaction strength $K_1$ (resp.~$K_2$) fixed.
    The values of $K_1$, $K_2$ explored are also indicated in the top-center panel by a vertical (resp.~horizontal) black line.
    Pink (resp.~brown) triangles pointing upward/downward (resp.~leftward/rightward) mark representative pairs $(K_1, K_2)$.
    The bottom panels display the distributions (integral normalized to $1$) of $R$ across the  $N_{\C{H}} = 300$ realizations of hypergraph, initial conditions, and frequencies. {This Figure should be compared with Figure~\ref{fig:distributions_order_parameter_uniform} of the Main Text, for which simulations were performed with uniformly distributed frequencies.}}
    \label{fig:distributions_order_parameter_gaussian}
\end{figure}

\begin{figure}[p]
    \centering
    \includegraphics[scale=1]{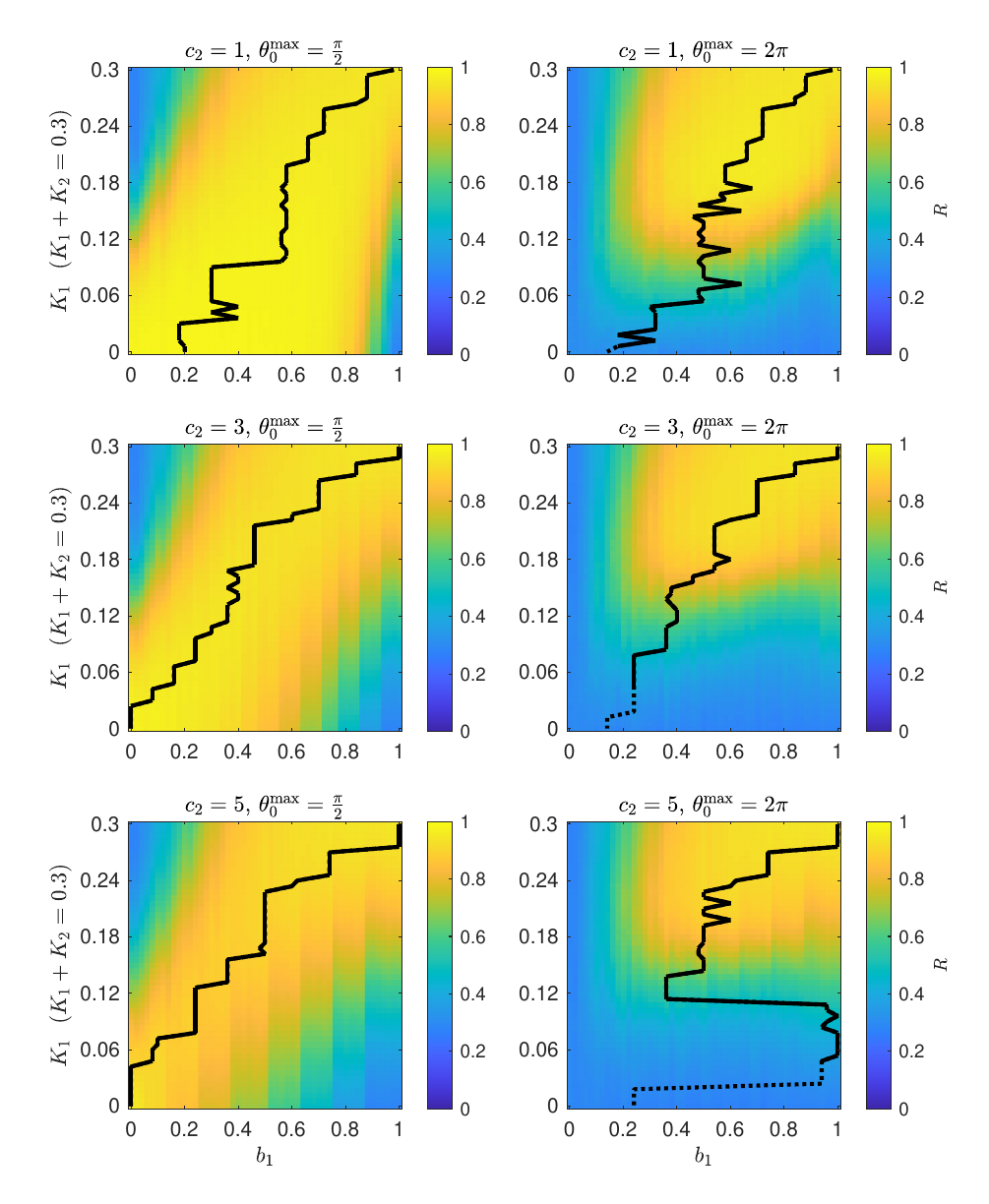}
    \caption{
    Average Kuramoto order parameter $R$ for different combinations of links and triangles.
    $K_1$ and $K_2$ are the coupling strengths of pairwise and $3$-body interactions, respectively.
    The grid $(b_1,K_1)$ is $51\times51$. 
    For each value of the link allocation fraction $b_1$, $N_{\C{H}} = 500$ random hypergraphs were generated, with $N_0=10$ nodes, frequencies distributed normally with average $0.15$ and standard deviation $0.1$ and initial phases drawn uniformly from $[0, \theta_0^\R{max}]$---$\theta_0^\R{max} = \frac{\pi}{2}$ in the left panels, and $\theta_0^\R{max} = 2\pi$ in the right panels.    
    The relative link cost is fixed at $c_1 = 1$;
    triangle costs $c_2$ are $1$ in top panels, $3$ in middle panels, and $5$ in bottom panels.
    For each $K_1$ (y-axis), the black line marks the $b_1$ (x-axis) yielding the highest $R$ (color). {The line is dotted when the 95\textsuperscript{th}–5\textsuperscript{th} percentile spread over the maximization domain is below 0.05. {This Figure should be compared with Figure~\ref{fig:alloc_study_standard_coupl_omega03_R} of the Main Text, for which simulations were performed with uniformly distributed frequencies}}.
    }\label{fig:alloc_study_standard_coupl_gaussian_R1}
\end{figure}

In Figures
\ref{fig:couplings_study_standard_coupl_gaussian_R1}, 
\ref{fig:distributions_order_parameter_gaussian}, \ref{fig:couplings_study_standard_coupl_100_nodes_R1}
(for the coupling strength analysis) and
\ref{fig:alloc_study_standard_coupl_gaussian_R1}, 
\ref{fig:alloc_study_standard_coupl_gaussian_100_nodes_R1}
(for the hyperedge allocation analysis),
we show the order parameter $R$ obtained for all the simulations considered, in the case that, for each hypergraph realization, oscillator frequencies $\omega_j$ are drawn randomly from a Gaussian distribution with mean $0.15$ and standard deviation $0.1$.

The results are qualitatively analogous to those reported in the Main Text, where a uniform frequency distribution was considered instead.

\subsection{Larger hypergraphs ($N_0=100$)}\label{sec:larger_networks}

\begin{figure}[p]
    \centering
    \includegraphics[scale=1]{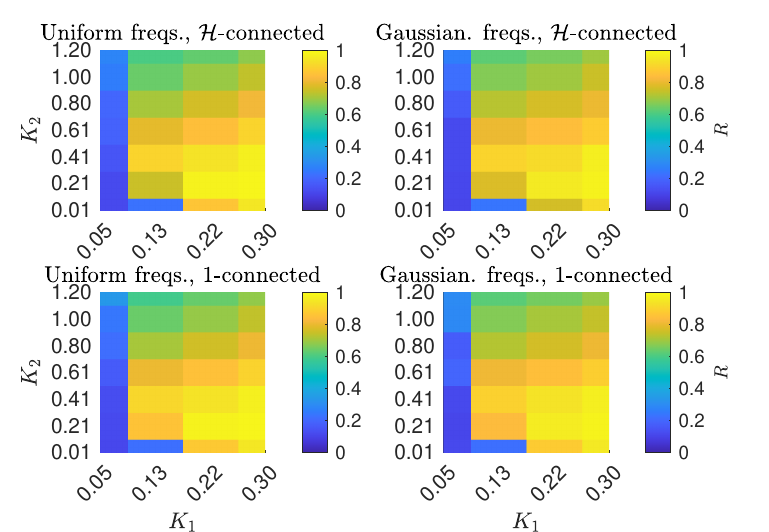}
    \caption{
    Average Kuramoto order parameter $R$ as a function of the coupling strengths of pairwise ($K_1$) and higher-order interactions ($K_2$), computed over $N_{\C{H}}=100$ randomly generated hypergraphs. 
    The grid $K_1,K_2$ is $4 \times 7$.
    Each hypergraph has $N_0=100$ nodes, $N_1 = 2200$ links and $N_2 = 13\,475$ triangles.
    On the left panel{s}, oscillator frequencies $\omega_{j}$ are uniformly distributed in $[0, 0.3]$, while on the right panel{s} they are distributed normally with mean $0.15$ and standard deviation $0.1$.
    Initial phases are drawn uniformly in $[0, \theta_0^\R{max}=2\pi]$. {The left panels of this Figure should be compared with Figure~\ref{fig:couplings_study_standard_coupl_R1} of the Main Text, while the right panels with Figure~\ref{fig:couplings_study_standard_coupl_gaussian_R1}, where analogous simulations were performed with $N_0=10$ nodes}.}\label{fig:couplings_study_standard_coupl_100_nodes_R1}
\end{figure}

\begin{figure}[p]
    \centering
    \includegraphics[scale=1]{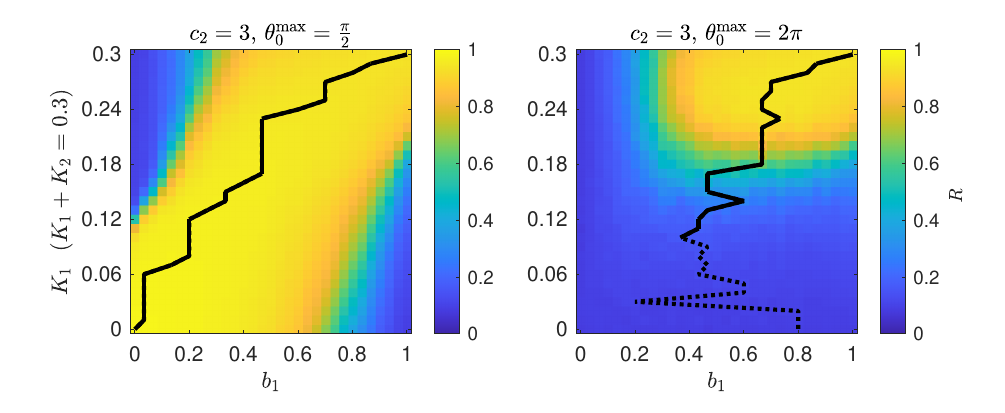}
    \caption{
    Average Kuramoto order parameter $R$ for different combinations of links and triangles.
    $K_1$ and $K_2$ are the coupling strengths of pairwise and $3$-body interactions, respectively.
    The grid $(b_1,K_1)$ is $31\times31$. 
    For each value of the link allocation fraction $b_1$, $N_{\C{H}} = 100$ random hypergraphs were generated, with $N_0=100$ nodes, frequencies drawn uniformly at random from $[0, 0.3]$, and initial phases from $[0, \theta_0^\R{max}]$---$\theta_0^\R{max} = \frac{\pi}{2}$ in the left panel, and $\theta_0^\R{max} = 2\pi$ in the right panel.    
    The relative link cost is fixed at $c_1 = 1$, while 
    triangle cost is $c_2 = 3$.
    For each $K_1$ (y-axis), the black line marks the $b_1$ (x-axis) yielding the highest $R$ (color). {The line is dotted when the 95\textsuperscript{th}–5\textsuperscript{th} percentile spread over the maximization domain is below 0.05}. {This Figure should be compared with Figure~\ref{fig:alloc_study_standard_coupl_omega03_R} of the Main Text, where analogous simulations were performed with $N_0=10$ nodes}.   }\label{fig:alloc_study_standard_coupl_uniform_100_nodes_R1}
\end{figure}

\begin{figure}[p]
    \centering
    \includegraphics[scale=1]{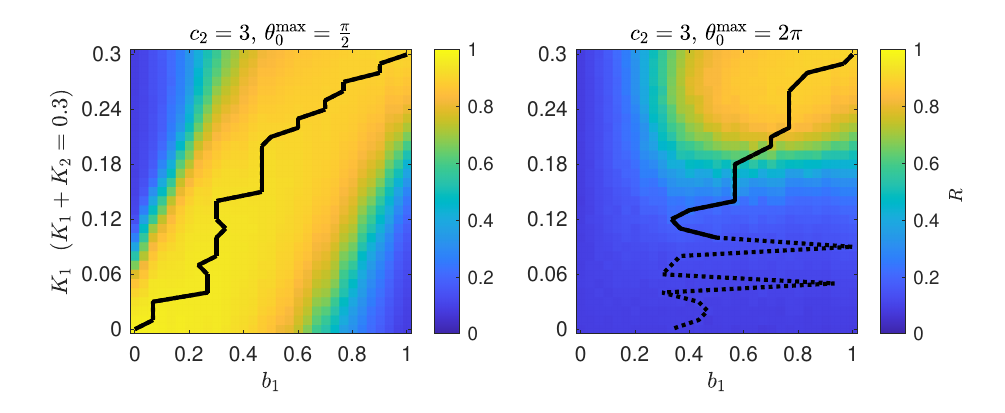}
    \caption{
    Average Kuramoto order parameter $R$ for different combinations of links and triangles.
    $K_1$ and $K_2$ are the coupling strengths of pairwise and $3$-body interactions, respectively.
    The grid $(b_1,K_1)$ is $31\times31$. 
    For each value of the link allocation fraction $b_1$, $N_{\C{H}} = 100$ random hypergraphs were generated, with $N_0=100$ nodes, frequencies distributed normally with mean $0.15$ and standard deviation $0.1$, and initial phases drawn uniformly from $[0, \theta_0^\R{max}]$---$\theta_0^\R{max} = \frac{\pi}{2}$ in the left panel, and $\theta_0^\R{max} = 2\pi$ in the right panel.    
    The relative link cost is fixed at $c_1 = 1$, while 
    triangle cost is $c_2 = 3$.
    For each $K_1$ (y-axis), the black line marks the $b_1$ (x-axis) yielding the highest $R$ (color). {The line is dotted when the 95\textsuperscript{th}–5\textsuperscript{th} percentile spread over the maximization domain is below 0.05}. {This Figure should be compared with Figure~\ref{fig:alloc_study_standard_coupl_uniform_100_nodes_R1}, for which simulations were performed with uniformly distributed frequencies.}}\label{fig:alloc_study_standard_coupl_gaussian_100_nodes_R1}
\end{figure}

In Figures
\ref{fig:couplings_study_standard_coupl_100_nodes_R1}
(for the coupling strength analysis) and
\ref{fig:alloc_study_standard_coupl_uniform_100_nodes_R1},
\ref{fig:alloc_study_standard_coupl_gaussian_100_nodes_R1}
(for the hyperedge allocation analysis),
we show the value of the order parameter $R$ obtained for simulations of larger hypergraphs of $N_0 = 100$, with both uniform and normal frequency distributions. 

To illustrate the choice of the simulation parameters, we recall the expressions for 
(i) the minimum number $N_1^\R{min}$ of links required for $1$-connectivity,
(ii) the minimum number $N_2^\R{min}$ of triangles required for $2$-connectivity,
(iii) the maximum number $N_1^\R{max}$ of distinct links, and 
(iv) the maximum number $N_2^\R{max}$ of distinct triangles.
For a structure of $N_0$ nodes, these are
\begin{equation*}
N_1^{\mathrm{min}}(N_0) = N_0 - 1,
\qquad
N_2^{\mathrm{min}}(N_0) = \left\lceil \frac{N_0 - 1}{2} \right\rceil,
\qquad
N_1^{\mathrm{max}}(N_0) = \frac{N_0 (N_0 - 1)}{2},
\qquad
N_2^{\mathrm{max}}(N_0) = \frac{N_0 (N_0 - 1)(N_0 - 2)}{6}.
\end{equation*}

For the coupling strength analysis, we set the numbers of links $N_1$ and triangles $N_2$ so as to preserve the same interaction densities used for the smaller hypergraphs ($N_0 = 10$). 
Preserving link and triangle densities requires keeping the ratios $N_1 / N_1^{\mathrm{max}}$ and $N_2 / N_2^{\mathrm{max}}$ constant.
For $N_0 = 10$, we have
$N_1^{\mathrm{max}}(10) = 45$ and
$N_2^{\mathrm{max}}(10) = 120$,
and we set
$N_1 = 20$ and $N_2 = 10$.
Then, for $N_0 = 100$, we compute
$N_1^{\mathrm{max}}(100) = 4\,950$ and
$N_2^{\mathrm{max}}(100) = 161\,700$;
thus, matching the previous densities yields
$N_1 = 2\,200$ and $N_2 = 13\,475$.
Owing to the very high computational cost associated with such a large number of triangles, in Figure~\ref{fig:couplings_study_standard_coupl_100_nodes_R1} we adopted a coarser $4 \times 7$ parameter grid and a reduced number of hypergraph realizations $N_\mathcal{H} = 100$. 
In this case, the higher-order Kuramoto model was integrated for $1\,000$ time units (instead of $600$), as convergence required longer times; {we observed no qualitative changes in terms of convergence when the simulation time was extended to $3\,000$ time units}.
Despite the coarser grid, the non-monotonicity of $R$ with respect to $K_2$ is observed also in this case.

For choosing the budget value $J$ for the hyperedge allocation analysis, we proceeded as follows.
Let $c_1^\mathrm{min}$ (resp.~$c_2^\mathrm{min}$) be the minimum cost associated to links (resp.~triangles), and let $c_1^\mathrm{max}$ (resp.~$c_2^\mathrm{max}$) denote the corresponding maximum cost.
In our study, $c_1^\mathrm{min} = c_1^\mathrm{max} = 1$, $c_2^\mathrm{min} = 1$, and $c_2^\mathrm{max} = 5$.
The budget $J$ was then chosen to satisfy two conditions:
(i) $J \ge \max \{c_1^\mathrm{max} N_1^\R{min}, c_2^\mathrm{max} N_2^\R{min}\}$, ensuring $\C{H}$-connected hypergraphs could always be generated;
(ii) $J \le \min \{c_1^\mathrm{min} N_1^\R{max}, c_2^\mathrm{min} N_2^\R{max}\}$, so that no portion of the budget is wasted.
For $N_0 = 100$, we have
$N_1^{\mathrm{min}}(100) = 99$,
$N_2^{\mathrm{min}}(100) = 50$,
$N_1^{\mathrm{max}}(100) = 4\,950$, and
$N_2^{\mathrm{max}}(100) = 161\,700$.
Hence, conditions (i) and (ii) yield
$J \ge \max\{ 1 \cdot 99, 5 \cdot 50\} = 250$ and 
$J \le \min\{ 1 \cdot 4\,950, 1 \cdot 161\,700\} =  4\,950$.
This defines the admissible range $J \in [250, 4\,950]$, from which we selected $J = 600$.
In this case, we used a $31 \times 31$ parameter grid and $N_\mathcal{H} = 100$ hypergraph realizations. 
The qualitative findings reported in the Main Text are confirmed also in this setting.

\subsection{Kuramoto-Daido order parameter}\label{sec:kuramoto_daido}

\begin{figure}[p]
    \centering
    \includegraphics[scale=1]{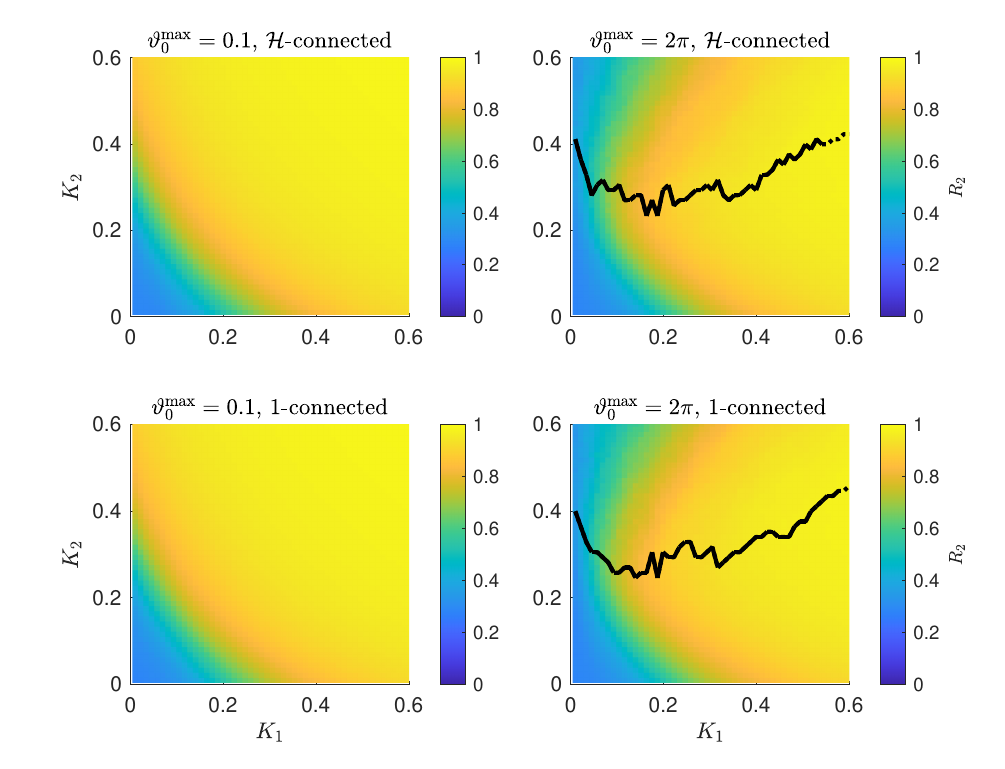}
    \caption{Average Kuramoto-Daido order parameter $R_2$ as a function of the coupling strengths of pairwise ($K_1$) and higher-order interactions ($K_2$), computed over $N_{\C{H}}=300$ randomly generated hypergraphs. 
    The grid $K_1,K_2$ is $51\times51$ (see Appendix~\ref{app:B}). 
    Each hypergraph has $N_0=10$ nodes, $N_1 = 20$ links and $N_2 = 10$ triangles, with oscillator frequencies $\omega_{j}$ uniformly distributed in $[0, 0.3]$, and initial phases $[0, \theta_0^\R{max}]$.
    Left panels show the cases of initial phases close to synchronization, i.e., $\theta_0^\R{max} = 0.1$;
    right panels show the case of incoherent initial states, i.e., $\theta_0^\R{max} = 2\pi$.
    On the right panels, the black line indicates, for each $K_1$, the $K_2$ yielding the maximum $R_2$. {The line is dotted when the 95\textsuperscript{th}–5\textsuperscript{th} percentile spread over the maximization domain is below 0.05}. {This Figure should be compared with Figure~\ref{fig:couplings_study_standard_coupl_R1} of the Main Text, which displays the Kuramoto order parameter $R$.}}    \label{fig:couplings_study_standard_coupl_R2}
\end{figure}

\begin{figure}[p]
    \centering
    \includegraphics[scale=1]{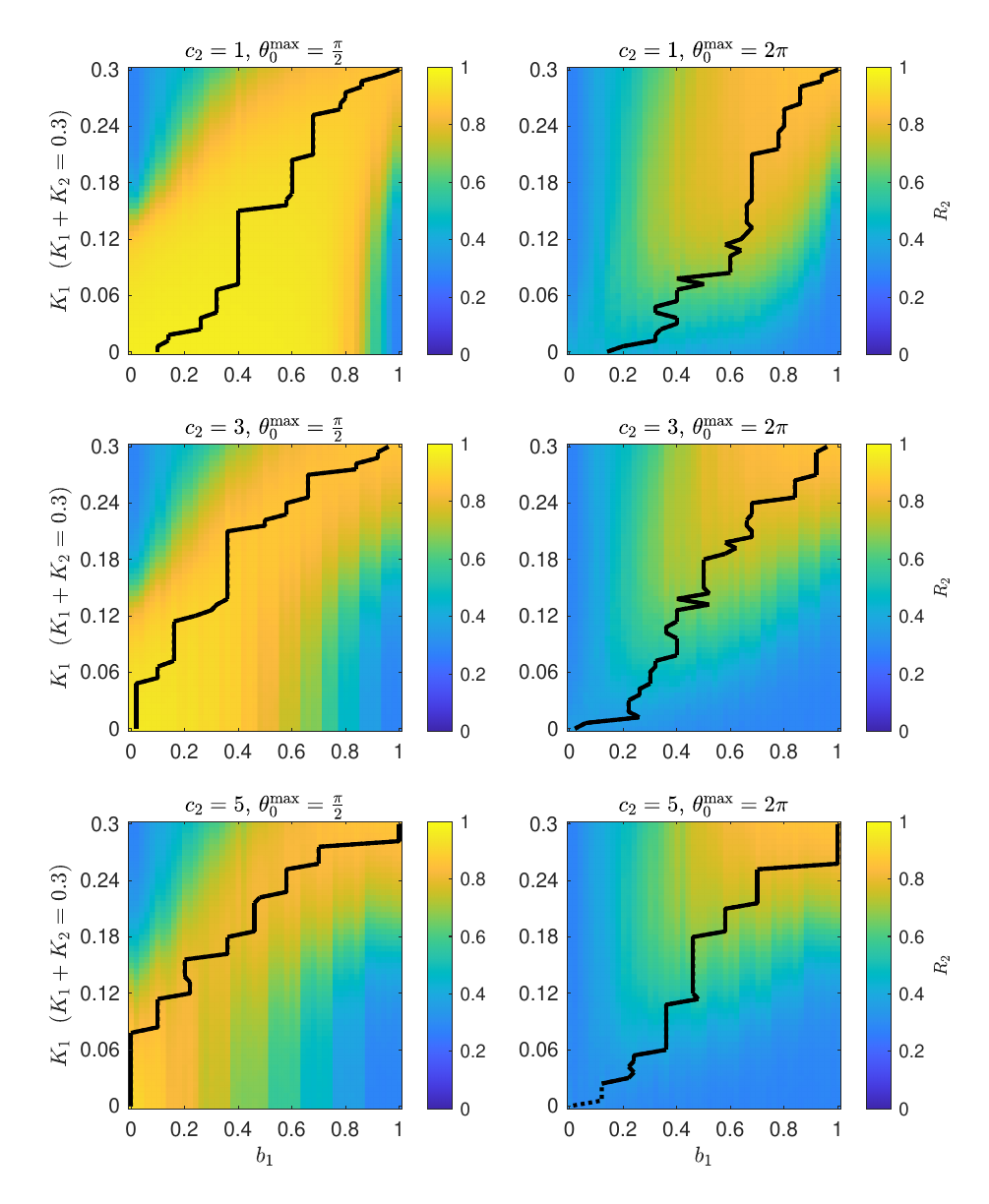}
    \caption{
    Average Kuramoto-Daido order parameter $R_2$ for different combinations of links and triangles.
    $K_1$ and $K_2$ are the coupling strengths of pairwise and $3$-body interactions, respectively.
    The grid $(b_1,K_1)$ is $51\times51$. 
    For each value of the link allocation fraction $b_1$, $N_{\C{H}} = 500$ random hypergraphs were generated, with $N_0=10$ nodes, frequencies drawn uniformly at random from $[0, 0.3]$, and initial phases from $[0, \theta_0^\R{max}]$---$\theta_0^\R{max} = \frac{\pi}{2}$ in the left panels, and $\theta_0^\R{max} = 2\pi$ in the right panels.    
    The relative link cost is fixed at $c_1 = 1$;
    triangle costs $c_2$ are $1$ in top panels, $3$ in middle panels, and $5$ in bottom panels.
    For each $K_1$ (y-axis), the black line marks the $b_1$ (x-axis) yielding the highest $R_2$ (color). {The line is dotted when the 95\textsuperscript{th}–5\textsuperscript{th} percentile spread over the maximization domain is below 0.05}. {This Figure should be compared with Figure~\ref{fig:alloc_study_standard_coupl_omega03_R} of the Main Text, which displays the Kuramoto order parameter $R$.} }   \label{fig:alloc_study_standard_coupl_omega03_R2}
\end{figure}

\begin{figure}[p]
    \centering
    \includegraphics[scale=1]{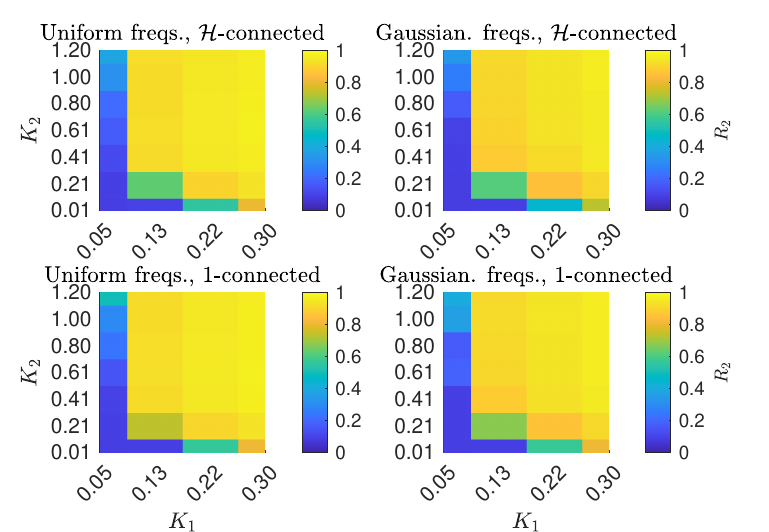}
    \caption{
    Average Kuramoto-Daido order parameter $R_2$ as a function of the coupling strengths of pairwise ($K_1$) and higher-order interactions ($K_2$), computed over $N_{\C{H}}=100$ randomly generated hypergraphs.
    The grid $K_1,K_2$ is $4 \times 7$.
    Each hypergraph has $N_0=100$ nodes, $N_1 = 2200$ links and $N_2 = 13\,475$ triangles.
    On the left panel, oscillator frequencies $\omega_{j}$ are uniformly distributed in $[0, 0.3]$, while on the right panel they are distributed normally with mean $0.15$ and standard deviation $0.1$.
    Initial phases are drawn uniformly in $[0, \theta_0^\R{max}=2\pi]$. {This Figure should be compared with Figure~\ref{fig:couplings_study_standard_coupl_100_nodes_R1}, which displays the Kuramoto order parameter $R$.}}\label{fig:couplings_study_standard_coupl_100_nodes_R2}
\end{figure}

\begin{figure}[p]
    \centering
    \includegraphics[scale=1]{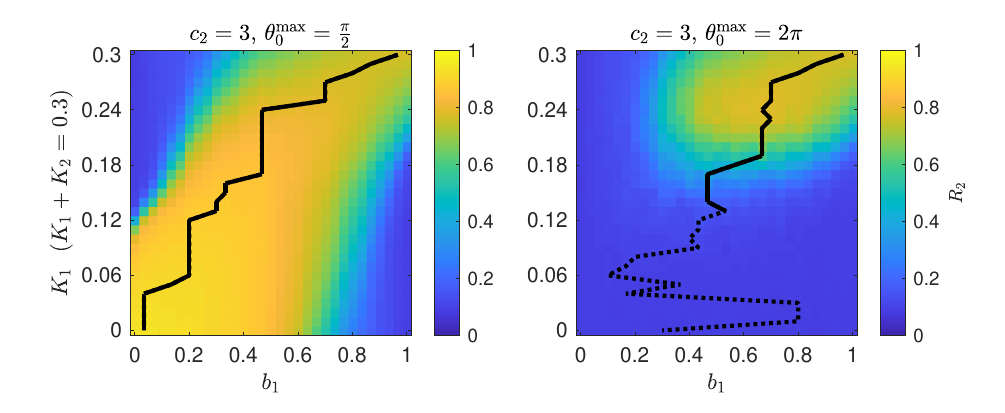}
    \caption{
    Average Kuramoto-Daido order parameter $R_2$ for different combinations of links and triangles.
    $K_1$ and $K_2$ are the coupling strengths of pairwise and $3$-body interactions, respectively.
    The grid $(b_1,K_1)$ is $31\times31$. 
    For each value of the link allocation fraction $b_1$, $N_{\C{H}} = 100$ random hypergraphs were generated, with $N_0=100$ nodes, frequencies drawn uniformly at random from $[0, 0.3]$, and initial phases from $[0, \theta_0^\R{max}]$---$\theta_0^\R{max} = \frac{\pi}{2}$ in the left panel, and $\theta_0^\R{max} = 2\pi$ in the right panel.    
    The relative link cost is fixed at $c_1 = 1$, while 
    triangle cost is $c_2 = 3$.
    For each $K_1$ (y-axis), the black line marks the $b_1$ (x-axis) yielding the highest $R_2$ (color). {The line is dotted when the 95\textsuperscript{th}–5\textsuperscript{th} percentile spread over the maximization domain is below 0.05}. {This Figure should be compared with Figure~\ref{fig:alloc_study_standard_coupl_omega03_R2}, where analogous simulations were performed with $N_0=10$ nodes.} }\label{fig:alloc_study_standard_coupl_uniform_100_nodes_R2}
\end{figure}

{In Figures
\ref{fig:couplings_study_standard_coupl_R2},
\ref{fig:couplings_study_standard_coupl_100_nodes_R2} 
(for the coupling strength analysis) and
\ref{fig:alloc_study_standard_coupl_omega03_R2},
\ref{fig:alloc_study_standard_coupl_uniform_100_nodes_R2},
(for the hyperedge allocation analysis),}
we show the value of the Kuramoto-Daido order parameter $R_2 = \abs{\frac{1}{N_0}\sum_{j=1}^{N_0} e^{2i\theta_j}}$ obtained for the simulations considered. 
The optimal $K_2$ remains nonzero for $R_2$ as well.
On random hypergraphs with $N_0 = 10$ the two order parameters behave
similarly, so that the decrease of $R$ at large $K_2$ reflects genuine
incoherence; for $N_0 = 100$, $R_2$ remains appreciably above $R$ in the
same regime, consistently with the partial clustering discussed in
Section~\ref{sec:results1a}.

\subsection{All-to-all hypergraphs}
\label{sec:alltoall_vary_couplings}

\begin{figure*}[tb]
    \centering
    \includegraphics[scale=1]{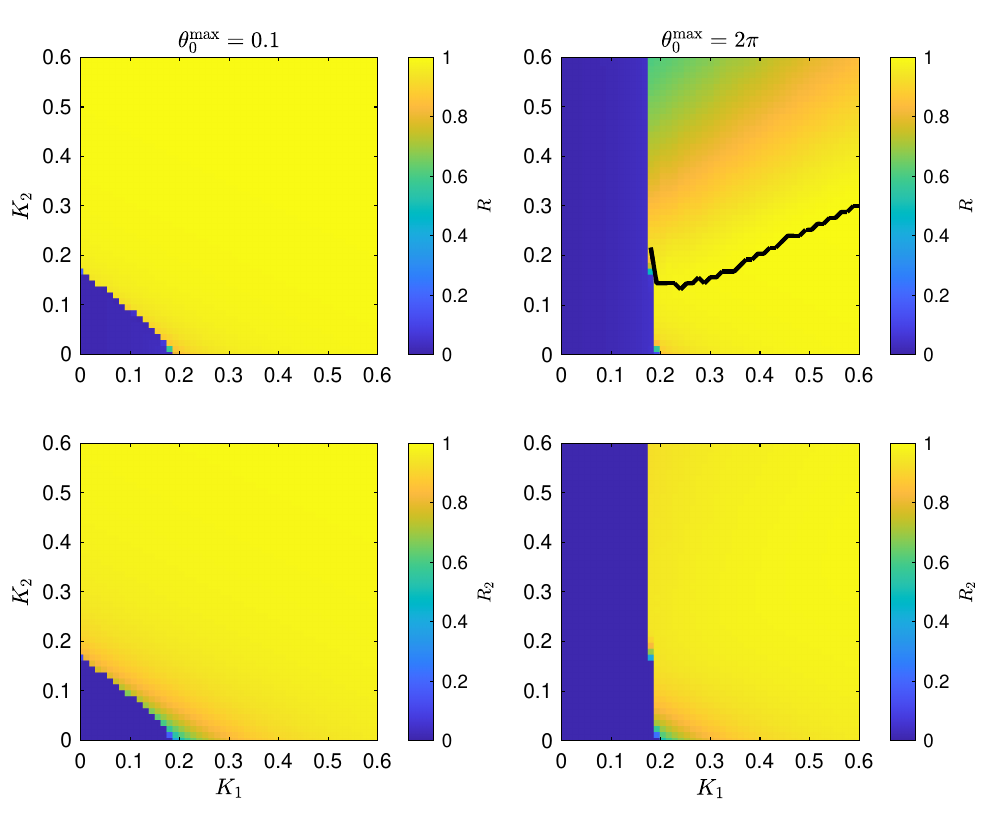}
    \caption{%
    {Average Kuramoto order parameter $R$ (top panels) and Kuramoto-Daido order parameter $R_2$ (bottom panels) as functions of the coupling strengths of pairwise ($K_1$) and higher-order $3$-body interactions ($K_2$), computed over an all-to-all hypergraph with $N_0 = 10\,000$ nodes.
    The grid $K_1,K_2$ is $51\times51$.  
    Oscillator frequencies $\omega_{j}$ are uniformly distributed in $[0, 0.3]$, and initial phases in $[0, \theta_0^\R{max}]$.
    Left panels show the case of initial phases close to synchronization, i.e., $\theta_0^\R{max} = 0.1$;
    right panels show the case of incoherent initial states, i.e., $\theta_0^\R{max} = 2\pi$.
    In the top right panel, the black line indicates, for each $K_1$, the $K_2$ yielding the maximum $R$: that value is never zero. 
    This Figure, which is the $2$-dimensional version of \ref{fig:alltoall}, should be compared with Figure~\ref{fig:alltoall} of the Main Text which shows a different representation of the same system condition (restricted to the case $\theta_0^\R{max} = 2\pi$).}
    \label{fig:alltoall_vary_both_couplings}}
\end{figure*}

{In Figure \ref{fig:alltoall_vary_both_couplings} we report the values of $R$ and $R_2$ for the all-to-all hypergraph with $N_0 = 10\,000$ nodes (as in Figure \ref{fig:alltoall}), inspecting a full $51 \times 51$ grid of $K_1$ and $K_2$ values.

The results confirm that when starting from incoherent phases, a weak higher-order coupling enhances synchronization in terms of the Kuramoto order parameter $R$.
Figure \ref{fig:alltoall_vary_both_couplings} (in accordance with Figure \ref{fig:alltoall}) also shows that $R$ starts to degrade for larger $K_2$, while $R_2$ remains high, pointing at the emergence of a two-cluster state.
The $K_1$--$K_2$ grid values are chosen as in Figure \ref{fig:couplings_study_standard_coupl_R1} to also allow comparison with the latter.
Indeed, we find that for this large all-to-all hypergraph, while the bistability is more prominent, the benefit of low $K_2$ (in terms of $R$) compared to $K_2 = 0$ is less prominent than for the smaller ($N_0 = 10$) random hypergraphs of Figure \ref{fig:couplings_study_standard_coupl_R1}. Nonetheless, also in this setting, we can observe the non-monotonic effect, i.e., low (non-zero) values of $K_2$ are beneficial for synchronization. The main difference with the cases of lower $N_0$ is that $K_1$ must grow beyond a certain threshold before the system exits the bistability region.}

\subsection{Independently varied coupling strengths in allocation analysis}\label{sec:allocation_study_free_gains}

\begin{figure}[p]
    \centering
    \includegraphics[scale=1]{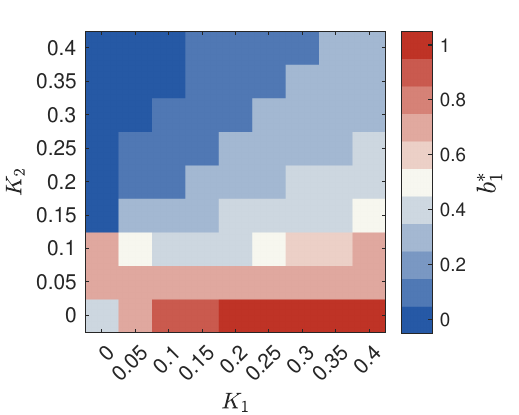}
    \caption{
    {Optimal value $b_1^*$ of the link allocation fraction $b_1$, in terms of maximizing $R$.
    $K_1$ and $K_2$ are the coupling strengths of pairwise and $3$-body interactions, respectively.
    The grid $(K_1,K_2)$ is $9\times9$.
    For each combination of gains, $N_{\C{H}} = 100$ random hypergraphs were generated, with $N_0=10$ nodes, for each value of $b_1 \in \{0, 0.1, \dots, 1\}$ (11 values).    
    Frequencies were drawn uniformly at random from $[0, 0.3]$, and initial phases from $[0, \theta_0^\R{max}]$---$\theta_0^\R{max} = \frac{\pi}{2}$.
    The relative link cost is fixed at $c_1 = 1$, while 
    triangle cost is $c_2 = 3$. 
    For $(K_1, K_2) = (0, 0)$, $R$ is independent of $b_1$, thus $b_1^*$ is meaningless in that case.
    This Figure is obtained in the same setup considered for Figure~\ref{fig:alloc_study_standard_coupl_omega03_R} of the Main Text.
    }}\label{fig:alloc_study_free_couplings}
\end{figure}

{For the hyperedge allocation analysis of Section~\ref{sec:results2}, Figure~\ref{fig:alloc_study_free_couplings} shows the optimal link allocation fraction $b_1^*$ when $K_1$ and $K_2$ vary freely, rather than being constrained to a constant sum. 
The results confirm that maximizing $R$ almost always calls for an intermediate $b_1$, i.e., splitting the budget between links and triangles.%
\footnote{We verified that the non-monotonicity observed for $K_2 = 0.1$ is due to small fluctuations in $R$ due to finite sampling of the hypergraphs.}
Allocating the entire budget to a single interaction type is optimal only when the coupling strength of the other type is zero or negligibly small in comparison.}


\end{document}